\documentclass[journal,comsoc]{IEEEtran}

\usepackage[T1]{fontenc}

\ifCLASSINFOpdf
\else
\fi

\usepackage{amsmath}
\usepackage{comment} %
\usepackage{cite}
\usepackage{graphicx}
\usepackage{epsfig,graphics,subfigure,psfrag,amsmath,amssymb}
\usepackage{booktabs}
\usepackage{amsfonts}
\usepackage{epstopdf}
\usepackage{amssymb}
\usepackage{array}
\usepackage{amsmath}
\usepackage{makecell}
\usepackage{multirow}

\begin{document}
%
\title{A Survey of Beam Management for mmWave and THz Communications Towards 6G}

\author{Qing~Xue,~\IEEEmembership{Member,~IEEE,}
	Chengwang~Ji,
	Shaodan~Ma,~\IEEEmembership{Senior Member,~IEEE,}
	Jiajia~Guo,~\IEEEmembership{Member,~IEEE,}
	Yongjun~Xu,~\IEEEmembership{Senior Member,~IEEE,}
	Qianbin~Chen,~\IEEEmembership{Senior Member,~IEEE,}
	and Wei~Zhang,~\IEEEmembership{Fellow,~IEEE}
	
    \thanks{Manuscript received XXX XX, 2023; revised XXX XX, 202X. This work was supported in part by the National Natural Science Foundation of China under Grants U23A20279, 62001071, 62061007, and 62271094, and in part by the Science and Technology Research Program of Chongqing Municipal Education Commission under Grant KJQN202200617. The work of S. Ma was supported in part by the Science and Technology Development Fund, Macau SAR under Grants 0087/2022/AFJ and 001/2024/SKL, in part by the National Natural Science Foundation of China under Grant 62261160650, and in part by the Research Committee of University of Macau under Grants MYRG-GRG2023-00116-FST-UMDF and MYRG2020-00095-FST. The work of W. Zhang was supported in part by the Key Area Research and Development Program of Guangdong Province under Grant 2020B0101110003. \emph{(Corresponding author: Shaodan Ma.)}}
	
    \thanks{Qing Xue is with the School of Communications and Information Engineering, Chongqing University of Posts and Telecommunications, Chongqing 400065, China, and was also  with the State Key Laboratory of Internet of Things for Smart City, University of Macau, Macao SAR, China (e-mail: xueq@cqupt.edu.cn).}
	
	\thanks{Chengwang Ji and Shaodan Ma are with the State Key Laboratory of Internet of Things for Smart City and the Department of Electrical and Computer Engineering, University of Macau, Macao SAR, China (e-mails: yc27411@connect.um.edu.mo; shaodanma@um.edu.mo).}
	
	\thanks{Jiajia Guo is with the State Key Laboratory of Internet of Things for Smart City, University of Macau, Macao SAR, China (e-mail: jiajiaguo@um.edu.mo).}
	
	\thanks{Yongjun Xu and Qianbin Chen are with the School of Communications and Information Engineering, Chongqing University of Posts and Telecommunications, Chongqing 400065, China (e-mails: xuyj@cqupt.edu.cn; chenqb@cqupt.edu.cn).}
		
	\thanks{Wei Zhang is with School of Electrical Engineering and Telecommunications, The University of New South Wales, Sydney, NSW 2052, Australia (e-mail: w.zhang@unsw.edu.au).}
}

\maketitle

\begin{abstract}
	Communication in millimeter wave (mmWave) and even terahertz (THz) frequency bands is ushering in a new era of wireless communications. Beam management, namely initial access and beam tracking, has been recognized as an essential technique to ensure robust mmWave/THz communications, especially for mobile scenarios. However, narrow beams at higher carrier frequency lead to huge beam measurement overhead, which has a negative impact on beam acquisition and tracking. In addition, the beam management process is further complicated by the fluctuation of mmWave/THz channels, the random movement patterns of users, and the dynamic changes in the environment. For mmWave and THz communications toward 6G, we have witnessed a substantial increase in research and industrial attention on artificial intelligence (AI), reconfigurable intelligent surface (RIS), and integrated sensing and communications (ISAC). The introduction of these enabling technologies presents both open opportunities and unique challenges for beam management. In this paper, we present a comprehensive survey on mmWave and THz beam management. Further, we give some insights on technical challenges and future research directions in this promising area.
\end{abstract}


\IEEEpeerreviewmaketitle

\section{Introduction}
In order to meet the expected growth of wireless data traffic, millimeter wave (mmWave) technology, operating in about 30-100 GHz Radio Frequency (RF) band, has become one of the promising candidates for indoor and outdoor wireless communications. This is evident from the emergence of standards, which mainly include IEEE 802.11ad \cite{IEEE-Std-802.11ad} and its evolution IEEE 802.11ay \cite{IEEE-Std-802.11ay} for wireless local area networks (WLANs), IEEE 802.15.3c \cite{IEEE-Std-802.15.3c} for wireless personal area networks (WPANs), and a series of Releases standardized by the 3rd Generation Partnership Project (3GPP) for the 5th generation (5G) New Radio (NR) access networks. Since 3GPP completed the first release of 5G NR in its Release 15 \cite{3GPP-TR-21.915-Release15}, the 5G evolution has progressed swiftly in Releases 16 and 17 \cite{3GPP-TR-21.916-Release16,3GPP-TR-21.917-Release17}. Now 3GPP is entering the second stage of the 5G NR standardization, which is also known as 5G-Advanced and will be specified in Release 18 \cite{Overview-5G-Advanced-3GPP-Release18}. Although mmWave communications can greatly improve network speed and capacity, it may still not meet the future growth demand of wireless data traffic. It is because that, each year, various new devices in different form factors with increased capabilities and intelligence are introduced and adopted in the market. In order to meet the growing demand of the wireless communication industry and support Tbps-level data rates, the exploration and research on the RF spectrum with higher speed and greater bandwidth have been stimulated. Among the available frequency bands, the Terahertz (THz) band, which nominally occupies the spectrum of 0.1-10 THz, has attracted extensive attention. The first standard for wireless communications over the THz band, IEEE 802.15.3d \cite{IEEE-Std-802.15.3d}, was officially approved in Fall 2017, where the defined switched point-to-point links operate at the sub-THz frequency around 300 GHz and enable data rates of up to 100 Gbit/s. The THz communications are envisioned as a key technology to fulfill future demands for beyond 5G such as the 6th generation (6G) wireless systems. In order to provide seamless high-quality services, \emph{beam management}, which collectively encompasses initial beam training/alignment, monitoring and tracking, as well as recovery from beam failures, is crucial for 6G mmWave and THz communications.

\subsection{Review: 6G Vision}
In response to the tremendous emergence of smart devices and the rapid expansion of Internet of Things (IoT) networks, 5G has made tradeoffs in terms of throughput, latency, energy efficiency, deployment cost, hardware complexity, and end-to-end reliability. Facing 2030 and beyond, 6G will be developed to jointly fulfill strict network requirements in a holistic fashion. 6G is envisioned to provide autonomous, ultra-large-scale, extremely dynamic, and fully intelligent services with high quality of experience (QoE). By now, there are a number of surveys on 6G networks, e.g.,  \cite{6G-Wireless-Networks-MVT-2019,A-Vision-6G-MNET-2020,Toward-6G-Networks-MCOM-2020,6G-Wireless-Systems-JPROC-2021,On-the-Road-to-6G,DCN-CHEN2020312}. We here briefly review the core requirements and enabling technologies of 6G.

For 5G, the usage scenarios recommended by International Telecommunication Union (ITU)-Radiocommunication Sector (ITU-R) can be classified as enhanced mobile broadband (eMBB), ultra-reliable low-latency communications (URLLC), and massive machine-type communications (mMTC) \cite{IMT2020-2015}. With the advent of new technologies and the continuous evolution of existing technologies, many unprecedented applications can be cultivated in 6G era. The potential usage scenarios in 6G communication systems are expected to be featured by ubiquitous mobile ultra-broadband, extremely reliable and low-latency communications, ultra-mMTC, ultra-high data density, and extremely low-power communications. The key performance indicators (KPIs) for evaluating 6G wireless networks include:

$\circ$ \emph{Peak data rate: $\ge 1 Tbit/s$}

$\circ$ \emph{User experienced data rate: $10\sim100 Gbit/s$}

$\circ$ \emph{Over-the-air latency: $0.01\sim0.1 ms$}

$\circ$ \emph{Mobility: $\ge 1000 km/h$}

$\circ$ \emph{Reliability: $ 99.99999\%$}

$\circ$ \emph{Connection density: $10^8\sim10^9 devices/km^3$}

$\circ$ \emph{Network energy efficiency: $10\sim100$ times that of 5G}

$\circ$ \emph{Area traffic capacity: $0.1\sim10 Gbit/s/m^3$}

$\circ$ \emph{Spectrum efficiency: at least $2\sim3$ times that of 5G}

$\circ$ \emph{Positioning accuracy: $10 cm$ indoor and $1 m$ outdoor}

To well achieve these KPIs, 6G is expected to integrate several new capabilities currently not covered by wireless communication systems, such as

\hangafter 1
\hangindent 0.9em
\noindent
$\bullet$ \emph{Ubiquitous three-dimensional (3D) connectivity}: Integrating terrestrial, airborne (e.g., unmanned aerial vehicles, UAVs), and satellite networks into a single wireless system will be essential for 6G \cite{Space-Air-Ground-Survey-2018}. The network coverage will be globally ubiquitous and will be shifted from two-dimensional in traditional terrestrial networks to 3D in a space-air-ground integrated network.

\hangafter 1
\hangindent 0.9em
\noindent
$\bullet$ \emph{Collective network intelligence}: As mobile networks are increasingly sophisticated and heterogeneous, many optimization tasks become intractable, which offers an opportunity for artificial intelligence (AI) \cite{Edge-Artificial-Intelligence-6G-2022}, more specifically machine learning (ML) techniques. Additionally, 6G will bring intelligence from centralized computing facilities to end terminals to provide distributed autonomy, while improving security, secrecy, and privacy.

\hangafter 1
\hangindent 0.9em
\noindent
$\bullet$ \emph{Integration of sensing and communication}: The function of mobile network is evolving from communication only towards integrated sensing and communications (ISAC) (a.k.a joint communication and radar/radio sensing, JCAS) \cite{Integrated-Sensing-Communications-2022}. 6G networks will exploit a unified interface for sensing/localization and communications to improve control operations relying on context information.

\hangafter 1
\hangindent 0.9em
\noindent
$\bullet$ \emph{Communication with large intelligent surfaces}: Reconfigurable intelligent surface (RIS) \cite{Smart-Radio-Environments-RIS-2020}, also known under the names intelligent reflecting surface (IRS) and software-controlled metasurface, has positive significance in solving the pain points of small cells such as non-line-of-sight (NLoS) transmissions and reducing coverage holes. Large-scale RIS is envisaged as the massive multiple-input multiple-output (MIMO) technology 2.0 in 6G.

\hangafter 1
\hangindent 0.9em
\noindent
$\bullet$ \emph{Other new paradigms} empowered by blockchain, digital twin, quantum computing and communications, etc.

6G wireless network is promising to significantly improve QoE and support a sustainable future. The unprecedented new technological trends will shape its performance goals.

\subsection{Motivation}
For 6G mmWave and THz communication systems, ultra-large-scale antenna arrays may be equipped at both base station (BS) and user equipment (UE) sides, such that high path loss can be compensated by generating narrow beams with strong beamforming gains. This leads to the fact that mmWave and THz communications rely heavily on beam management (beam training/alignment/tracking) to select the appropriate beam(s) during intra/inter-cell mobility in a quick manner to avoid any beam misalignment (transmit-receive beams do not point to each other) or beam/link failure. However, beam management is challenging because of the unfavorable propagation characteristics of mmWave/THz signals. Traditional schemes such as exhaustive search suffer from severe overhead, increased communication delay, and degraded spectral efficiency. To address these issues, various beam management mechanisms have been proposed to reduce beam training overhead or to enhance beam alignment accuracy \cite{Beamspace-SU-MIMO-2017,Beamspace-MU-MIMO-2019,User-Centric-Association-Ultra-Dense-mmWave-2021,XUE20221115,Beam-Management-Ultra-Dense-mmWave-Network-2023,xue2023aiml}. The typical mechanisms for conventional mmWave networks have been investigated by several survey papers, which will be introduced in detail later. But after these surveys were published, a number of new beam management solutions based on 6G technology enablers have sprung up like mushrooms. In addition, the survey on THz beam management is now extremely lacking. Motivated by these facts, we conduct in-depth research on the latest 6G beam management schemes.

For mmWave and THz communications toward 6G, the most powerful tools are the AI, RIS, and ISAC. In this regard, four key trends are as follows.

\hangafter 1
\hangindent 0.9em
\noindent
$\bullet$ \emph{Empowered by AI.} AI-empowered beam management can quickly adjust beam parameters to continuously provide considerable QoE and maintain network health by monitoring real-time network dynamics. In addition to the popular deep learning (DL) \cite{Deep-Learning-Physical-Layer}, a few cutting-edge AI techniques represented by federated learning (FL) and transfer learning (TL) are beginning to show strong potential in 6G beam management because they pose a chance to distributed collaborative learning tasks and improve training speed.

\hangafter 1
\hangindent 0.9em
\noindent
$\bullet$ \emph{Enabled by Sensing.} In 6G, sensing function will be tightly integrated with communications to support autonomous systems, which motivates the recent research theme of ISAC. The sensing function can be used to track UE for beam prediction/tracking in mmWave/THz communications. We expect that ISAC will become more popular in beam management in the near future, especially in practical dynamic environments where wireless channels vary fast.

\hangafter 1
\hangindent 0.9em
\noindent
$\bullet$ \emph{Enhanced by RIS.} Among all potential candidates to alleviate blockage (channel drop caused by obstacles, device movement, or rotation) and expand coverage of mmWave/THz communications, RIS is widely considered promising for 6G as it offers a cost-effective solution. In addition to the benefits, the deployment of RIS complicates the system architecture and poses a significant challenge for beam management which coordinates BS-RIS to jointly manage two-hop transmission.

\hangafter 1
\hangindent 0.9em
\noindent
$\bullet$ \emph{Driven by Combination.} In some scenarios, AI, RIS and sensing functions may be coupled with each other, bringing new opportunities and challenges to beam management.

As new technologies constantly emerge to improve the future generation communication networks, some existing techniques in beam management may become under-performing which directly challenges the overall network performance. The research interest in beam management will be renewed to seek improved or novel solutions combined with emerging technologies. This area is concerned by several interesting research work to date but remains a very hot topic. In this context, we survey the state-of-the-art by pursuing the above four avenues.

\subsection{Our Survey Scope and Contributions}
This paper conducts a comprehensive and dedicated survey on the emerging research achievements related to tackling a set of beam management problems in 6G mmWave and THz networks and applications, and discusses the unsolved challenges and open research avenues. The key contributions are highlighted as follows.

1) \emph{This review surveys the advancements in beam management for mmWave and THz communication systems, where the integration of AI, sensing, and RIS is thoroughly examined.} Whereas most existing survey papers on the subject tend to focus solely on mmWave communications, with only cursory attention to THz communications, our review delves into both. Furthermore, while these surveys typically summarize some AI-driven technologies, they often overlook the burgeoning research on cutting-edge technologies in the fields of ISAC and RIS-assisted systems. To our knowledge, our review is pioneering in offering a comprehensive examination of the latest research aimed at enhancing beam management performance for both mmWave and THz communications. Moreover, it illuminates emerging trends that exploit AI, ISAC, and RIS methodologies for these advanced communication systems.

2) \emph{In this review, we extend our scope beyond single-agent configurations to explore collaborative beam management strategies involving multiple agents or tasks.} AI techniques, recognized for their capacity to discern and adapt to intricate environmental dynamics, have emerged as effective solutions for beam management. While previous papers have summarized AI-based beam management schemes, they predominantly address single-agent contexts. Building upon this, our survey not only probes new developments for the single agent but also delves into in-depth analysis of multi-agent, multitasking collaboration scenarios that leverage FL, TL, and split learning. Given the increasing complexity of wireless communication environments, collaborative learning presents a potentially invaluable approach to addressing the challenges of achieving high-precision and low-latency beam management.

3) \emph{This paper marks the first-ever review of predictive beam management solutions for ISAC systems found in the literature.} The precise alignment and tracking of beams, particularly those that are highly focused, pose a significant challenge in mmWave/THz communication for high-mobility scenarios. Predictive beam management that leverages ISAC (sensing) capabilities has demonstrated considerable promise, outperforming traditional communication-only protocols, especially in meeting stringent latency demands. A variety of research endeavors have been pursued in this domain. We dedicate ourselves to the thorough research and analysis of existing efforts in ISAC-enabled beam management, a subject that, until now, has not been explored in existing works.

4) \emph{This survey represents a pioneering exploration of beam management within RIS-assisted mmWave and THz communication systems.} Positioned as a key enabler for 6G technology, RIS has the potential to mitigate the issues of signal and channel blockage prevalent in mmWave/THz networks by intelligently reshaping the propagation environment. As an emerging paradigm, RIS-assisted systems offer novel solutions to overcome blockage challenges, yet they also introduce distinct and unprecedented challenges in the context of beam management. This survey provides an examination of the latest beam management methodologies for RIS-assisted mmWave/THz systems, highlighting their respective strengths and limitations. It pays special attention to the role of AI-driven and sensing-aided algorithms, uncovering the exciting new directions emerging in this research area. To date, very few surveys have broached this topic.

5) This review meticulously examines over 150 scholarly papers in the domain of beam management. \emph{It encapsulates the lessons learned, delineates the persistent challenges that remain open, and charts potential avenues for future research.} Our goal is to furnish readers with a comprehensive synopsis of the accomplishments to date, as well as to outline the prevailing research trajectories involving the application of AI, RIS, and ISAC (sensing) to enhance beam management efficacy. We are confident that this scholarly survey will serve as a valuable resource for both novices and seasoned professionals in the sphere of 6G beam management. It will undoubtedly inspire readers to generate innovative ideas within this evolving field of research.

\begin{figure}[t]
	\centering
	\includegraphics[width=8.8cm]{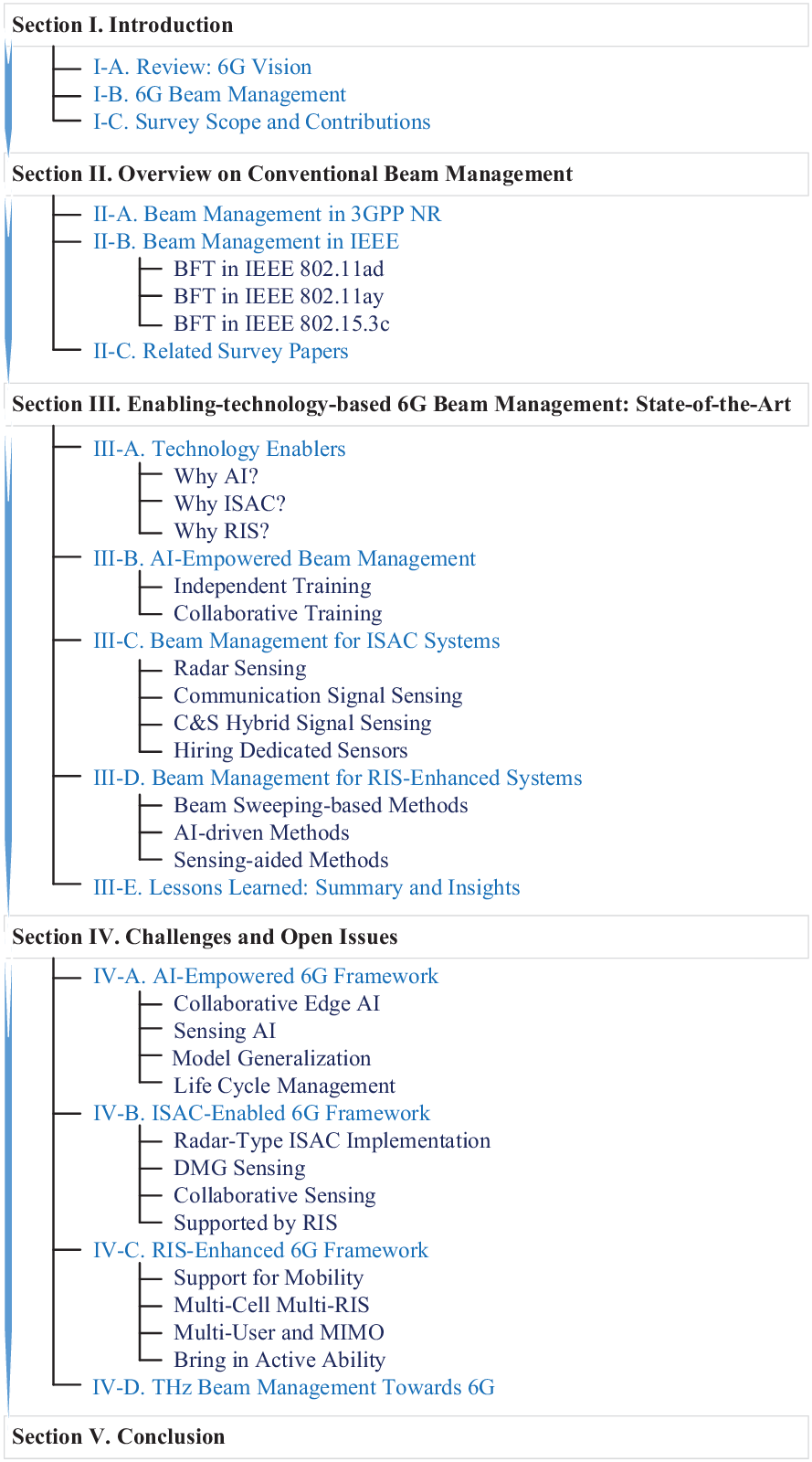}
	\caption{The outline of this survey paper.}
	\label{fig:Survey-outline}
\end{figure}

The rest of this paper is organized as follows. Section II provides an overview of the beam management procedures in 3GPP NR and IEEE standards and summarizes the related survey papers. Section III carries out an extensive literature review on the state-of-the-art 6G beam management approaches based on emerging technologies, including AI, RIS, and ISAC. The possible avenues for future work are discussed in Section IV. Finally, we conclude the paper in Section V. An overview of the outline of this survey paper is illustrated in Fig.~\ref{fig:Survey-outline}.

\begin{table*}[htbp]
	\centering
	\caption{Key Terms and Descriptions in Beam Management.\label{view-beam-Management}}
	\scalebox{1.0}{
		\begin{tabular}{ll}
			\Xhline{0.5pt}
			\Xhline{0.5pt}
			\textbf{Term} &\textbf{Description}\\
			\Xhline{0.5pt}
			Beam &The main lobe of the radiation pattern of an antenna array.\\
			Beam pair &A pair of beams (transmit and receive patterns) for transmission and corresponding reception.\\
			Beam management ({\tiny 3GPP}) &A set of beam-related procedures for fine alignment of the transmitter and receiver beams.\\
			Beamforming  &A signal processing technique used in antenna arrays for directional signal transmission or reception.\\
			Beam alignment &The operations of searching for candidate beam directions at the transmitter and/or receiver to find the optimal beam pair.\\
			Beam training &A set of operations to estimate the beam steering directions for beam alignment.\\
			Beam tracking &The priori-aided beam training to track beams (or channels) in mobile environment.\\
			Beam steering &A technique for changing the direction of the main lobe of a beam (radiation pattern).\\
			\Xhline{0.5pt}
			\Xhline{0.5pt}
		\end{tabular}
	}
\end{table*}

\section{Overview on Conventional Beam Management}
Several standardized organizations are involved in the development and standardization of beam management techniques within wireless communication systems. These organizations play a critical role in establishing interoperability, defining best practices, and facilitating the adoption of beam management solutions across diverse industry stakeholders. Some of the key organizations and their focus on beam management include: \emph{3GPP:} Develops protocols for mobile telecommunications, including those for 5G NR which incorporates sophisticated beam management techniques to enhance network capacity and efficiency. \emph{IEEE:} Through its 802.11 Working Group, it has developed standards such as 802.11ad and 802.11ay for Wi-Fi networks, which include beamforming to improve signal quality and data rates. The IEEE 802.15 Working Group has also developed standards that utilize beamforming for personal area networks and short-range communication. \emph{International Telecommunication Union (ITU):} Plays a crucial role in allocating global radio spectrum and satellite orbits, ensuring the compatibility of systems and standards for telecommunications and information technologies, including those using beamforming technologies. \emph{European Telecommunications Standards Institute (ETSI):} Contributes to the standardization of Information and Communication Technologies (ICT) within Europe, with global applicability. It has technical committees working on standards that encapsulate beam management. \emph{Internet Engineering Task Force (IETF):} While not specifically focused on beam management for wireless networks, it plays a significant role in the broader context of network management and could influence aspects related to the management of beams in communication infrastructures. These organizations work collaboratively and sometimes in competition to develop and refine standards that can accommodate the latest technologies, with AI-empowered beam management becoming increasingly important as wireless technologies evolve.

Beam management aims to acquire and maintain a set of transmitter and/or receiver beams which can be used for downlink and/or uplink transmission/reception. To better understand the terminology, descriptions of the key terms in beam management are given in Table~\ref{view-beam-Management}. In this section, we first provide the overview on beam management procedure for 3GPP NR cellular networks and further present some expected enhancements considered for the future NR standard. We then elaborate on the beam training and tracking scheme designed for WLANs and WPANs in IEEE 802.11ad, 802.11ay, and 802.15.3c. After that, we make a comparative analysis of the existing survey papers on beam management.

\begin{figure}[t]
	\centering
	\includegraphics[width=9cm]{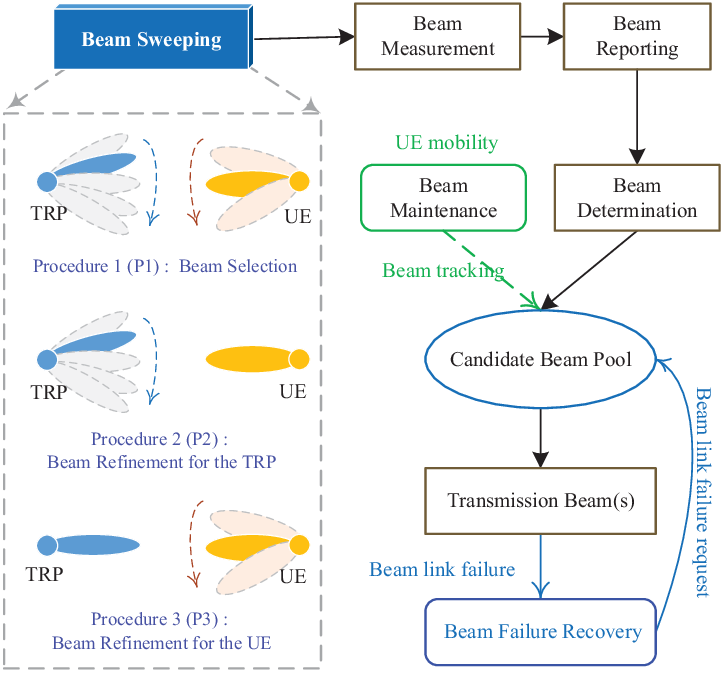}
	\caption{3GPP beam management procedure.}
	\label{fig:3GPP-BM}
\end{figure}

\subsection{Beam Management in 3GPP NR}
In 3GPP NR, beam management is defined as a set of basic beam-related procedures in the physical (PHY)/medium access control (MAC) layer, as shown in Fig.~\ref{fig:3GPP-BM}, including at least the following four operations \cite{3GPP-TR-38802}.

\hangindent 2.0em
\checkmark \emph{Beam sweeping}: an operation of covering a spatial area, with beams transmitted and/or received during a time interval in a predetermined way.

\hangindent 2.0em
\checkmark \emph{Beam measurement}: an operation for transmission-reception points (TRPs) or UE to measure characteristics of received and/or transmitted beamformed signals.

\hangindent 2.0em
\checkmark \emph{Beam reporting}: an operation for UE to report information of beamformed signal(s) based on beam measurement.

\hangindent 2.0em
\checkmark \emph{Beam determination}: an operation for TRP(s) or UE to select at least one of its own transmit/receive beam(s).

The measurement process is carried out with an exhaustive search, also known as beam sweeping, at pre-specified intervals and directions. During beam sweeping, the TRP/UE measures the received power of beamformed reference signals (RSs) to derive the beam quality, which is typically expressed in terms of reference signal received power (RSRP). Measurement of the signal-to-interference-plus-noise ratio (SINR) of RSs is also supported in Release 16 of 5G \cite{Six-Key-Challenges-Beam-Management-2021}. The measurement results (beam quality and beam decision information) of the UE will be sent to the BS. Based on the measurement, the TRP/UE selects the optimal beam (or set of beams) to set up a directional communication link. These beam management operations are periodically repeated to update the optimal beam pair(s) over time. Further details of the beam management procedure in 5G NR can be found, for example, in \cite{Beam-Management-2020}. During communication, the environment may change, which may lead to beam failure. A beam failure event (link blockage or beam misalignment) occurs when the quality of beam-pair link(s) of an associated control channel falls low enough (e.g., comparison with a threshold, time-out of an associated timer). In general, beam misalignment between the transmitter and receiver is most likely to occur with narrow beamwidth, especially in a high mobile network. 3GPP NR supports that UE can trigger mechanism to recover from beam failure. This procedure is known as \emph{beam recovery} which can be broken down into four steps: beam failure detection, candidate beam identification, recovery request transmission, and monitoring response for recovery request. To accomplish this, the NR UE continuously monitors whether any beam failure condition is triggered. Once beam failure is detected, the UE proceeds to identify the candidate beam(s) that can restore the connectivity. When one or more candidate beams are found, the UE transmits a beam failure recovery request carrying an identifier of the new beam to the TRP. After that, the UE  monitors the corresponding control channel to receive a response to the recovery request, which is sent by the TRP with the new beam identified by the UE. If the response is successfully received, the beam recovery procedure is successful and a new beam pair link is established for subsequent communication. Otherwise, the UE may perform additional beam recovery requests. If it still fails, the UE initiates this beam recovery procedure, which possibly includes cell re-selection.

For an idle UE, who accesses the network for the first time, beam management is the basis of designing a directional initial access strategy. After the initial access, beam management is required to maintain the link by updating the steering direction and the shape of beam as the channel changes dynamically when the UE moves. This procedure is generally known as \emph{beam tracking}. Thereby, the beam management can be categorized as initial beam establishment and beam tracking (beam maintenance). At present, 3GPP continues to evolve 5G NR. The beam management related procedures that can be further studied in the current 3GPP technical specifications are presented as follows.

1) The requirements for beam determination procedure related to selecting at least one of its own transmit/receive beam(s) need to be further studied by the 3GPP technical specification group (TSG) radio access network (RAN) work group (WG) 4, which is commonly abbreviated to RAN4 \cite{3GPP-TR-38803}. Meanwhile, the detailed beam measurement requirements, e.g., beam measurement period and accuracy, need to be further studied since the physical layer design has not been finalized. The potential beam reporting requirements including at least reporting delay also need to be studied by RAN4. Moreover, whether the requirements of radio resource management for beam sweeping at both NR UE and TRP sides are needed needs further study.

2) Larger antenna arrays are expected to support certain link budgets in high frequencies. Methods for managing narrower beams and a larger number of beams should be studied \cite{3GPP-TR-38807}. Meanwhile, enhancement on discovery and tracking to support various beam assumptions should be investigated, as the number of supported beams and the beam codebook space may vary depending on the form factor and device.

3) Beam management is one of the use cases for applying AI/ML to NR air interface in 3GPP Release 18 \cite{3GPP-TR-38843}. In this case, beam management utilizing AI/ML is used for beam prediction in the time and/or spatial domain  to reduce overhead and latency, and to improve the precision of beam selection. The details of the finalized 3GPP Release 18 have not yet been publicly disclosed. In \cite{xue2023aiml}, we offer some general insights into what beam management might entail in the context of 5G-Advanced, based on previous releases and ongoing trends in the industry.

4) For NR supporting non-terrestrial networks, the beam management procedures may need to be modified \cite{3GPP-TR-38811}. NR beam management for mobility between spot-beams on the same BS cannot be ported to satellite to minimize the handoff overhead, because it usually assumes that adjacent beams of the same BS can use the same frequency, but for adjacent beams of the same satellite, different frequencies or different polarizations may be used.

5) For NR Integrated Access and Backhaul (IAB), beam failure recovery and radio link failure procedures are beneficial and should be supported and enhanced \cite{3GPP-TR-38874}. It mainly includes two enhancements that should be considered for IAB nodes: Enhancement to support interaction between beam failure recovery success indication and radio link failure; Enhancement to existing beam management procedures for faster beam switching/coordination/recovery to avoid backhaul link outages.

\subsection{Beam Management in IEEE}
MmWave frequency, particularly at unlicensed 60 GHz band, has received great attention in the last decade from WLAN and WPAN communities. Specifically, three IEEE standards of mmWave have been designed, i.e., IEEE 802.11ad, 802.11ay, and 802.15.3c, where the 802.11ay is an enhancement of the 802.11ad standard. In these IEEE standards, the procedure for beam alignment is known as beamforming training (BFT). The BFT aims at selecting the best beamforming/combining vectors from a pre-defined codebook without explicit estimation of the channel. A two-stage BFT operation is utilized: coarse-grained beam training (named sector\footnote{A sector is a specific wide antenna radiation pattern generated by changing the antenna weights applied to the phased array elements.} sweeping in 802.11ad/ay and low-resolution beam training in 802.15.3c) and fine-grained beam training (named beam refinement in 802.11ad/ay and high-resolution beam training in 802.15.3c).

$\bigstar$ \textbf{BFT in IEEE 802.11ad} \cite{IEEE-Std-802.11ad}: To achieve the necessary directional multi-gigabit (DMG) link budget for subsequent communication between two stations\footnote{IEEE 802.11-2007 formally defines STA as: Any device that contains an IEEE 802.11-conformant MAC and PHY interface to the wireless medium.} (STAs), the BFT in 802.11ad WLAN is split into two phases: sector-level sweep (SLS) and beam refinement protocol (BRP). The BFT starts with a SLS from the initiator, who refers to the DMG STA that commences the beamforming. A BRP may follow, if requested by either the initiator or the responder (the DMG STA that is the recipient). After the successful completion of BFT, the best refined beam pair is used for communication between the two participating DMG STAs. During data transmission, \emph{beam tracking} is optionally employed to adjust for channel variations. In 802.11ad, the beam tracking is performed by appending training fields for channel estimation to data packets.

\hangafter 1
\hangindent 1.0em
\noindent
$\bullet$ \emph{SLS phase}: During this phase, the initiator periodically transmits sector-sweep frames from pre-defined sectors, while the responder remains in omnidirectional mode and provides feedback to the initiator on the sector received with the highest quality during the initiator sector-sweep. Subsequently, the two DMG STAs swap roles. It is worth noting that SLS is mandatory and only trains the transmit sectors of the initiator and the responder. Since imperfect quasi-omni antenna patterns are used to find the best sectors, SLS can only provide coarse-grained beam training.

\hangafter 1
\hangindent 1.0em
\noindent
$\bullet$ \emph{BRP phase}: After SLS, the transmit and receive antenna configurations can be optionally improved in BRP using an iterative refinement procedure. To further refine the beam pattern/direction, the BRP phase is implemented through three main subphases, i.e., BRP setup, multiple sector identifier detection (MID), and beam combining (BC), as well as one or more beam refinement transactions. More specifically, the MID reverses the scanning roles from the transmit sector sweep, and the BC tests the transmit and receive beams in pairwise combinations.

$\bigstar$ \textbf{BFT in IEEE 802.11ay} \cite{IEEE-Std-802.11ay}: The basic procedure of BFT in 802.11ay is almost the same as that in 802.11ad standard. However, compared with 802.11ad, 802.11ay has further enhanced and optimized the original DMG beamforming, making it more flexible and efficient, and named it enhanced DMG (EDMG) beamforming. One of the major advancements is that the EDMG beamforming enables MIMO operations, including both single-user MIMO (SU-MIMO) and downlink multi-user MIMO (MU-MIMO), thus achieving both beamforming and multiplexing gain. To enable MIMO communication between an initiator and one or more responders, the SU/MU-MIMO BFT protocol comprises two consecutive phases: single-input single-output (SISO) phase and MIMO phase. Here the initiator and responder are all EDMG STAs that are SU/MU-MIMO capable. In addition, there are two types of beam tracking in 802.11ay. One is analog beam tracking for DMG STAs, which is similar to that in 802.11ad. The other is digital baseband channel tracking using digital beamforming for EDMG STAs.

\hangafter 1
\hangindent 1.0em
\noindent
$\bullet$ \emph{SU-MIMO BFT}: The SISO phase comprises either a MIMO BRP transmit sector sweep (TXSS) procedure or a SISO feedback procedure, where the former is optional, and the latter is mandatory. The purpose is to enable the initiator to collect feedback of the last initiator TXSS (I-TXSS) from the responder, and also enable the responder to collect feedback of the last responder TXSS (R-TXSS) from the initiator. The MIMO phase enables the
simultaneous training of transmit and receive sectors and DMG antennas to determine best combinations of transmit and receive sectors and DMG antennas for SU-MIMO operation.

\hangafter 1
\hangindent 1.0em
\noindent
$\bullet$ \emph{MU-MIMO BFT}: The SISO phase starts with an optional I-TXSS subphase, followed by a mandatory feedback subphase. During this phase, the initiator and the responders who intend to participate in a MU group need to train transmit and receive DMG antennas and sectors and collect feedback. Based on SISO feedback, the initiator is required to send a number of action frames to the responders to perform the MIMO phase, which is comprised of four consecutive subphases, namely, beamforming setup, BFT, beamforming feedback, and beamforming selection.

$\bigstar$ \textbf{BFT in IEEE 802.15.3c} \cite{IEEE-Std-802.15.3c}: To find the best pair of beam patterns between two devices (DEVs) with a given beam resolution, a two-level beam training is used in 802.15.3c WPAN, which consists of a sector level and beam level training. The space area of interest will be limited by sector level training, and then the best pair of sectors (or low-resolution beams) will be sliced into high-resolution beams to prepare for beam level training. Since the channel characteristics are time-varying, 802.15.3c, similar to 802.11ad/ay, also contains a beam tracking phase, which is used to track the best beam patterns between the two DEVs and to improve connectivity.

\hangafter 1
\hangindent 1.0em
\noindent
$\bullet$ \emph{Sector level training}: During this phase, the transmitter (DEV1) sends sector training sequences over all possible sector directions sequentially, while the receiver (DVE2) attempts to listen to the training sequences using different receive directions/sectors. In this way, DVE2 is able to identify which transmit sector of DVE1 has the highest quality. The procedure is then reversed to determine the best transmit-receive sector pair.

\hangafter 1
\hangindent 1.0em
\noindent
$\bullet$ \emph{Beam level training}: This phase explores beams within the best sectors to find the best beam pair for the two DEVs. For that, DEV1 transmits repetitions of a beam training sequence over each possible beam direction within the obtained best transmit sector, while DEV2 monitors the training sequences using different receive beams and then selects the optimal beam pair, i.e., DVE1's best transmit beam and DVE2's best receive beam. After exchanging roles, a similar beam training from DVE2 to DVE1 takes place. Through beam feedback, the two DVEs can know their optimal beam directions.

\begin{table*}[t]
	\centering
	\caption{Existing surveys on beam-management-related topics and our new contributions.}\label{Related-surveys}
	\scalebox{0.9}{
		\begin{tabular}{|c|c|c|c|c|c|c|c|c|}
			\Xhline{0.5pt}
			\Xhline{0.5pt}
			\multicolumn{1}{|c|}{\multirow{2}{*}{\textbf{Ref.}}} & \multicolumn{1}{c}{\multirow{2}{*}{\textbf{Year}}} & \multicolumn{1}{|c|}{\multirow{2}{*}{\textbf{Focus of Discussion}}} & \multicolumn{2}{c|}{\textbf{Frequency}} & \multicolumn{3}{c|}{\textbf{Advanced Approaches}} & \multicolumn{1}{c|}{\multirow{2}{*}{\textbf{Type}}}\\
			\cline{4-5}\cline{6-8}
			\multicolumn{1}{|c|}{} & \multicolumn{1}{c}{} & \multicolumn{1}{|c|}{} &mmWave &THz &AI &RIS &Sensing &\multicolumn{1}{c|}{} \\
			\Xhline{0.5pt}
			{\makecell[c]{\cite{Beamforming-Survey-2016}}}	&2016 &{\makecell[lp{8.5cm}]{Tracked the evolution of beamforming for mmWave communications and focused on mmWave channel characteristics and indoor use.}} &\checkmark &  & & & &{\makecell[lp{2cm}]{Beamforming}}\\
			\Xhline{0.5pt}
			{\makecell[c]{\cite{Hybrid-Beamforming-Massive-MIMO-Survey-2017}}}	&2017 &{\makecell[lp{8.5cm}]{Provided a survey of hybrid beamforming in massive MIMO systems, and focused on hybrid beamforming structures on the basis of the CSI types (instantaneous or average).}} &\checkmark &  & & & &{\makecell[lp{2cm}]{Hybrid beamforming}}\\
			\Xhline{0.5pt}
			{\makecell[c]{\cite{Survey-Hybrid-Beamforming-2018}}}	&2018   &{\makecell[lp{8.5cm}]{Provided a review of hybrid beamforming in mmWave massive MIMO systems, and focused on the implementation, signal processing, and application aspects.}} &\checkmark &  & & & &{\makecell[lp{2cm}]{Hybrid beamforming}} \\
			\Xhline{0.5pt}
			{\makecell[c]{\cite{Modular-CSI-Beam-Management-2018}}}	&2018 &{\makecell[lp{8.5cm}]{Provided an overview of key features pertaining to CSI reporting and beam management for the 5G NR standardized in 3GPP Release 15.}} &\checkmark & & & & &{\makecell[lp{2cm}]{Beam measurement and recovery}}\\
			\Xhline{0.5pt}
			{\makecell[c]{\cite{802.11ay-2018}}}	&2018 &{\makecell[lp{8.5cm}]{Presented some technical challenges for mmWave communications in IEEE 802.11ay standardization activities, especially the enhancements of beam training to 802.11ad.}} &\checkmark & & & & &{\makecell[lp{2cm}]{Beam training and tracking}} \\
			\Xhline{0.5pt}
			{\makecell[c]{\cite{Tutorial-Beam-Management-3GPP-NR-2019}}}	&2019  &{\makecell[lp{8.5cm}]{Presented a tutorial on beam management frameworks for 3GPP NR and focused on the measurement techniques for beam and mobility management in Release 15.}} &\checkmark & & & & &{\makecell[lp{2cm}]{Initial access and beam tracking}}\\
			\Xhline{0.5pt}
			{\makecell[c]{\cite{Standalone-Non-Standalone-Beam-Management-3GPP-NR-2019}}}	&2019  &{\makecell[lp{8.5cm}]{Compared the performance of standalone and non-standalone deployments for the management of the beams of users in both connected and idle modes.}} &\checkmark & & & & &{\makecell[lp{2cm}]{Initial access and beam tracking}}\\
			\Xhline{0.5pt}
			{\makecell[c]{\cite{Beam-Management-2020}}} &2020  &{\makecell[lp{8.5cm}]{Provided an overview on  the standardized MIMO framework supporting beam management and CSI acquisition in 5G NR and focused on the beam management procedures standardized in 3GPP Releases 15 and 16.}} &\checkmark & &Partially & & &{\makecell[lp{2cm}]{Initial access and beam maintenance}}\\
			\Xhline{0.5pt}
			{\makecell[c]{\cite{Beam-Based-Mobility-V2X-Survey-2021}}}  &2021   &{\makecell[lp{8.5cm}]{Provided an overview of the beam-level and cell-level mobility management in 5G mmWave V2X communications and focused on overcoming mobility interruption.}} &\checkmark &  &Partially &Partially & &{\makecell[lp{2cm}]{Beam switching}}\\
			\Xhline{0.5pt}
			{\makecell[c]{\cite{Millimeter-Wave-Hardware-Beam-Management-2021}}}  &2021   &{\makecell[lp{8.5cm}]{Surveyed the mmWave system hardware technologies, and investigated the hierarchical search and Bayesian filter-based beam management.}} &\checkmark &  &Partially & & &{\makecell[lp{2cm}]{Beam alignment}} \\
			\Xhline{0.5pt}
			{\makecell[c]{\cite{Initial-Access-Beam-Alignment-mmWave-Terahertz-2022}}}  &2022  &{\makecell[lp{8.5cm}]{Presented a survey of beam alignment and initial access in mmWave and THz 5G/6G systems and highlighted the new trends towards deploying RISs and DL approaches.}}  &\checkmark &\checkmark &Partially &Partially & &{\makecell[lp{2cm}]{Initial access and beam alignment}} \\
			\Xhline{0.5pt}
			{\makecell[c]{\cite{Enabling-JCAS-survey-2022}}}  &2022  &{\makecell[lp{8.5cm}]{Provided a comprehensive review on PMN using the ISAC/JCAS techniques and investigated the sensing-assisted beamforming.}}  &\checkmark & & & &Partially &{\makecell[lp{2cm}]{Predictive beamforming}}\\
			\Xhline{0.5pt}
			{\makecell[c]{\cite{Machine-Learning-Beam-management-survey-2023}}}  &2023  &{\makecell[lp{8.5cm}]{Provided an overview of the existing ML-based mmWave/THz beam management techniques.}}  &\checkmark &\checkmark &Partially & & &{\makecell[lp{2cm}]{Beam prediction}} \\
			\Xhline{0.5pt}
			\multicolumn{2}{|c|}{\textbf{This work}} &{\makecell[lp{8.5cm}]{Presents a comprehensive survey on the state-of-the-art beam management schemes for 6G communications.}} &\checkmark &\checkmark &\checkmark &\checkmark &\checkmark &{\makecell[lp{2cm}]{Beam alignment, tracking, and prediction}}\\
			\Xhline{0.5pt}
			\Xhline{0.5pt}
		\end{tabular}
	}
\end{table*}

\subsection{Related Survey Papers}
In the early years, there were some reviews on beam management, especially beamforming, but they did not cover the new technologies that have emerged in recent years. At present, although there are some surveys on specific fields that involve beam management issues, the content is relatively limited. There are very few surveys dedicated to beam management, but most are only for AI, and some of the latest achievements such as multi-agent collaboration are not included. With the rapid development of technology, in addition to AI, some new advanced technologies have emerged, such as ISAC and RIS, for which no comprehensive review on beam management has been found.

Table~\ref{Related-surveys} displays a comparison between the survey paper at hand and a collection of other surveys covering beam-management-related literature. The findings of this study are analyzed below in more detail.

One of the key techniques of beam management is antenna beamforming, which is essentially a spatial filtering operation typically using an array of radiators to radiate or capture energy in a specific direction over its aperture. The gain realized through beamforming can compensate for the high free-space path loss and penetration loss of mmWave signals. Shajahan Kutty \emph{et al.} \cite{Beamforming-Survey-2016} presented a holistic view of antenna beamforming techniques till mid 2015 for mmWave communications. The scope includes description of mmWave beamforming architectures, signal processing algorithms, and RF system design and implementation aspects with particular emphasis on beamforming for IEEE 802.15.3c WPAN, IEEE 802.11ad WLAN and some outdoor applications. Beamforming architectures are grouped into three categories: analog, digital, and hybrid beamforming. To keep a low complexity and implementation cost, hybrid analog-digital beamforming has emerged as a leading contender for outdoor mmWave communications, especially in massive MIMO systems \cite{CLNet,CSI-Feedback-Massive-MIMO,Antenna-Selection}. Andreas F. Molisch \emph{et al.} \cite{Hybrid-Beamforming-Massive-MIMO-Survey-2017} provided a comprehensive survey of the various incarnations of such hybrid multiple-antenna transceiver structures that have been proposed in the literature till 2016. Irfan Ahmed \emph{et al.} \cite{Survey-Hybrid-Beamforming-2018} also provided a review of hybrid beamforming methods, both existing and proposed till the first quarter of 2017. In comparison, the previous survey on hybrid beamforming \cite{Hybrid-Beamforming-Massive-MIMO-Survey-2017} tends to focus on beamforming structures on the basis of the channel state information (CSI) types (instantaneous or average), while the latter paper \cite{Survey-Hybrid-Beamforming-2018} focuses on the implementation, signal processing, and application aspects of the hybrid beamforming.

Besides mmWave beamforming \cite{Successive-Localization-Beamforming}, some papers surveyed the beam management process in the aspects of 5G standardization. Eko Onggosanusi \emph{et al.} \cite{Modular-CSI-Beam-Management-2018} provided a good overview of modular and high-resolution CSI acquisition and beam management procedures such as measurement, reporting, and recovery for 5G NR globally standardized by the 3GPP in Release 15. In \cite{Tutorial-Beam-Management-3GPP-NR-2019}, which is a tutorial on the design and dimensioning of beam management frameworks for 5G cellular networks, Marco Giordani \emph{et al.} reviewed the most relevant downlink and uplink measurement signals supported by 3GPP Release 15 for beam management purposes and presented three measurement collection frameworks for both initial access and tracking purposes. The three considered measurement schemes are standalone-downlink scheme, non-standalone-downlink scheme, and non-standalone-uplink scheme, where the standalone/non-standalone architecture is a deployment configuration of NR networks. In the same year, they compared the performance of standalone and non-standalone deployments for the beam management of users in both connected and idle modes \cite{Standalone-Non-Standalone-Beam-Management-3GPP-NR-2019}. It showed that a non-standalone configuration exploiting multi-connectivity can offer more potential, including improving end-to-end performance in mmWave networks, ensuring higher resilience and improved reactiveness in case of link failure, and reducing the impact of beam reporting overhead. Meanwhile, not only 5G cellular networks, but also WLANs are widely concerned about mmWave communications. Several mmWave WLAN standards have been designed such as IEEE 802.11ad and its evolution standard IEEE 802.11ay. Pei Zhou \emph{et al.} \cite{802.11ay-2018} highlighted some MAC-related technologies in IEEE 802.11ad and presented some technical challenges for mmWave communications in the IEEE 802.11ay standardization activities, especially the beamforming training for both SU-MIMO and MU-MIMO.

These above papers provide an excellent survey on mmWave beam management, but they are not recent achievements and do not address emerging 6G technologies. Typically, AI-based methods are drawing unparalleled research interest. AI including ML/DL algorithms \cite{DCN-LIU2021589} is a tool to help network to make a quicker and wiser decision based on training data in the past. Researchers are exploring the potential of AI to solve problems specific to the mobile networking domain. For beam management, AI can be applied in the two aspects, beamforming training and beam tracking, as investigated in \cite{Beam-Management-2020,Beam-Based-Mobility-V2X-Survey-2021,Millimeter-Wave-Hardware-Beam-Management-2021,Initial-Access-Beam-Alignment-mmWave-Terahertz-2022}. Yu-Ngok Ruyue Li \emph{et al.} \cite{Beam-Management-2020} provided overview of beam management procedure in 3GPP NR Releases 15 and 16 and particularly considered the topic of AI-based beam management for future beam prediction in 3GPP standard. Abdulkadir Kose \emph{et al.} \cite{Beam-Based-Mobility-V2X-Survey-2021} summarized the ML-based beam mobility management approaches proposed in the literature until 2020, and mainly surveyed the cell-level and beam-level mobility management in 5G mmWave vehicle-to-everything (V2X in short) communications focusing on the URLLC requirements. Jihoon Bang \emph{et al.} \cite{Millimeter-Wave-Hardware-Beam-Management-2021} provided a well balanced understanding of hardware technologies for prototyping and signal processing techniques for beam management in mmWave communication systems, but only a small part of the paper involves the ML/DL-based algorithms. Wissal Attaoui \emph{et al.} \cite{Initial-Access-Beam-Alignment-mmWave-Terahertz-2022} highlighted the ML/DL viewpoint to enable fast initial access and reliable beam alignment in 5G communications and beyond. These papers have more or less investigated the ML/DL-based beam management mechanism, but they have not yet covered the new promising technologies such as FL, TL, and split learning. For WLANs empowered by IEEE 802.11 (Wi-Fi), Szymon Szott \emph{et al.} \cite{Wi-Fi-Meets-ML-2022} provided an overview of the general future research directions in applying AI for improving Wi-Fi performance. The research has shown that AI methods can have a positive impact on the beamforming in indoor Wi-Fi scenarios, and the new  techniques, FL and TL, are discussed as future trends because they provide opportunities to distribute learning tasks and improve training speed. In \cite{Machine-Learning-Beam-management-survey-2023}, M. Qurratulain Khan \emph{et al.} provided an overview of the existing ML-based mmWave/THz beam management techniques, which includes the existing FL-based studies.

Not only AI technologies, but also RIS has attracted extensive attention in recent years \cite{Joint-Optimization-Beamforming-Multi-IRS-2022,Refracting-RIS-Hybrid-Satellite-Terrestrial-Relay,IRS-Aided-Wireless-Powered-MEC-2023,CSI-Feedback}, especially in overcoming blockage issues in mmWave/THz communication networks for future generation such as 6G. However, the use of RIS poses a challenge to beam management because of the need to jointly coordinate and manage the transmission of two hops. This challenge will be more severe in the RIS-assisted mobile communication scenarios, in which multi-point beam switching will become a major problem and new techniques and analytical approaches are required to maintain connectivity between various network connection points. The aforementioned survey \cite{Beam-Based-Mobility-V2X-Survey-2021} described in detail only one beam-based mobility management method proposed by Chenglu Jia \emph{et al.} \cite{ML-beam-management-IRS} in 2020. In \cite{Initial-Access-Beam-Alignment-mmWave-Terahertz-2022}, the authors provided the solutions proposed to solve the beam alignment in THz communications with the importance of using RIS. More recently, sensing-assisted beam management is emerging as an attractive solution for the ISAC systems with harmonized and integrated communication and sensing functions. Some related work has been investigated by J. Andrew Zhang \emph{et al.} \cite{Enabling-JCAS-survey-2022}, who also discussed some major problems that have not yet been  tackled to make the sensing-aided beamforming practical. In \cite{Enabling-JCAS-survey-2022}, AI and RIS are also mentioned, but they are not described for beam management.

Although the existing surveys have done a good work in the basics, techniques, features, and challenges of beam management from different perspectives, the new approaches proposed to tackle the beam management for 6G mmWave and THz communications have not been cited or are not exhaustive, especially for ISAC and systems integrated with RIS, and more investigations are required.

\begin{figure*}[t]
	\centering
	\includegraphics[width=18cm]{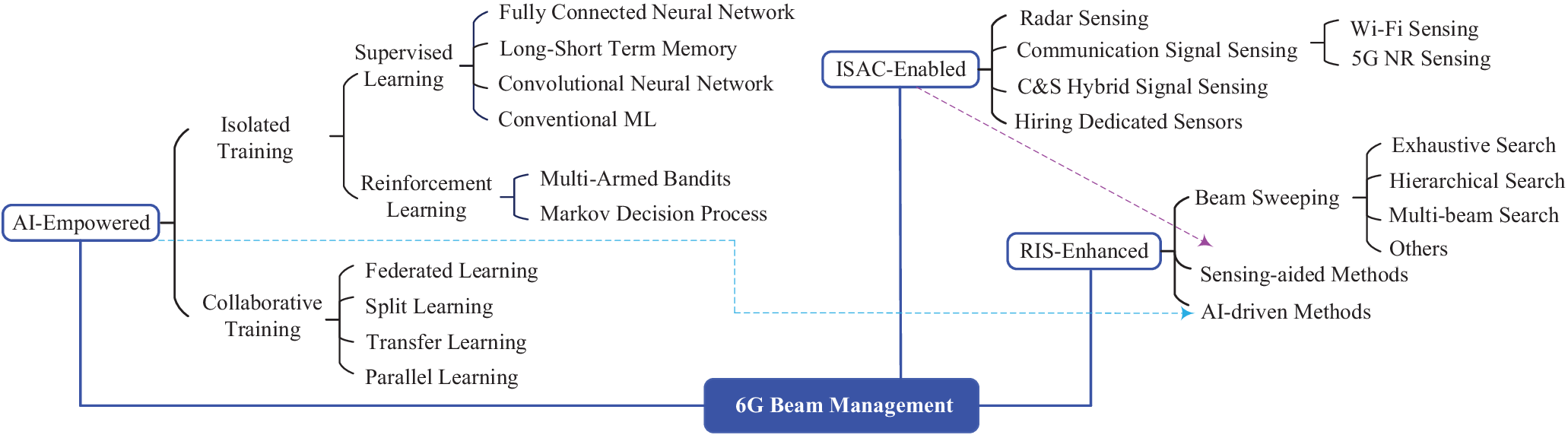}
	\caption{Classification of existing algorithms for beam management.}
	\label{fig:BM-Classification}
\end{figure*}

\section{Enabling-Technology-based 6G Beam Management: State-of-the-Art}
In this section, we provide a comprehensive survey on the state-of-the-art 6G beam management mechanisms. More specifically, we first present the potential benefits of the expected enabling technologies for mmWave and THz communications, i.e., AI, ISAC, and RIS. Then, according to the enabling technologies, we roughly classify the research work on beam management into the AI-based methods, sensing-based methods for ISAC, and methods for RIS-assisted networks, which is revealed in Fig.~\ref{fig:BM-Classification}. The visions of their combination are also discussed. Finally, summary and insights are provided at the end of this section.

\begin{figure*}[t]
	\centering
	\includegraphics[width=18cm]{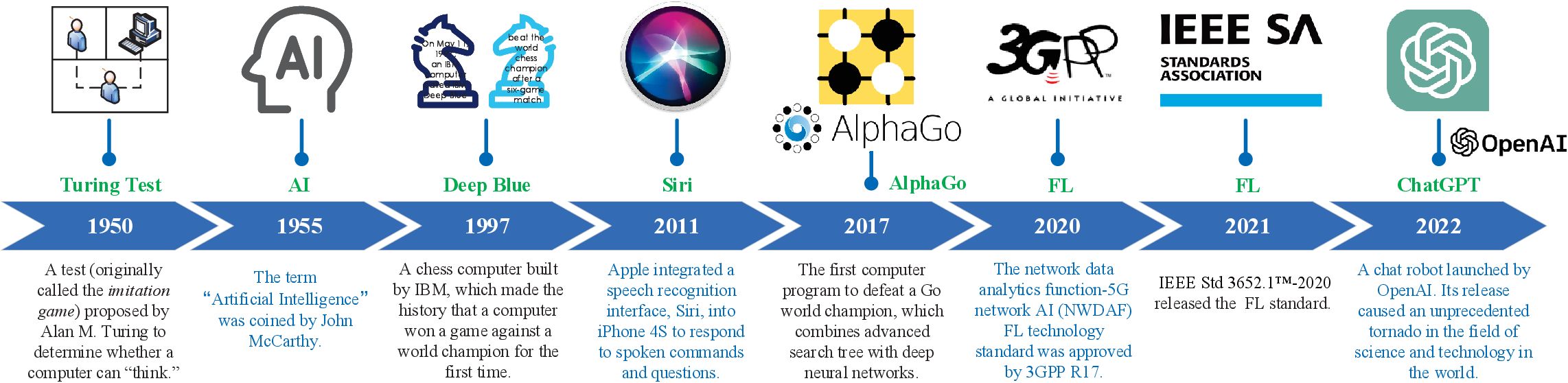}
	\caption{Roadmap to edge AI. In particular, 3GPP R17 introduced AI's FL technology in the global 5G communication field for the first time, and IEEE Std 3652.1$^{\text{TM}}$-2020 \cite{IEEE-Guide-FL} approved the first FL standard.}
	\label{fig:AI-history}
\end{figure*}

\subsection{Technology Enablers}
$\bigstar$ \emph{Why AI?} AI is an interdisciplinary science with multiple approaches that allow machines to model, and even improve upon, the capabilities of the human mind. With the advent of the big data era, the growth of various software and hardware computing resources, and the advancements in ML and DL, AI has become a field with active research topics and practical applications thanks to its agility and effectiveness, especially in dealing with dynamics and uncertainty of large-scale problems. Fig.~\ref{fig:AI-history} shows the history and key timelines of AI. In recent years, AI techniques have been broadly adopted to enhance the performance of wireless communications, in particular, in highly dynamic networks (e.g., V2X), and have achieved unprecedented success in many aspects such as mobility management, network load and resource management, channel estimation, etc. For mmWave/THz communications, integration with AI has three main advantages \cite{Deep-Learning-Beam-Management-Arxiv-2021}.

\hangindent 2.0em
$\bullet$ The complex relationships between diverse nonlinear sources in practical scenarios can be accurately modeled by AI, which can be exploited to promote effective beam management. By contrast, in order to facilitate processing, traditional mathematical methods usually idealize some conditions and ignore the nonlinear factors in the real world.

\hangindent 2.0em
$\bullet$ Due to the mobility of users and scatterers, it is necessary to dynamically adjust the optimal beam direction(s) to prevent blockage events, but the conventional methods based on beam (re-)sweeping face high overhead. Since AI can adapt to the dynamic fluctuations of the environment well, AI-based approaches can detect blockages and find bypassing beam pairs by using the location and sensory information of the surrounding environment.

\hangindent 2.0em
$\bullet$ The optimal beam direction for mmWave/THz communication is usually determined by the high-dimensional feature space formed by numerous scatterers with different positions, sizes and shapes in the wireless environment. AI may be used to extract high-dimensional environmental features to assist beam management.

$\bigstar$ \emph{Why ISAC?} With the proliferation of the IoT and Tactile Internet, a large number of sensors and actuators with different functions, such as radar, light detection and ranging (LiDAR), and cameras, are connected to mobile networks. This promotes the evolution of mobile network capability from mere communication to ISAC. In order to efficiently provide both communication and sensing (C\&S) services, various designs on ISAC system architecture, ISAC waveform, and transmit beamforming are in full swing. At present, to help wireless communication decision-making, ISAC has aroused interest in different use cases, such as autonomous driving, drone, and virtual/augmented/extended reality (VR/AR/XR). ISAC is envisioned to be an integral part of future wireless systems, especially when operating at the mmWave and THz frequency bands. In particular, leveraging sensory data to guide mmWave/THz beam management problem has gained increasing interest for the following reasons:

\hangindent 2.0em
$\bullet$ Sensing could provide environment-awareness to communications, with potentials to improve security and performance. Sensing information about the surrounding environment, such as the geometry of scatterers and the locations/directions of transmitters/receivers, can help with beam tracking, blockage prediction or proactive handoff.

\hangindent 2.0em
$\bullet$ The sensing information can be used to identify promising beamforming directions and avoid extensive (blind) beam training, and can also potentially help the mmWave/THz transceiver manage its beams to avoid interference with adjacent users. By periodically monitoring the location/direction of the target user, the AP/BS can track the user's beams and model its movement pattern.

\hangindent 2.0em
$\bullet$ In highly mobile communication scenarios, like V2X, UAVs \cite{Unmanned-Aerial-Vehicle-Wireless-Networks} and high-speed trains, mmWave/THz channels are highly dynamic, and the communication link needs to be reconfigured frequently, which requires a large training overhead. The sensing-assisted strategy can significantly reduce this overhead, which is particularly useful in NLoS scenarios.

$\bigstar$ \emph{Why RIS?} RIS, which is generally a planar surface consisting of a large number of low-cost and passive reflecting elements, can help reshape the wireless channel/radio propagation environment in a programmable manner. By densely deploying RISs in wireless network and skillfully coordinating their reflections, the radio signal propagation between transmitters and receivers can be flexibly reconfigured to achieve the desired realization and/or distribution. Recent research progress shows that RIS \cite{DCN-DAJER202287,DCN-ZHENG2023603}, as an auxiliary device, has great potential in improving spectral and energy efficiency, reducing interference, enhancing physical security, and so on, which thus improves the capacity and reliability of the RIS-assisted system. Due to the increasing number of active antennas/RF chains, mmWave/THz system incurs significantly higher energy consumption and hardware cost than sub-6 GHz wireless systems. Meanwhile, the mmWave/THz communication links are more susceptible to blockage and suffer from higher propagation loss in general. These critical issues can be efficiently tackled by properly deploying RISs in the system.

\hangindent 2.0em
$\bullet$ On the one hand, since RIS can passively reflect impinging signals to the desired direction by phase shifters, it can be operated in a green and low-carbon way without involving any sophisticated signal processing, and thus the energy consumption can be reduced by orders of magnitude compared with traditional active antenna arrays.

\hangindent 2.0em
$\bullet$ On the other hand, when the direct link between an access point (AP)/BS and its served user is not feasible due to severe pathloss or blockage, an RIS-aided virtual line-of-sight (LoS) link (i.e., AP/BS-RIS-user) can be created, which can bypass major obstacles, and thereby significantly enhance their communication performance.

\hangindent 2.0em
$\bullet$ Compared to conventional systems, the implementation of RIS in 6G is particularly suitable for beam alignment, because the signals reflected from all the elements in the RIS to the receiver can be phase aligned.

Overall, these three enabling technologies each have their own advantages. AI-empowered beam management is critical for intelligently automating and optimizing beamforming processes, leading to improved network efficiency and user experience. In sensing-enabled systems, beam management is essential for fine-tuning the system's capabilities to accurately gather data about the environment while maintaining communication quality. For RIS-enhanced communication systems, beam management is harnessed to control the wavefronts with precision, enabling better coverage and signal fidelity even in complex environments. The relevance of these advanced beam management techniques lies in their ability to greatly enhance the reliability, capacity, and performance of future wireless communication networks.

\begin{table}[t]
	\centering
	\caption{Classification of AI-based beam management solutions for mmWave and THz communications.}\label{Classification-AI-BM}
	\scalebox{0.9}{
		\begin{tabular}{|c|c|c|}
			\Xhline{0.5pt}
			\Xhline{0.5pt}
			\textbf{Model Training Mode} &\textbf{AI Techniques} &\textbf{References}\\
			\Xhline{0.5pt}
			{\multirow{2}{*}{\textbf{Independent Traning}}} &Supervised Learning &\cite{Data-Driven-Analog-Beam-Selection-2018,Deep-Learning-Beam-Management-Interference-Coordination-2019,Deep-Learning-Aided-Fingerprint-Beam-Alignment-2019,Machine-Learning-Analog-Beam-Selection-2020,Machine-Learning-Beam-Alignment-MIMO-2020,Deep-Learning-mmWave-Beam-Blockage-Prediction-2020,Learning-Predictive-Beamforming-UAV-2020,A-Deep-Learning-Low-Overhead-Beam-Selection-2021,Learning-Predictive-Transmitter-Receiver-Beam-Alignment-2021,Machine-Learning-Beam-Selection-Low-Complexity-Hybrid-Beamforming-2021,Machine-Learning-Assisted-Beam-Alignment-2021,Deep-Learning-Calibrated-Beam-Training-2021,Deep-Learning-Beam-Tracking-2021,KF-LSTM-Beam-Tracking-UAV-2022,A-Deep-Learning-Approach-Location-2022,Novel-Design-User-Scheduling-Analog-Beam-Selection-2022,A-Deep-Learning-Framework-Beam-Selection-2022,Deep-Learning-Beam-Training-Large-Scale-2023,Deep-Learning-Multi-Finger-Beam-Training-2023}\\
			\cline{2-3}
			\multicolumn{1}{|c|}{} &Reinforcement Learning &\cite{User-Centric-Association-Ultra-Dense-mmWave-2021,Multi-Armed-Bandit-Beam-Alignment-Tracking-2019,Fast-Beam-Alignment-Correlated-Bandit-2019,Beam-Alignment-Tracking-Bandit-Learning-2020,Beam-Alignment-Non-Stationary-Bandits-2020,Intelligent-Interactive-Beam-Training-2021,Design-Implementation-Deep-Learning-Adjustable-Beamforming-Training-2021,Joint-Deep-Reinforcement-Learning-Unfolding-2021,Context-and-Social-Aware-Online-Beam-Selection-2021,Beam-Drift-2021,Millimeter-Wave-Communications-Overhead-Messenger-Wire-2021,Learning-Adaptation-Millimeter-Wave-Beam-Tracking-Training-2022,Zero-Shot-Adaptation-Beam-Tracking-2022,Cost-Efficient-Beam-Management-Resource-Allocation-2022,Deep-Reinforcement-Learning-Blind-Alignment-2022,Deep-Recurrent-Q-Network-Beam-Tracking-2022,Training-Beam-Sequence-Design-2022}\\
			\cline{2-3}
			\Xhline{0.5pt}
			{\multirow{4}{*}{\textbf{Collaborative Training}}} &Federated Learning & \cite{Beam-Management-Ultra-Dense-mmWave-Network-2023,Federated-Learning-Hybrid-Beamforming-2020,Federated-mmWave-Beam-Selection-LIDAR-2021,Fully-Decentralized-Federated-Learning-On-Board-Mission-2021,Learning-Rate-Optimization-Federated-Learning-2021,Federated-Deep-Reinforcement-Learning-THz-Beam-Search-2022,Backdoor-Federated-Learning-2022,Self-Supervised-Deep-Learning-Beam-Steering-2022,Over-the-Air-Federated-Multi-Task-Learning-2023} \\
			\cline{2-3}
			\multicolumn{1}{|c|}{} &Split Learning & \cite{A-Privacy-Preserved-Split-Learning-2022} \\
			\cline{2-3}
			\multicolumn{1}{|c|}{} &Transfer Learning & \cite{Deep-Learning-mmWave-Beam-Blockage-Prediction-2020,Beams-Selection-MmWave-Multi-Connection-2021,Deep-Transfer-Learning-Location-Aware-2021,Intelligent-Analog-Beam-Selection-Beamspace-Channel-Tracking-2023} \\
			\cline{2-3}
			\multicolumn{1}{|c|}{} &Parallel Learning &\cite{Optimal-Beam-Association-High-Mobility-2021,Joint-Beam-Training-Data-Transmission-Control-2022,Decentralized-Beamforming-Cell-Free-Massive-MIMO-2022} \\
			\Xhline{0.5pt}
			\Xhline{0.5pt}
		\end{tabular}
	}
\end{table}

\subsection{AI-Empowered Beam Management}
Beam management based on AI algorithms can automatically extract and apply relevant information from previous results to limit the subsequent search area. As a result, empowering the beam management process with intelligence is a potential technique for reducing overhead. Meanwhile, the prediction accuracy of such methods is superior to conventional technologies. In this subsection, we give an overview of AI-empowered beam management techniques for mmWave and THz communications. Various studies on AI-empowered beam management have been reported, and their classification is revealed in Table~\ref{Classification-AI-BM}. The training modes of AI models in existing studies can be roughly divided into two categories, namely isolated training and collaborative training. 

Isolated training refers to a process of training an AI model using a specific single dataset or on a single task, without combining it with other datasets or tasks for simultaneous training. This approach focuses on enabling the model to master a particular skill or knowledge area extensively rather than attempting to learn from multiple domains or tasks at the same time. Such training is beneficial for developing specialized models with strong performance on their intended applications but can result in limited flexibility or generalization to other tasks not represented in the training data. The AI technologies used include supervised learning (SL) and reinforcement learning (RL). 

In contrast, collaborative training in the context of ML and AI refers to a process where multiple models, systems, or agents work together in training, either by sharing data, insights, or learning strategies. This collaboration can occur in various forms and has different applications such as FL and TL. Collaborative training leverages the collective strengths and insights of multiple models or data sources, leading to enhanced learning efficiency, improved performance, and better generalization capabilities. It is especially useful in complex scenarios where a single model or approach may not adequately capture the nuances of the data or task.

{\bf{Isolated Training.}} To realize the intelligence in beam management, various AI techniques have been developed and adopted. AI \cite{Blockchain-Artificial-Intelligence} inherits the three basic ML paradigms including SL, unsupervised learning, and RL. The common paradigms introduced in beam management are SL and RL. Specifically, SL uses labeled data for training and is usually applied to applications with enough historical data. Its goal is to learn a function that maps feature vectors (input) to labels (output), based on example input-output pairs. RL is a dynamic learning approach that maximizes the outcome through trial and error, with a focus on finding a balance between exploration (of uncharted territory) and exploitation (of current knowledge). To perform SL and RL tasks for single-agent settings, artificial neural networks (ANNs), which can be simply called NNs or neural nets, are arguably the most important frameworks because they are capable of mimicking human intelligence \cite{Artificial-Neural-Networks-2019}. All the NNs that have multiple hidden layers are known as deep neural networks (DNNs).

\begin{table*}[t]
	\centering
	\caption{AI models employed for SL-based beam management mechanisms.}\label{SL-BM}
	\scalebox{0.9}{
		\begin{tabular}{|c|c|c|c|c|c|l|}
			\Xhline{0.5pt}
			\Xhline{0.5pt}
			\textbf{Models} &\textbf{Ref.} & \textbf{Year} &\textbf{Focus} &\textbf{Model Input}  &\textbf{Model Output} &\textbf{Performance Metrics}\\
			\Xhline{0.5pt}
			{\multirow{11}{*}{\textbf{{\makecell[c]{Fully\\ Connected\\Neural\\ Network}}}}} &\cite{Deep-Learning-Beam-Management-Interference-Coordination-2019} &2019 &{\makecell[c]{Beam management,\\ Interference coordination}} &SNR &{\makecell[c]{Binary association matrix,\\beamwidth matrix,\\power allocation matrix}} &{\makecell[l]{$\bullet$ Sum-rate ratio\\$\bullet$ Computation time ratio}} \\
			\cline{2-7}
			\multicolumn{1}{|c|}{} &\cite{Deep-Learning-Aided-Fingerprint-Beam-Alignment-2019} &2019  &Beam alignment &{\makecell[c]{User location, traffic density,\\received signal strength}} &{\makecell[c]{Probabilities associated with\\AoA-AoD pairs}} &{\makecell[l]{$\bullet$ Received signal strength}} \\
			\cline{2-7}
			\multicolumn{1}{|c|}{} &\cite{Deep-Learning-mmWave-Beam-Blockage-Prediction-2020} &2020  &{\makecell[c]{Beam and blockage\\ prediction}} &Sub-6GHz channel vectors &{\makecell[c]{Probability distribution over\\ all the available classes}} &{\makecell[l]{$\bullet$ Prediction accuracy  \\ $\bullet$ Spectral efficiency\\ $\bullet$ Average achievable rate}} \\
			\cline{2-7}
			\multicolumn{1}{|c|}{} &\cite{Machine-Learning-Beam-Selection-Low-Complexity-Hybrid-Beamforming-2021} &2021  &{\makecell[c]{Iterative\\beam-user selection}} &Channel matrix &Selected beam-user pairs &{\makecell[l]{$\bullet$ Energy efficiency\\$\bullet$ Spectral efficiency}} \\
			\cline{2-7}
			\multicolumn{1}{|c|}{} &\cite{A-Deep-Learning-Approach-Location-2022} &2022  &{\makecell[c]{Probabilistic\\ beam selection}}  &{\makecell[c]{Coordinates and orientations\\ of user}} &All possible beam pairs &{\makecell[l]{$\bullet$ Misalignment probability\\$\bullet$ Effective spectral efficiency}}\\
			\Xhline{0.5pt}
			{\multirow{11}{*}{\textbf{{\makecell[c]{Long-Short\\Term\\Memory}}}}} &\cite{Learning-Predictive-Beamforming-UAV-2020} &2020 &Predictive beamforming & Angle sequences &Predicted altitude angle &{\makecell[l]{$\bullet$ CDF of angle estimation error \\ $\bullet$ Communication rate}}\\
			\cline{2-7}
			\multicolumn{1}{|c|}{} &\cite{Deep-Learning-Calibrated-Beam-Training-2021} &2021  &Calibrated beam training &{\makecell[c]{Received signals of current\\and prior beam training}}  &Predicted probabilities  &{\makecell[l]{$\bullet$ Narrow beam prediction loss\\ $\bullet$ Normalized beamforming gain \\$\bullet$ CDF of predicted beam gain}} \\
			\cline{2-7}
			\multicolumn{1}{|c|}{} &\cite{Deep-Learning-Beam-Tracking-2021} &2021  &Beam tracking &{\makecell[c]{Previous channel estimates,\\sensor measurements,\\ user's situation }} &Channel estimate & {\makecell[l]{$\bullet$ Bit error rate\\$\bullet$ Normalized mean square error}}\\
			\cline{2-7}
			\multicolumn{1}{|c|}{} &\cite{KF-LSTM-Beam-Tracking-UAV-2022} &2022  &Beam tracking  &Results of Kalman filtering &Cell state  &{\makecell[l]{$\bullet$ Beam tracking accuracy\\$\bullet$ Spectrum efficiency}} \\
			\cline{2-7}
			\multicolumn{1}{|c|}{} &\cite{A-Deep-Learning-Framework-Beam-Selection-2022} &2022  &Beam selection &{\makecell[c]{Signal strength,\\user location and velocity}}  &The selected beam &$\bullet$ Relative ratio of signal strength\\
			\Xhline{0.5pt}
			{\multirow{8}{*}{\textbf{{\makecell[c]{Convolutional\\Neural\\Network}}}}}  &\cite{Machine-Learning-Beam-Alignment-MIMO-2020} &2020  &Beam alignment &Received signal vectors &Beam distribution vector &{\makecell[l]{$\bullet$ Spectral efficiency}} \\
			\cline{2-7}
			\multicolumn{1}{|c|}{} &\cite{A-Deep-Learning-Low-Overhead-Beam-Selection-2021} &2021  &Beam quality estimation &{\makecell[c]{Low-resolution image of \\received pilot signal power}} &High-resolution beam image &{\makecell[l]{$\bullet$ CDF of SNR}}\\
			\cline{2-7}
			\multicolumn{1}{|c|}{} &\cite{Deep-Learning-Calibrated-Beam-Training-2021} &2021  &Calibrated beam training &{\makecell[c]{Normalized\\ received signal vector}}   &Predicted probabilities  &{\makecell[l]{$\bullet$ Narrow beam prediction loss\\ $\bullet$ Normalized beamforming gain \\$\bullet$ CDF of predicted beam gain}}\\
			\cline{2-7}
			\multicolumn{1}{|c|}{} &\cite{Deep-Learning-Beam-Training-Large-Scale-2023} &2023  &Beam training &Signal vector  &{\makecell[l]{Estimated probability of the\\ optimal near-field codeword}} &{\makecell[l]{$\bullet$ Normalized SNR\\$\bullet$ Effective achievable rate}} \\
			\cline{2-7}
			\multicolumn{1}{|c|}{} &\cite{Deep-Learning-Multi-Finger-Beam-Training-2023} &2023  &Multi-finger beam training &Training measurements &{\makecell[c]{Prediction probability of\\the optimal beam}} &{\makecell[l]{$\bullet$ Misalignment probability\\$\bullet$ Spectrum efficiency}} \\
			\Xhline{0.5pt}
			{\multirow{11}{*}{\textbf{{\makecell[c]{Conventional\\ML}}}}} &\cite{Data-Driven-Analog-Beam-Selection-2018} &2018 &Analog beam selection &Channel feature vector &Optimal analog beam &{\makecell[l]{$\bullet$ Average uplink sum-rate \\$\bullet$ Complexity}} \\
			\cline{2-7}
			\multicolumn{1}{|c|}{} &\cite{Machine-Learning-Analog-Beam-Selection-2020} &2020  &Analog beam selection &{\makecell[c]{Data samples based on\\ propagation paths}} &{\makecell[c]{Hyperplane separation \\coefficients between codewords}}  &{\makecell[l]{$\bullet$ Average sum rate\\$\bullet$ Computational complexity}}\\
			\cline{2-7}
			\multicolumn{1}{|c|}{} &\cite{Learning-Predictive-Transmitter-Receiver-Beam-Alignment-2021} &2021  &Beam alignment &{\makecell[c]{Experiences with beam \\index and coordinates}} &Beam confidence interval  &{\makecell[l]{$\bullet$ Effective achievable rate\\$\bullet$ Probability of alignment success}}\\
			\cline{2-7}
			\multicolumn{1}{|c|}{} &\cite{Machine-Learning-Assisted-Beam-Alignment-2021} &2021  &Beam prediction &Coordinates of users &Optimal AP and beam pair  &{\makecell[l]{$\bullet$ Beam prediction accuracy\\$\bullet$ Robustness of beam prediction\\$\bullet$ Sensitivity of beam prediction}} \\
			\cline{2-7}
			\multicolumn{1}{|c|}{} &\cite{Novel-Design-User-Scheduling-Analog-Beam-Selection-2022} &2022  &Analog beam selection &{\makecell[c]{Samples based on \\channel model}} &Index of analog beam  &{\makecell[l]{$\bullet$ Achievable sum rate\\$\bullet$ Computational complexity\\$\bullet$ Effectiveness (CDF)}}\\
			\Xhline{0.5pt}
			\Xhline{0.5pt}
			\multicolumn{7}{l}{CDF: cumulative distribution function}\\
		\end{tabular}
	}
\end{table*}

$\bullet$ \emph{Supervised Learning.} Due to its simplicity, SL is the most frequently used AI technique for mmWave and THz beam management. Table~\ref{SL-BM} provides a summary of the existing AI models employed for SL-based beam management, as well as performance metrics for evaluating the surveyed efforts. The most popular AI models utilized are fully connected neural network (FCNN), long short-term memory (LSTM), convolutional neural network (CNN), and conventional ML, among which FCNN, LSTM and CNN are different types of NNs.

\emph{1) Fully Connected Neural Network.} FCNN is an NN consisting solely of fully-connected neural network layers, where a fully-connected layer refers to a neural network in which each input node is connected to each output node. In \cite{Deep-Learning-Beam-Management-Interference-Coordination-2019}, P. Zhou \emph{et al.} proposed to use DNN for performing the beam management and interference coordination in dense mmWave WLAN, and designed a BFT information aided algorithm based on the BFT mechanism in IEEE 802.11ay to generate the training data for the DNN model. Considering a vehicular scenario, K. Satyanarayana \emph{et al.} \cite{Deep-Learning-Aided-Fingerprint-Beam-Alignment-2019} employed a fingerprint-based database for low-complexity beam alignment, where the fingerprint is comprised of a set of possible beam pairs (i.e., the angle-of-arrival and the angle-of-departure pairs, AoA-AoD pairs) for a given location, and invoked a feedforward neural network, namely the softmax linear classifier, for intelligently selecting the fingerprints. Experimental results in \cite{Steering-eyes-closed,Out-of-Band-Millimeter-Beamforming,Deep-Learning-Assisted-Beam-Prediction-2020,Deep-Learning-mmWave-Beam-Prediction-2021} have verified that some knowledge of sub-6 GHz channels can be used to assist system and network operations under mmWaves, mainly due to the spatial correlation between the two frequency bands. M. Alrabeiah \emph{et al.} \cite{Deep-Learning-mmWave-Beam-Blockage-Prediction-2020} exploited DNN to learn the mapping functions that can predict the optimal mmWave beam and blockage status directly from the sub-6 GHz channel. In addition to sub-6 GHz CSI, user location is often utilized to maintain a database and train AI models to map this information for beam prediction. S. Rezaie \emph{et al.} \cite{A-Deep-Learning-Approach-Location-2022} presented a DNN-based initial beam alignment method, which uses only the location and orientation information of the user as input. A fast DL-driven hybrid beamforming scheme was proposed in \cite{DCN-CHEN2022}. By carefully observing the DNN structures in the algorithms proposed in the above studies, it can be seen that they are all FCNNs. It is important to note that although side/contextual information such as the geometric property of the environment and user location and orientation can greatly reduce beam management overhead, obtaining this information comes at a cost, which may mean additional overhead or implementation costs for the system. Fortunately, the upcoming ISAC systems for 6G will facilitate the context-aware beam management.

\emph{2) Long Short-Term Memory.} As an extended version of recurrent neural networks (RNNs), LSTM is widely used for modeling time series data. An LSTM contains LSTM units, each of which has a memory cell for storing features extracted from the sequence data using recurrent connections. The key to a common LSTM unit is the cell state, which is protected and controlled via three gates: an input gate, an output gate, and a forget gate. LSTM has been successfully applied to enhance the beam management process due to its ability to learn long-term dependencies. For a UAV communication scenario, W. Yuan \emph{et al.} \cite{Learning-Predictive-Beamforming-UAV-2020} developed an LSTM-based predictive beamforming algorithm to address the beam misalignment caused by UAV jittering, where the prediction model is trained by exploiting the temporal features from the sequential angle data. To reduce the beam training overhead of mmWave MIMO system, K. Ma \emph{et al.} \cite{Deep-Learning-Calibrated-Beam-Training-2021} first leveraged CNN to implement the optimal narrow beam prediction based on the instantaneous received signals of wide beam training, and then utilized LSTM to track the movement of UE and further calibrated the predicted beam direction. In \cite{Deep-Learning-Beam-Tracking-2021}, S. H. Lim \emph{et al.} proposed an enhanced beam tracking method that models rapidly-varying mmWave channels due to the motion of UE, where an LSTM was employed to describe the temporal evolution of the AoAs and AoDs based on the sequence of the previous channel estimates and inertial measurement unit measurements. Meanwhile, the proposed LSTM-based prediction model was incorporated into a sequential Bayesian filtering framework to update the channel estimate. To balance the beam tracking accuracy and overheads for UAV-assisted high-speed railway wireless communications, L. Yan \emph{et al.} \cite{KF-LSTM-Beam-Tracking-UAV-2022} proposed to combine Kalman filtering with LSTM. In the double high-mobility scenario, varying Kalman filtering update periods were used under different beam angular variations, where a long update period was used in the slow beam angle changing areas to reduce the beam tracking overheads, and a short update period was used in the fast beam angle changing areas to enhance tracking accuracy. The authors then employed LSTM to further improve the beam tracking accuracy by taking the estimation results of Kalman filtering as training data. In \cite{A-Deep-Learning-Framework-Beam-Selection-2022}, T. T. Nguyen \emph{et al.} investigated an LSTM-based beam selection framework with a capability of missing data imputation, where a DNN with signal strength and user location as inputs was used to predict the missing data, and the predicted signal strength and the geographical information were used as the LSTM inputs to select the suitable beam.

\emph{3) Convolutional Neural Network.} Although FCNN can be used for learning features and classifying data, this architecture is generally impractical for larger inputs that require massive numbers of neurons. Some modern NNs, such as CNNs, have used partial connections to streamline their connection manners. CNNs use a mathematical operation called convolution in place of general matrix multiplication in at least one of their layers. Convolution reduces the number of free parameters, thus allowing the network to be deeper. Some CNN-based approaches have been exploited for more efficient beam management purposes. In \cite{Machine-Learning-Beam-Alignment-MIMO-2020}, W. Ma \emph{et al.} proposed a beam alignment method using CNN for an uplink multi-user mmWave massive MIMO system, where the CNN was trained using simulated environments according to the mmWave channel model and then deployed to predict the beam distribution vector using partial beams. Based on wide beam measurements, H. Echigo \emph{et al.} \cite{A-Deep-Learning-Low-Overhead-Beam-Selection-2021} reduced the training overhead of narrow beams by utilizing a CNN, in which the spatial correlation in beam quality was applied to improve estimation accuracy. Meanwhile, to reduce the frequency of beam training, they designed a beam quality prediction model to capture spatiotemporal correlations with a convolutional LSTM network. Instead of coarse scanning with wide beams, Y. Liu \emph{et al.} \cite{Deep-Learning-Multi-Finger-Beam-Training-2023} proposed a multi-finger training scheme to take a limited number of initial measurements, wherein a customized CNN was leveraged to extract hidden features from the received measurements for angular direction prediction and candidate beam selection. In \cite{Deep-Regularized-Waveform-Learning}, leveraging CNN architecture, H. Huang \emph{et al.} proposed a mixed regularization training method for training the beam prediction neural network under limited training samples. In large-scale massive MIMO system, codebook-based beam training is widely adopted to enhance the beamforming gain, but the overhead is very high. This is because the near-field region \cite{Fraunhofer-Fresnel-Distances} of the system is extended, so that both the angle and distance from the BS to the user need to be considered in the codebook design, which is different from the far-field region. To reduce the overhead, W. Liu \emph{et al.} \cite{Deep-Learning-Beam-Training-Large-Scale-2023} designed two NNs with convolutional modules to estimate the angle and distance of the optimal near-field beam based on the received signals of the far-field wide beams.

\emph{4) Conventional ML.} In addition to the advanced AI algorithms such as LSTMs and CNNs, several conventional AI/ML techniques have also been employed for beam management, \cite{Data-Driven-Analog-Beam-Selection-2018,Machine-Learning-Analog-Beam-Selection-2020,Learning-Predictive-Transmitter-Receiver-Beam-Alignment-2021,Machine-Learning-Assisted-Beam-Alignment-2021,Novel-Design-User-Scheduling-Analog-Beam-Selection-2022}, with support vector machine (SVM) being the most favored one. Compared to newer AI algorithms like NNs, SVM has the advantages of higher speed and better performance with a limited number of samples. The main idea behind SVM is to utilize orthogonal vectors to construct a hyperplane or a set of hyperplanes in a high or infinite-dimensional space that maximally separates the different classes in the training data. SVM excels in analyzing data for both classification and regression tasks, but generally, it works best in classification problems. Taking this as an opportunity, some efforts have been made to model analog beam selection as a classification problem and train the classifier by SVM algorithm. In \cite{Data-Driven-Analog-Beam-Selection-2018}, Y. Long \emph{et al.} considered an uplink massive MIMO system with hybrid beamforming, and exploited SVM to classify the uplink channels of each user to several different types, with each type corresponding to a candidate for the analog beam. In general, each user's analog beam has more than two candidates, which leads to an imbalance in the training data for a one-vs-the-rest classifier. To overcome it, they proposed a biased-SVM algorithm, where the major and minor training data used different error penalties, and then obtained a statistical classification model that maximizes the sum rate. In \cite{Novel-Design-User-Scheduling-Analog-Beam-Selection-2022}, the analog beam selection and user scheduling in a downlink MU-MIMO hybrid mmWave system were jointly studied to maximize the achievable sum rate. According to the channel correlation, the users were clustered based on the \emph{K}-means algorithm. After user clustering, the cluster-and-beam mapping problem was reformulated as a multi-class classification problem, in which the multi-class classifier was trained via the biased-SVM algorithm. In \cite{Machine-Learning-Analog-Beam-Selection-2020}, Y. Yang \emph{et al.} proposed an SVM-based analog beam selection approach to obtain the average sum rate of mmWave vehicle-to-vehicle (V2V) concurrent communications, using an iterative one-to-one SVM classifier to combat the imbalanced training samples to improve the accuracy of beam prediction for each vehicle user. These research results indicate that SVM algorithms can choose the analog beams with low complexity and produce significant accuracy with less computation power, which is why they are highly preferred.

\begin{table*}[t]
	\centering
	\caption{AI models employed for RL-based beam management mechanisms.}\label{RL-BM}
	\scalebox{0.9}{
		\begin{tabular}{|c|c|c|c|c|c|l|}
			\Xhline{0.5pt}
			\Xhline{0.5pt}
			\textbf{Models} &\textbf{Ref.} & \textbf{Year} &\textbf{Focus} & \textbf{Model Action} & \textbf{Model Reward} &\textbf{Performance Metrics}\\
			\Xhline{0.5pt}
			{\multirow{15}{*}{\textbf{{\makecell[c]{Multi-Armed\\Bandit}}}}} &\cite{Multi-Armed-Bandit-Beam-Alignment-Tracking-2019} &2019 &{\makecell[c]{Beam alignment\\and tracking}}  &Optimal training beamformer  &Receive SNR &{\makecell[l]{$\bullet$ Normalized beamforming gain\\$\bullet$ Physical angle for tracking}} \\
			\cline{2-7}
			\multicolumn{1}{|c|}{} &\cite{Fast-Beam-Alignment-Correlated-Bandit-2019} &2019 &Beam alignment &Current optimal beam &{\makecell[c]{Noisy\\received signal strength}}  &{\makecell[l]{$\bullet$ Cumulative regret\\$\bullet$ Measurement complexity\\$\bullet$ Beam detection accuracy/latency}}\\
			\cline{2-7}
			\multicolumn{1}{|c|}{} &\cite{Beam-Alignment-Tracking-Bandit-Learning-2020} &2020 &{\makecell[c]{Beam alignment\\and tracking}} &{\makecell[c]{A pair of integers consisting\\of beam index difference\\ and number of beams}} &Effective achievable rate &{\makecell[l]{$\bullet$ Effective achievable rate\\$\bullet$ Probability of alignment success}}\\
			\cline{2-7}
			\multicolumn{1}{|c|}{} &\cite{Beam-Alignment-Non-Stationary-Bandits-2020} &2020 &Beam alignment &Beamforming vector index &Received SNR &{\makecell[l]{$\bullet$ Outage probability\\$\bullet$ Regret (Reward difference/loss)\\$\bullet$ Minimum number of iterations}} \\
			\cline{2-7}
			\multicolumn{1}{|c|}{} &\cite{Context-and-Social-Aware-Online-Beam-Selection-2021} &2021 &Beam selection &Subset of beams &{\makecell[c]{Expected amount of\\data received}}  &{\makecell[l]{$\bullet$ Cumulative/aggregate received data\\$\bullet$ Average contact time\\$\bullet$ Convergence speed and learning time}}\\
			\cline{2-7}
			\multicolumn{1}{|c|}{} &\cite{Beam-Drift-2021} &2021  &{\makecell[c]{Beamwidth\\optimization}}  &Codebook and beamwidth &Effective achievable rate &{\makecell[l]{$\bullet$ Probability of successful alignment\\$\bullet$ Average effective achievable rate}}\\
			\Xhline{0.5pt}
			{\multirow{16}{*}{\textbf{{\makecell[c]{Markov\\Decision\\Process}}}}}
			&\cite{User-Centric-Association-Ultra-Dense-mmWave-2021} &2021 &{\makecell[c]{User-centric\\association}} &BSs/beams &{\makecell[c]{User achievable rate,\\network load balance}}& {\makecell[l]{$\bullet$ Convergence\\$\bullet$ User achievable rate\\$\bullet$ User load on each BS}}\\
			\cline{2-7}
			\multicolumn{1}{|c|}{} &\cite{Intelligent-Interactive-Beam-Training-2021} &2021 &Beam training &Beam subset &Effective achievable rate & {\makecell[l]{$\bullet$ Average effective achievable sum-rate\\$\bullet$ Probability of successful alignment}}\\
			\cline{2-7}
			\multicolumn{1}{|c|}{} &\cite{Joint-Deep-Reinforcement-Learning-Unfolding-2021} &2021  &{\makecell[c]{Beam selection and\\digital precoding}} &Beam &{\makecell[c]{Sum-rate, beam energy,\\approximation of SINR,\\average energy of each user}} &{\makecell[l]{$\bullet$ Convergence\\$\bullet$ Achievable system sum-rate\\$\bullet$ Generalization ability}}\\
			\cline{2-7}
			\multicolumn{1}{|c|}{} &\cite{Learning-Adaptation-Millimeter-Wave-Beam-Tracking-Training-2022} &2022  &Beam training &Set of beam pair indexes &\emph{Null}&{\makecell[l]{$\bullet$ Average spectral efficiency\\$\bullet$ Average total data delivered}}\\
			\cline{2-7}
			\multicolumn{1}{|c|}{} &\cite{Zero-Shot-Adaptation-Beam-Tracking-2022} &2022  &Beam tracking &Beam direction &Received signal power &{\makecell[l]{$\bullet$ Robustness\\$\bullet$ Average received signal power}}\\
			\cline{2-7}
			\multicolumn{1}{|c|}{} &\cite{Deep-Reinforcement-Learning-Blind-Alignment-2022} &2022 &Beam alignment  &Beam alignment configuration &Average rate  &{\makecell[l]{$\bullet$ Average sum rate evolution\\$\bullet$ CDF of observed rates}}\\
			\cline{2-7}
			\multicolumn{1}{|c|}{} &\cite{Deep-Recurrent-Q-Network-Beam-Tracking-2022} &2022  &Beam tracking &Index of angular bin &{\makecell[c]{Correlation between decision\\and associated optimum point}}&{\makecell[l]{$\bullet$ Mean square error of AoA}} \\
			\cline{2-7}
			\multicolumn{1}{|c|}{} &\cite{Training-Beam-Sequence-Design-2022} & 2022 &Beam/AoD tracking &Training beams &Estimation error rate&{\makecell[l]{$\bullet$ Estimation error rate\\$\bullet$ Average achievable throughput}}\\
			\Xhline{0.5pt}
			\Xhline{0.5pt}
		\end{tabular}
	}
\end{table*}

$\bullet$ \emph{Reinforcement Learning.} Although simple, SL lacks interaction with the environment, requiring a huge number of training samples to ensure excellent performance, which limits its application. RL \cite{Reinforcement-Learning-2018}, by contrast, is more versatile and suitable for general application scenarios because it does not need labeling and can operate in a truly autonomous manner. The typical framing of a RL scenario: an \emph{agent} takes \emph{actions} in an environment, which is interpreted into a \emph{reward} and a representation of the \emph{state}, which are fed back into the agent. Simply put, RL is about learning the optimal action in an environment to obtain maximum reward. Basic RL can be frequently modeled as a multi-armed bandit (MAB) problem or a Markov decision process (MDP), which has been widely studied to enhance the performance of beam management, as shown in Table~\ref{RL-BM}. In fact, the bandit problem is formally equivalent to a one-state MDP.

\emph{1) Multi-Armed Bandit.} MAB \cite{Introduction-Bandits}, a lightweight RL technique, is simple but very powerful for making decisions over time under uncertainty. The MAB problem is sometimes referred to as the \emph{K-armed bandit} problem. In the basic model, an algorithm has \emph{K} possible actions to choose from, a.k.a. \emph{arms}, and \emph{T} rounds. In each round, the algorithm chooses an arm and observes some reward for the choice it made. The goal is to maximize its cumulative reward by dynamically balancing the exploitation of those arms that have yielded high rewards in the past with the exploration of untried arms. In the literature, beam management problems have been addressed using different MAB variants, e.g., in \cite{Multi-Armed-Bandit-Beam-Alignment-Tracking-2019,Fast-Beam-Alignment-Correlated-Bandit-2019,Beam-Alignment-Tracking-Bandit-Learning-2020,Beam-Alignment-Non-Stationary-Bandits-2020,Context-and-Social-Aware-Online-Beam-Selection-2021,Beam-Drift-2021}.

More specifically, most efforts leverage the MAB capability to improve beam alignment performance for mmWave point-to-point communication systems. M. B. Booth \emph{et al.} \cite{Multi-Armed-Bandit-Beam-Alignment-Tracking-2019} developed a beam alignment and tracking algorithm for time-varying mmWave channels using a Bayesian MAB beam selection policy, and tracked the channel through sparse Bayesian learning integrated with a Kalman filter. W. Wu \emph{et al.} \cite{Fast-Beam-Alignment-Correlated-Bandit-2019} proposed a hierarchical beam alignment scheme utilizing the correlation structure among beams and the prior knowledge on the channel fluctuation to accelerate the process  of identifying the optimal beam pair. They formulated the beam alignment problem as a stochastic MAB problem for stationary environments, with the goal of sequentially selecting beams to maximize the cumulative received signal strength over a certain period of time. In \cite{Beam-Alignment-Tracking-Bandit-Learning-2020}, J. Zhang \emph{et al.} also formulated the problem of beam alignment and tracking into a stochastic bandit problem to maximize the expected cumulative effective achievable rate. Instead of defining each codeword/beam as an arm, they defined each arm based on the beam index difference/offset of the two optimal beams in adjacent time slots. Unlike the stationary bandit suitable for time-invariant channels \cite{Fast-Beam-Alignment-Correlated-Bandit-2019,Beam-Alignment-Tracking-Bandit-Learning-2020}, R. Gupta \emph{et al.} \cite{Beam-Alignment-Non-Stationary-Bandits-2020} designed two beam alignment algorithms using a non-stationary MAB model that is especially suited for time-varying channels, with the aim of minimizing outage probability. In the first algorithm, a discount factor was used for accounting the timeliness of the state, while the second one used a state-window that slides over time.

In conventional MAB setting, the algorithm observes only the reward for the selected arm, and nothing else, which in practice may result in a large performance loss and even fail the algorithm. In contrast, the contextual bandit, an extension of the MAB, incorporates some external environment information (called the \emph{context}) into decision-making, which enables it to be applicable to real-world applications. In contextual bandits, the reward in each round depends on both the context and the chosen arm. Introducing some contextual information (e.g., location information) represents a natural approach to adaptive beam management in complex systems and environments. On the basis of this point, the contextual bandits have been employed to model the problems in beam management. Based on a contextual MAB model, D. Li \emph{et al.} \cite{Context-and-Social-Aware-Online-Beam-Selection-2021} studied a context- and social-aware online beam allocation scheme for mmWave vehicular communications, where the coarse direction-of-arrival information (i.e., north, south, east, and west) was chosen as the context of a vehicle. Meanwhile, the social structure of preferences between the neighboring vehicles and their passengers was leveraged to improve the beam coverage efficiency. For communications that require beam alignment, the link may be subject to beam drift caused by the rapidly changing environments and the non-ideal features inherent in practical beams. To mitigate the effect of the beam drift, J. Zhang \emph{et al.} \cite{Beam-Drift-2021} proposed a design philosophy for beam training and data transmission in mmWave communications, where multi-resolution beams with varying beamwidths were employed for data transmission while narrow beams were used for beam training. Furthermore, Bayesian contextual bandit was utilized to optimize the beamwidth for data transmission, where information reflecting the state of the system, including the equivalent channel coefficient, beam training overhead, transmit power, and beamforming (or array) gain, was used as the context.

MAB has been recognized for its effectiveness in the field of beam management, given that it is a useful tool for dealing with sequential decision-making problems. It must be noted that balancing exploration and exploitation is crucial for making optimal decisions.

\begin{figure}[t]
	\centering
	\includegraphics[width=8cm]{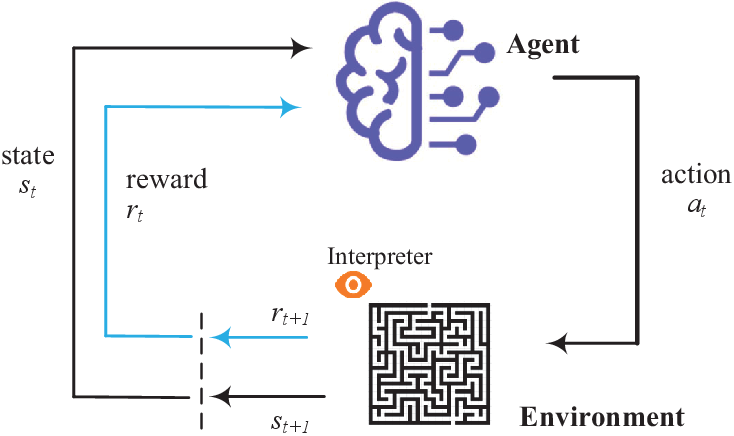}
	\caption{The agent-environment interaction in an MDP.}
	\label{fig:MDP}
\end{figure}

\emph{2) Markov Decision Process.} MDP represents a straightforward framing of the problem of learning from interactions to achieve a goal, as illustrated in Fig.~\ref{fig:MDP}. In bandit problems, it is to estimate the value of each arm/action. In MDPs, however, the value is estimated for each action in each state, or for each state given the optimal action selection \cite{Reinforcement-Learning-2018}. In the figure, $s_t$ denotes the state of the environment at time step $t$, $a_t$ denotes the action taken by the agent in this state, $r_t$ specifies the immediate reward for taking action $a_t$ in state $s_t$ and entering the next state $s_{t+1}$, and $r_{t+1}$ is the reward at time step $t+1$. At each time step $t$, the agent receives the current state $s_t$ and reward $r_t$. It then chooses an action $a_t$ from the set of available actions, which is subsequently sent to the environment. The environment moves to a new state $s_{t+1}$ and the reward $r_{t+1}$ associated with the transition $\left(s_t, a_t, s_{t+1} \right)$ is determined. In a classical MDP, agent can perceive the entire environment state fully. But for most real-world mobility applications, the environment is actually partially observable. This partial observability is described by a partially observable MDP (POMDP). The two most widely used RL algorithms to solve the beam management problems modeled as MDPs or POMDPs are as follows.

\emph{- Deep Q-network.} The deep Q-network (DQN) represents the optimal action-value (\emph{Q}-value) function as an NN, instead of a table as in \emph{Q}-learning. Combining \emph{Q}-learning with a DNN, DQN is considered the main algorithm used in environments with unlimited states and discrete actions. J. Zhang \emph{et al.} \cite{Intelligent-Interactive-Beam-Training-2021} formulated the beam training problem as an MDP, where each action was defined as a subset of the codebook for beam training and each state was defined as a matrix stacked by the real vector formed by the modulus of all components of the user's equivalent channel vector. Because the states are continuous while the actions are discrete, DQN was used to solve the MDP. Unfortunately, one drawback to DQN is that it sometimes substantially overestimates the values of actions. If an overestimation does occur, this negatively affects the performance in practice. As an improvement on DQN, double DQN (DDQN) with two NNs decomposes the max operation in the target into action selection and action evaluation, resulting in more stable and reliable learning \cite{Double-Q-Learning-2016}. In \cite{Joint-Deep-Reinforcement-Learning-Unfolding-2021}, the joint design of beam selection and digital precoding for mmWave MU-MIMO systems was investigated to maximize the sum-rate. The beam selection problem was formulated as an MDP and a DDQN algorithm was developed to solve it, where the BS was treated as an agent, the selection of each beam was modeled as an action, and the channel matrix and an indicator tensor used for beam selection composed the state space, of which the dimension was reduced by exploiting the sparsity of beamspace channel. Undoubtedly, real-world tasks more often feature incomplete and noisy state information resulting from partial observability. Considering a user-centric ultra-dense mmWave communication system, Q. Xue \emph{et al.} \cite{User-Centric-Association-Ultra-Dense-mmWave-2021} modeled the decision-making process of user association as a POMDP, applying DQN to intelligently select multiple mmWave BSs for simultaneous access, wherein the reward represents a compromise between the user achievable rate and network load balance. Seriously though, by replacing the first fully-connected layer of DQN with a recurrent LSTM layer of the same size, deep recurrent Q-network (DRQN) better handles the information loss in POMDPs than does DQN \cite{Deep-Recurrent-Q-Learning}. J. Park \emph{et al.} \cite{Deep-Recurrent-Q-Network-Beam-Tracking-2022} presented a DRQN approach for tackling the beam tracking task in mmWave massive MIMO system as a POMDP problem, in which the state was set to the channel matrix, an action was related to an angular bin in the search range, and the observation was chosen as the received signal matrix.

\emph{- Actor-Critic.} RL algorithms can be either model-free or model-based. Two major paradigms in model-free RL are value-based (e.g., \emph{Q}-learning) and policy-based (e.g., Monte Carlo policy gradient and deterministic policy gradient). The actor-critic (AC) framework combines the advantages of both value-based and policy-based approaches. AC consists of two components, i.e., actor and critic, who uses a value-based critic to improve updates to the actor (or policy). T. Zhang \emph{et al.} \cite{Cost-Efficient-Beam-Management-Resource-Allocation-2022} employed AC to solve the problem of joint beamwidth management and resource (transmit power, channel, and bandwidth) allocation in a mmWave backhaul heterogeneous network with hybrid energy supply aiming to maximize long-term cost efficiency. They first converted the discrete variables in the action space as continuous ones to get a continuous state-action space, and then adopted DNN as function approximator for the AC network. There were two critic DNNs in their algorithm to evaluate the given policies. One was an extrinsic critic DNN who used direct experience sampled from the environment to generate external advantage values, and the other was an intrinsic critic DNN that used exploration rewards to calculate internal advantage values. In intelligent beam management, one main AC RL algorithm to train deep RL (DRL) agent is deep deterministic policy gradient (DDPG), who works on continuous action space. In \cite{Deep-Reinforcement-Learning-Blind-Alignment-2022}, V. Raj \emph{et al.} explored a blind beam alignment method based on the RF fingerprints of users in a multi-BS multi-user scenario, and modeled the problem of BS selection and beam direction prediction as an MDP to improve the effective SINR experienced by the users. For handling the action space that mixes discrete (BS selection) and continuous (selection of beam alignment angles) actions, they proposed a new neural function approximator structure based on DDPG. The difference between their proposed method and vanilla DDPG is that the agent based on the proposed method has a user sub-net augmented actor-network. In \cite{Training-Beam-Sequence-Design-2022}, D. Zhang \emph{et al.} investigated the beam tracking problem for a time-varying mmWave multiple-input single-output (MISO) channel when the AoD transition function is unknown to the transmitter, i.e., model-free scenario, and rephrased the problem as a POMDP problem, where the transmit training beams in one beam tracking period was defined as the action in that period. To handle such a continuous and high-dimensional action space, they resorted to DDPG to gain an efficient training beam sequence design policy.

The above rich achievements indicate that it is feasible to transform the dynamic beam management problems into MDPs or POMDPs and solve them through various DRL algorithms. Generally, it is difficult to capture the changes of the true environment perfectly, exploiting the beam dynamics via POMDP is more practical. Therefore, towards 6G mmWave and THz communications, we should probably focus more on developing beam management scheme based on a  well-defined POMDP framework, and using the DRL algorithms with good adaptability to design an approximately optimal policy for the POMDP.

{\bf{Collaborative Training.}} AI/DL techniques enable automated analysis of the vast amount of data generated in wireless networks and subsequent optimization of highly dynamic and complex networks. However, the data collection and transmission not only entail heavy communication overhead, but also raise serious concerns about privacy breaches. An intuitive way to counteract these issues would be to train and inference directly at the network edges, such as BSs, roadside units (RSUs), and devices, using locally generated real-time data. As each edge may hold only a small training dataset, collaborative training and inference is a potential way to improve model accuracy and performance generalization \cite{Edge-Artificial-Intelligence-6G-2022,AI-in-6G-2022,Distributed-Artificial-Intelligence-End-Edge-Cloud-2023,Energy-Efficient-Mobile-Edge-Computing-2022}. FL and split learning are the current widely noticed distributed collaborative AI approaches for enhanced wireless networks. When training data is lacking, another promising solution is to introduce TL to improve the training efficiency through ``knowledge transfer'', which can actually be seen as a special type of collaboration between multiple agents or tasks. In addition to specific parallelism such as FL, there are several other attempts at collaborative training in the literature, which also involve parallel execution of multiple agents, collectively referred to as \emph{parallel learning} in this paper. Next, we will elaborate on the current contributions to beam management based on various collaborative training frameworks.

\begin{figure*}[t]
	\centering
	\includegraphics[width=18cm]{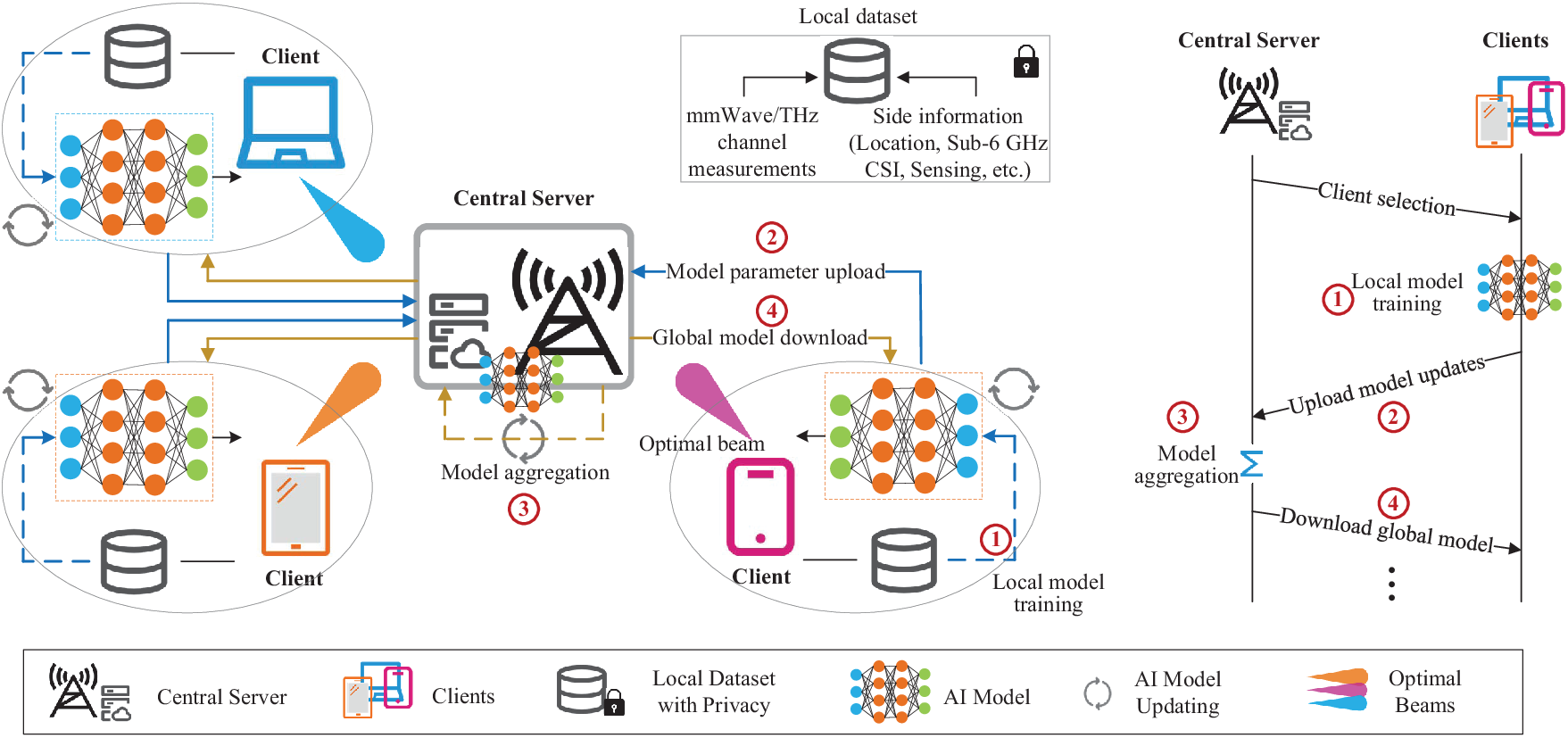}
	\caption{General architecture and communication process of centralized FL-based beam management.}
	\label{fig:Central-FL}
\end{figure*}

$\bullet$ \emph{Federated Learning.} Given the ever-increasing attention on data privacy at edge, FL \cite{FL-2021}, which is a distributed collaborative AI paradigm that trains a shared global model with locally stored data, is emerged. It has been applied to many scenarios, such as IoT networks \cite{Fairness-Awar-Federated-Learning}. References \cite{Beam-Management-Ultra-Dense-mmWave-Network-2023,Federated-Learning-Hybrid-Beamforming-2020,Federated-mmWave-Beam-Selection-LIDAR-2021,Fully-Decentralized-Federated-Learning-On-Board-Mission-2021,Learning-Rate-Optimization-Federated-Learning-2021,Federated-Deep-Reinforcement-Learning-THz-Beam-Search-2022,Backdoor-Federated-Learning-2022,Self-Supervised-Deep-Learning-Beam-Steering-2022,Over-the-Air-Federated-Multi-Task-Learning-2023} studied how to minimize the computing delay and energy consumption of AI model for beam management through FL, as well as how to improve the learning speed. Two typical FL structures have been investigated in them, namely centralized FL and decentralized FL, the only difference between the two is the presence or absence of a central server.

Till now, centralized FL, which contains a central server and a set of clients to perform an FL model, is the most popular FL architecture used in beam management, as shown in Fig.~\ref{fig:Central-FL}. In our previous work \cite{Beam-Management-Ultra-Dense-mmWave-Network-2023}, a dynamic beam configuration algorithm was proposed to improve beam utilization in a two-tier heterogeneous ultra-dense mmWave system by periodically sensing instantaneous user distributions. The beam management problem was formulated as an MDP with a large state-action space, which was solved using federated DDQN. Specifically, at the beginning of each round, the macro BS sent the current state of the global model to mmWave BSs. After that, each mmWave BS first performed a data cleaning technique to choose the users for participating in local training to ensure the quality and diversity of training data, and then separately trained its local beam configuration model based on the location information of the participants within its coverage range. Subsequently, the local updates were sent to the macro BS, who then applied these updates to its global state via FederatedAveraging (or FedAvg) \cite{Communication-Efficient-Learning-2023}. The process repeated. The proposed algorithm has greater advantages in privacy protection than traditional centralized DRL algorithms, due to the data cleaning and non-raw-data aggregation. In \cite{Federated-Learning-Hybrid-Beamforming-2020}, A. M. Elbir \emph{et al.}  introduced FL for hybrid beamforming in mmWave MU-MIMO systems, where a CNN architecture was exploited for the  model training. Each user computed its gradient information using its own available training data (a pair of the channel matrix and the corresponding beamformer index), then sent it to the BS. The BS collected the gradient data to train the CNN that outputs the analog beamformers. M. B. Mashhadi \emph{et al.} \cite{Federated-mmWave-Beam-Selection-LIDAR-2021} focused on LiDAR-aided beam selection for Vehicle-to-infrastructure (V2I) mmWave communication systems, in which connected vehicles trained a shared NN on their location and locally available LiDAR data by employing FedAvg, and the collaborative training was orchestrated by the BS. P.-C. Hsu \emph{et al.} \cite{Federated-Deep-Reinforcement-Learning-THz-Beam-Search-2022} presented a FL-based beam search scheme for THz communications to optimize the network throughput with limited CSI. All the BSs conducted a DDPG to obtain their THz beamforming policies and were controlled by an edge server to exchange the training models with hidden information thereby mitigating inter-cell interference. We observe that a separated communication-and-computation pattern is adopted for the model aggregation procedure in centralized FL. Such a pattern could result in a sharp rise in the consumption of wireless resources (e.g., the limited communication bandwidth), which becomes one of the key bottlenecks in deploying FL in the real world. As an alternative, by exploiting the waveform superposition property, \emph{over-the-air computation} (AirComp) has recently been integrated into FL to quickly aggregate local models so as to improve the communication efficiency \cite{Federated-Learning-Over-the-Air-2020,Learning-Rate-Optimization-Federated-Learning-2021,Over-the-Air-Federated-Multi-Task-Learning-2023}.

These above efforts yield the global model by aggregating updated information at a central server, improving diagnostic accuracy, while keeping training data secure and private. When the communication with the server is not available or the network topology is highly scalable, decentralized FL, which is an FL architecture that has no central server involved in coordinating the training process, can be employed to guarantee both the privacy preservation and communication efficiency. Unlike centralized FL, to perform AI collaborative training, clients in decentralized FL are connected together in a peer-to-peer manner. Y. Xiao \emph{et al.} \cite{Fully-Decentralized-Federated-Learning-On-Board-Mission-2021} studied noise-robust beamforming design in UAV communications by adopting extreme learning machine in a decentralized FL framework, by which a time-varying undirected topology corresponding to the dynamic multi-agent system was investigated. With its decentralized nature, decentralized FL scheme provides better privacy protection and lower communication cost for large-scale collaborative tasks than traditional centralized FL approaches.

The usefulness of FL opens new opportunities for mmWave and THz beam management. However, to realize existing work on FL in reality, the joint security, privacy, and incentive concerns for clients should be further studied and resolved. It is still an open and vital issue to secure the collaborative training while motivating the high-quality local model sharing in highly dynamic scenarios.

$\bullet$ \emph{Split Learning.} Split learning \cite{AI-in-6G-2022}, as the name suggests, splits an AI model between the edge/cloud server and edge clients. Each edge node trains only a portion of the full model. This model-splitting architecture enables a higher level of privacy and a better tradeoff between communication and computation. It is thus particularly suitable for large-sized DL. While FL helps mitigate data privacy concerns, most existing FL algorithms rely on that each client has sufficient storage and computing resources to locally update AI models, especially DNNs. Compared to FL, split DL provides a flexible way to train a DNN even under small memory and battery-limited clients (e.g. for mobile/IoT devices). In \cite{A-Privacy-Preserved-Split-Learning-2022}, M. Tian \emph{et al.} proposed a LiDAR and position-aided mmWave V2I beam selection method based on an improved split learning, where vehicles and a server jointly trained an NN in a private way to find the optimal beam pair, improving the probability of correct prediction while preserving user privacy. In the method, the input of the NN was a 2D matrix containing LiDAR and position information, and the output was the predicted top-\emph{k} beam pairs. During the training, the communication between the vehicles and the server was forwarded by a roadside BS. This work is an exploration of applying split learning to the beam selection task in mmWave V2I communications and presents a new option for AI-empowered beam management that requires both user information and privacy protection.

As opposed to FL that periodically exchanges model updates, split learning requires exchange of instantaneous model updates in forward and backward propagations \cite{Communication-Efficient-Distributed-Learning-2021}. In consequence, while effective in terms of accuracy and privacy, the communication efficiency of split learning in beam management is still questionable. Instead of being directly applied, split learning may need to be improved according to actual beam management scenarios to be more effective, have better generalization ability, or be more robust to non-independent and identically distributed (non-IID) data. Furthermore, the communication cost of split DL-based beam management depends on the architecture of the trained NN and how to cut its layers, calling for more investigation on NN architecture design and layer cutting.

$\bullet$ \emph{Transfer Learning.} Usually, a lot of data is needed to train an AI model from scratch but access to that data is not always available. This is where TL \cite{Survey-Transfer-Learning-2021} comes in handy. The general idea is to reuse the knowledge of a pre-trained model from one domain, usually with a large dataset, on a new task in another domain that doesn't have much data. In terms of beam management, \cite{Beams-Selection-MmWave-Multi-Connection-2021,Deep-Transfer-Learning-Location-Aware-2021,Intelligent-Analog-Beam-Selection-Beamspace-Channel-Tracking-2023} proposed beam selection algorithms based on TL. Specifically, In \cite{Beams-Selection-MmWave-Multi-Connection-2021}, H. Chen \emph{et al.} developed a parallel DNN with TL to speed up the beam search process of a mmWave multi-connection system, where the spatial correlation between sub-6 GHz and mmWave frequency band was exploited to map the sub-6 GHz channel information to the mmWave beam index. The two DNNs in the parallel DNN structure shared the common input, i.e., the sub-6 GHz channel information from the user to two BSs. They first trained one of the DNNs to output the best beam index of one BS, and then transferred the learned features to the other DNN for predicting the beam of the other BS. Clearly, TL was used to tackle the same task here, rather than a ``different'' but related task, in order to boost the system performance in reducing the training complexity. The major bottleneck of using DNNs in location-aided beam alignment procedure is the need for large datasets to tune their trainable parameters. S. Rezaie \emph{et al.} \cite{Deep-Transfer-Learning-Location-Aware-2021} showed that TL could be leveraged to significantly reduce the training data requirements of DNNs in performing location/orientation-aware mmWave beam alignment. In \cite{Intelligent-Analog-Beam-Selection-Beamspace-Channel-Tracking-2023}, H. Zarini \emph{et al.} focused on the problems of beamspace channel tracking and analog beam selection in a hybrid analog-digital THz beamspace massive MIMO system. First, a time-series-based DL approach was proposed to track and predict the beamspace channel over the sequences of time. Relying on the predicted beamspace channel, an analog beam selection strategy was presented to be learned as a classification task by fine-tuning an off-the-shelf pre-trained GoogleNet classifier based on TL. That is, the fine-tuned GoogleNet learned analog beam selection at the transceivers based on the beamspace channel feature space, while its internal weights, biases, and other parameters remained basically unchanged.

Due to its applicability across domains and tasks, the TL technique can reduce the time and cost related to collecting measurement data, thus promoting the use of NNs in practical beam management. It is worth mentioning that the transferred knowledge can only bring a positive impact on new domains/tasks when it has strong commonalities with the source domain/task.

$\bullet$ \emph{Parallel Learning.} We use the term parallel learning due to its objective, which is parallelizing the computation of model updates across multiple learning processes. For instance, A. V. Clemente \cite{Efficient-Parallel-Methods-2017} proposed a parallelization framework for DRL and demonstrated with an advantage AC algorithm on a graphics processing unit. In the framework, a set of actors can be trained synchronously on a single machine. Experiences were sampled from the distribution currently being observed from the environment, replacing sampling uniformly from previous experience. It thus can be motivated as an online experience memory. The proposed parallel framework reduces training time while maintaining state-of-the-art performance. Similarly, N. Van Huynh \emph{et al.} \cite{Optimal-Beam-Association-High-Mobility-2021} developed a parallel \emph{Q}-learning framework for beam association/handover in mmWave vehicular networks under high mobility. In their design, the vehicles in the coverage of a BS acted as active learners to help the system simultaneously collect data, based on what the BS can quickly learn the environment information. At each decision epoch and given a current state (i.e., the received signal strength indicator and the connected beam), each vehicle chose to connect the beam sent by the BS and observed the data rate of the connected beam as well as the next state. The observations were then sent to the BS to update a global \emph{Q}-table. The vehicles learned independently of each other, but shared the same global \emph{Q}-table. As such, the proposed parallel \emph{Q}-learning algorithm converges to the optimal policy much faster than the conventional \emph{Q}-learning. There are several other beam management approaches that utilize parallel learning. In \cite{Joint-Beam-Training-Data-Transmission-Control-2022}, W. Lei \emph{et al.} investigated the problem of joint beam training and data transmission control for mmWave communications. The considered problem was formulated as an MDP, and then solved via a parallel rollout method, in which multiple baseline policies were adopted simultaneously for computation. In \cite{Decentralized-Beamforming-Cell-Free-Massive-MIMO-2022}, H. Hojatian \emph{et al.} designed two local-DNN parallel architectures to perform decentralized hybrid beamforming cooperatively for a cell-free massive MIMO network, where the local DNNs were jointly trained during the training phase, and in the evaluation phase each AP had a local copy of a portion of the DNN. Parallel and distributed learning architectures have motivated innovative modifications to existing RL algorithms to efficiently make use of parallel execution.

{\bf{Summary.}} The burgeoning of data and advancements in computational power have sparked a remarkable surge in AI development over the last decade. At the forefront, AI has made extraordinary strides in mmWave and THz beam management, propelled by cutting-edge algorithms such as CNN, LSTM, MAB, and DRL, exemplified by DQN and AC methods. Despite their apparent simplicity, SL-based beam management schemes demand an extensive array of training samples, the acquisition of which is often expensive and necessitates frequent updates to reflect changing environments. In stark contrast, RL-based algorithms learn by direct interaction with the environment, gleaning training data through the learning agent's experiences and the subsequent environmental feedback. This interactive learning paradigm significantly alleviates the workload for supervisors responsible for the model training process, streamlining the path to efficient beam management. While RL has shown promise in addressing beam management issues within various simulated complex environments, its adoption in real-world scenarios is hindered by several challenges, including the need for rich experience by the RL agents, the occurrence of delayed rewards, and a general lack of interpretability. Furthermore, scenarios that involve large-scale data with requirements for low latency, efficiency, and scalability, or those that handle privacy-sensitive training data, may render traditional centralized AI frameworks inadequate. This has catalyzed interest in collaborative learning strategies. Techniques such as FL and split learning have been introduced to address concerns related to data privacy and communication overhead. TL is utilized to obviate the necessity for AI models to be trained from the ground up in situations with limited training data, thereby enhancing training efficiency. Nevertheless, these methods can sometimes compromise model performance or incur additional communication costs. The intricate trade-off among communication expenses, model accuracy, and privacy preservation is an area that warrants further exploration.

\begin{figure*}[t]
	\centering
	\includegraphics[width=18cm]{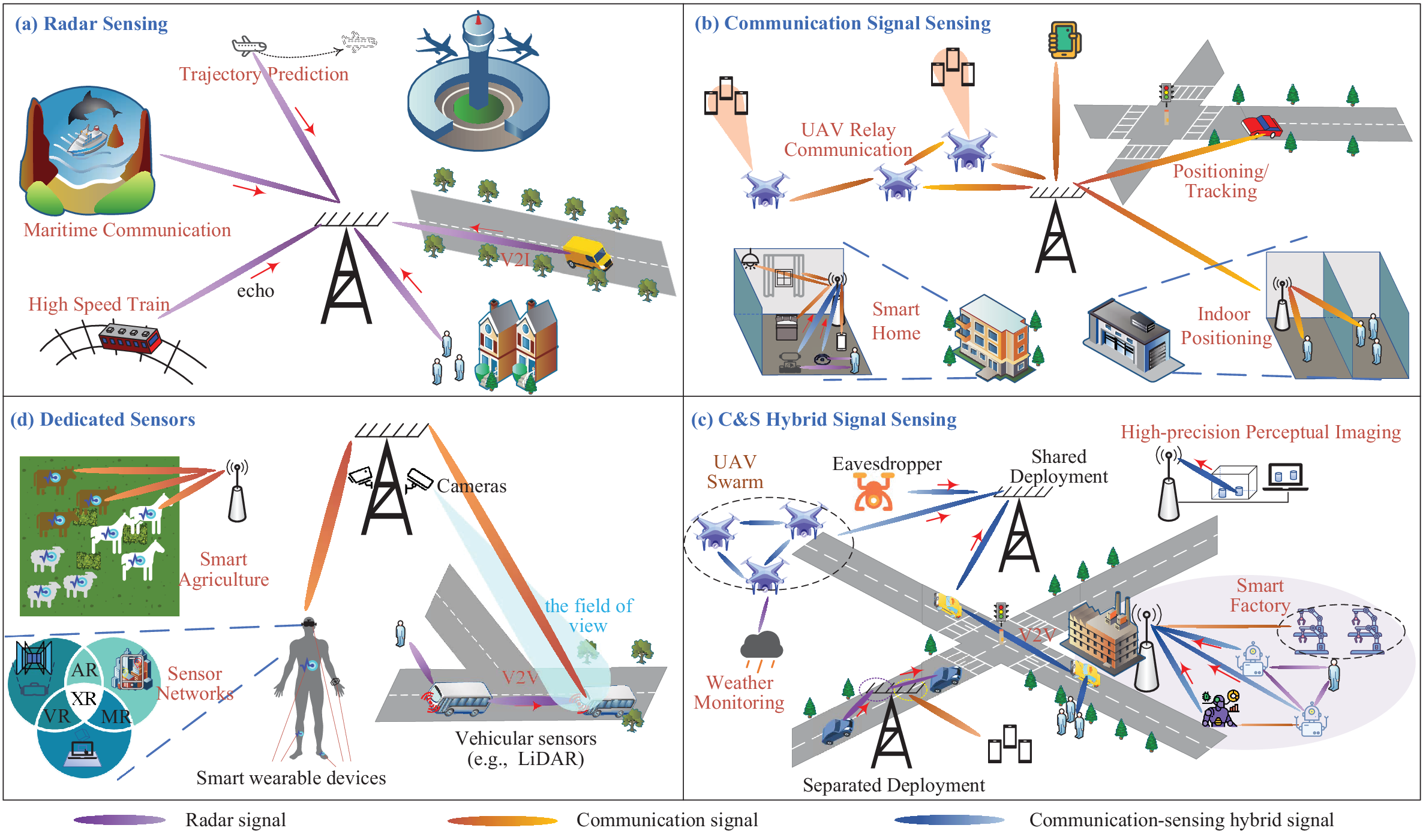}
	\caption{Typical use cases of ISAC systems with (a) radar sensing, (b) communication signal sensing, (c) hybrid signal sensing, and (d) dedicated sensors.}
	\label{fig:Use-Case-ISAC}
\end{figure*}

\begin{table*}[t]
	\centering
	\caption{Summary of the literature on beam management mechanisms for ISAC systems.}\label{ISAC-BM}
	\scalebox{0.8}{
		\begin{tabular}{|c|c|c|c|c|c|c|c|}
			\Xhline{0.5pt}
			\Xhline{0.5pt}
			\textbf{Techniques} &\textbf{Ref.} &\textbf{Year} &\textbf{Frequency} & \textbf{Scenarios} &\textbf{Sensing Source} &\textbf{Sensing Information} &\textbf{Focus}\\
			\Xhline{0.5pt}
			{\multirow{8}{*}{\textbf{Radar}}} &\cite{Radar-aided-beam-alignment-V2I-2016} &2016  &mmWave &V2I &BS &{\makecell[c]{Covariance estimates from\\radar signals}} &Beam alignment \\
			\cline{2-8}
			\multicolumn{1}{|c|}{} &\cite{Passive-Radar-Roadside-Unit-2020} &2020  &mmWave &V2I &RSU and vehicle &Radar signals' spatial covariance &Beam training \\
			\cline{2-8}
			\multicolumn{1}{|c|}{} &\cite{MIMO-Radar-mmWave-Channel-Estimation-2021} &2021  &mmWave &MU-MIMO V2X &RSU &Azimuth information of vehicles &Beamforming\\
			\cline{2-8}
			\multicolumn{1}{|c|}{} &\cite{Integrated-Scheduling-UAV-2022} &2022  &mmWave/THz &Cellular connected UAV &BS &UAV state &{\makecell[c]{Beam alignment \\for backhaul}}\\
			\cline{2-8}
			\multicolumn{1}{|c|}{} &\cite{DRL-Beam-Management-ISAC-HSR-2022} &2022  &mmWave &{\makecell[c]{Dual-band communications\\for high-speed railways}} &Remote radio units &SINR &Beam management\\
			\cline{2-8}
			\multicolumn{1}{|c|}{} &\cite{Radar-6G-Beam-Prediction-2022} &2022  &mmWave &BS serves a mobile user &BS &User's range, angles, and velocity &Beam prediction\\
			\Xhline{0.5pt}
			{\multirow{4}{*}{\textbf{Commu.}}} &\cite{An-Overview-80211bf-sensing-2022} &2022  &{\makecell[c]{Sub-7 GHz\\mmWave}} &WLANs (Wi-Fi) &Wi-Fi devices &{\makecell[c]{Channel measurement for IEEE 802.11ad/ay,\\range-Doppler-angular map (R-D-A map),\\ target-related parameters}} &Wi-Fi sensing\\
			\cline{2-8}
			\multicolumn{1}{|c|}{} &\cite{Enhancing-THz-mmWave-Beam-Alignment-2022} &2022  &mmWave/THz &SSB sensing based ISAC &BS &SSB measurement report &Beam alignment\\
			\cline{2-8}
			\multicolumn{1}{|c|}{} &\cite{An-ISAC-based-Beam-Alignment-2022} &2022 &THz &SSB-RS sensing based ISAC &BS  &{\makecell[c]{SSB measurement report,\\user speed}} &{\makecell[c]{Blockage detection,\\User tracking}}\\
			\Xhline{0.5pt}
			{\multirow{10}{*}{\textbf{Hybrid}}} &\cite{Radar-Assisted-Predictive-Beamforming-2020} &2020  &mmWave &ISAC V2I &RSU &Vehicle's angle, distance and velocity &Beam tracking \\
			\cline{2-8}
			\multicolumn{1}{|c|}{} &\cite{Bayesian-Predictive-Beamforming-2021} &2021  &mmWave &ISAC V2I &RSU &Vehicle's range, speed, angle, path loss  &Beam tracking\\
			\cline{2-8}
			\multicolumn{1}{|c|}{} &\cite{Predictive-Beam-Tracking-Cooperative-Sensing-2021} &2021  &mmWave &ISAC V2I  &RSU and vehicle &{\makecell[c]{AoD and distance at station,\\AoA and velocity from vehicle}} &Beam tracking \\
			\cline{2-8}
			\multicolumn{1}{|c|}{} &\cite{A-Novel-ISAC-2022} &2022  &mmWave &SS-OTFS-enabled ISAC  &BS &{\makecell[c]{AoA, delay and Doppler shifts}} &{\makecell[c]{Beam tracking,\\AoA estimation}}\\
			\cline{2-8}
			\multicolumn{1}{|c|}{} &\cite{ISAC-V2I-beamforming-2022} &2022  &mmWave &ISAC V2I &RSU &{\makecell[c]{Communication receiver's\\angle, distance and velocity}} &Beam tracking \\
			\cline{2-8}
			\multicolumn{1}{|c|}{} &\cite{Vehicular-Connectivity-Complex-Trajectories-2022} &2022  &mmWave &ISAC V2I &RSU &{\makecell[c]{System noise,\\the distance and velocity of vehicle}} &Beam tracking \\
			\Xhline{0.5pt}
			{\multirow{22}{*}{\textbf{{\makecell[c]{Dedicated\\Sensors}}}}} &\cite{Motion-Sensor-Beam-Tracking-2018} &2018  &mmWave &Motion beam misalignment &Sensors in mobile device &{\makecell[c]{UE movement, rotation deviation}}  &Beam alignment\\
			\cline{2-8}
			\multicolumn{1}{|c|}{} &\cite{LIDAR-Data-DL-2019} &2019  &mmWave &V2I &LiDAR on vehicle &LiDAR point cloud &Beam selection\\
			\cline{2-8}
			\multicolumn{1}{|c|}{} &\cite{3D-Scene-Beam-Selection-2020} &2020  &mmWave &BS serves a single mobile user &Camera &Point clouds of buildings &Beam selection\\
			\cline{2-8}
			\multicolumn{1}{|c|}{} &\cite{Millimeter-Wave-Cameras-2020} &2020  &mmWave &{\makecell[c]{Sub-6GHz and mmWave\\ dual-band system}} &Camera on BS &{\makecell[c]{RGB images}} &{\makecell[c]{Beam and blockage\\prediction}}\\
			\cline{2-8}
			\multicolumn{1}{|c|}{} &\cite{Environment-Aware-Beam-Alignment-2021} &2021  &mmWave &Downlink massive MIMO &Environment-sensing nodes &{\makecell[c]{Channel knowledge map,\\user location}} &Beam alignment\\
			\cline{2-8}
			\multicolumn{1}{|c|}{} &\cite{Vision-Aided-Beam-Tracking-DL-2021} &2021  &mmWave &BS serves a moving user &Camera &Features extracted from images &Beam tracking\\
			\cline{2-8}
			\multicolumn{1}{|c|}{} &\cite{Image-Index-2021} &2021  &mmWave &Internet of Vehicles  &Cameras on BS &RGB images &Beam tracking \\
			\cline{2-8}
			\multicolumn{1}{|c|}{} &\cite{Vision-Aided-6G-Blockage-2021} &2021  &mmWave/THz &BS serves mobile users &Camera on BS &RGB images &Blockage prediction\\
			\cline{2-8}
			\multicolumn{1}{|c|}{} &\cite{Testbed-Performance-Evaluation-2022} &2022  &mmWave &{\makecell[c]{Camera sensing enabled testbed}} &Camera at the transmitter &{\makecell[c]{Image features of receiver's\\phased array antenna}} &3D beam tracking\\
			\cline{2-8}
			\multicolumn{1}{|c|}{} &\cite{Camera-Sensing-Beam-Tracking-2022} &2022  &mmWave &Connected automated vehicles &Onboard camera &{\makecell[c]{Vehicles' speed and acceleration\\by analyzing the image frames}} &Beam tracking \\
			\cline{2-8}
			\multicolumn{1}{|c|}{} &\cite{Joint-Sensing-Communications-DRL-2022} &2022 &mmWave &BS serves a group of users &Satellite image &Pixel characteristic-based feature &BS beam management\\
			\cline{2-8}
			\multicolumn{1}{|c|}{} &\cite{Vision-Position-Multi-Modal-Beam-Prediction-2022} &2022  &mmWave &BS serves a mobile user &Camera on BS &RGB images &Beam prediction\\
			\cline{2-8}
			\multicolumn{1}{|c|}{} &\cite{Computer-Vision-Beam-Tracking-Real-World-2022} &2022  &mmWave &BS serves a mobile user &Camera on BS &RGB images &Beam tracking\\
			\cline{2-8}
			\multicolumn{1}{|c|}{} &\cite{Towards-Real-World-6G-Drone-Communication-2022} &2022  &mmWave//THz &BS serves a flying drone &Camera on BS &RGB images &Beam prediction\\
			\cline{2-8}
			\multicolumn{1}{|c|}{} &\cite{LIDAR-Position-Aided-mmWave-Beam-Selection-2022} &2022  &mmWave &V2I &LiDAR on vehicle &LiDAR point cloud &Beam selection\\
			\cline{2-8}
			\multicolumn{1}{|c|}{} &\cite{Machine-Learning-Based-Vision-Aided-Beam-Selection-2022} &2022  &mmWave &Multiuser MISO &Camera on BS &RGB images and the field of view  &Beam selection\\
			\cline{2-8}
			\multicolumn{1}{|c|}{} &\cite{A-Personalized-Solution-2023} &2023  &mmWave &V2I &LiDAR on vehicle &3D point cloud  &Beam selection\\
			\cline{2-8}
			\multicolumn{1}{|c|}{} &\cite{LiDAR-Future-Beam-Prediction-2023} &2023  &mmWave &V2I &LiDAR on BS &LiDAR point cloud &Beam tracking\\
			\Xhline{0.5pt}
			\Xhline{0.5pt}
		\end{tabular}
	}
\end{table*}

\subsection{Beam Management for ISAC Systems}
Early contributions for mmWave/THz beam management are generally based on the communication-only protocols. However, a good performance needs high training overhead to achieve. In highly mobile applications (e.g., V2X and drone), the beam alignment schemes should be able to \emph{predict} the beam directions in order to meet the critical latency requirements, as the beam training results may soon become outdated. Towards this goal, side information (e.g., the transmitter/receiver locations and the geometry/characteristics of surrounding environment) obtained through sensing has recently been introduced into wireless communication systems to reduce the beam training overhead and eliminate the feedback links. Research efforts on ISAC systems are well underway in both academia and industries. Broadly, the existing research on ISAC beam management is based on four types of sensing technologies, namely, radar sensing, communication signal sensing, C\&S hybrid signal sensing, and hiring dedicated sensors, as shown in Table~\ref{ISAC-BM}. Some typical usage scenarios are shown in Fig.~\ref{fig:Use-Case-ISAC}. Next, we will make a specific analysis and comparison.

{\bf{Radar Sensing.}} In the era of 6G, in addition to wireless communications, environmental awareness is also crucial. With the usage of high frequency (i.e., mmWave/THz), radar sensing function is expected to be realized in 6G wireless communication networks. Further, in order to unlock the potential for mmWave/THz in future mobile communication systems, some researchers are trying to use side information derived from radar functionality to assist in configuring mmWave/THz communication links \cite{Radar-aided-beam-alignment-V2I-2016,Passive-Radar-Roadside-Unit-2020,MIMO-Radar-mmWave-Channel-Estimation-2021,Integrated-Scheduling-UAV-2022,DRL-Beam-Management-ISAC-HSR-2022,Radar-6G-Beam-Prediction-2022}.

For instance, beam alignment in a V2I mmWave communication scenario is considered by N. Gonz\'{a}lez-Prelcic \emph{et al.} \cite{Radar-aided-beam-alignment-V2I-2016}, where a dedicated radar sensor is mounted on the BS to aid the communication. The key idea is to design the precoders and combiners on the vehicle and the infrastructure sides by using the covariance estimates obtained at the radar band as an estimation of the covariance of the communication signal in another frequency band. While reducing the beam training overhead, such a scheme makes the beam alignment more precise, but at the cost of extra hardware. A. Ali \emph{et al.} \cite{Passive-Radar-Roadside-Unit-2020} used the spatial covariance of a passive radar at the RSU to help establish the mmWave V2I communication link, where the radar and communication operate at different frequencies. The proposed solution relied on classical calibration techniques for the radar and communication systems, which may be expensive and hard to implement in reality. In \cite{MIMO-Radar-mmWave-Channel-Estimation-2021}, MIMO radar deployed on the RSU is used to measure the azimuth information of moving vehicles to assist in uplink channel estimation in V2X scenarios. The RSU antenna array is split into two modules: one for MIMO radar and the other for uplink wireless communications. For a cellular-connected UAV network, focusing on the backhaul transmission from UAV to the ground BS, B. Chang \emph{et al.} \cite{Integrated-Scheduling-UAV-2022} used radar sensing to track UAV for beam tracking and motion control of UAV, where BS can work as a sensing radar to monitor the real-time state of UAV with constant periodical time. In \cite{Radar-6G-Beam-Prediction-2022}, U. Demirhan \emph{et al.} developed a radar-aided beam prediction algorithm and evaluated its performance using a real-world dataset in a realistic vehicular communication scenario, where the BS adopts a frequency-modulated continuous wave radar to provide observations of the moving objects in the environment. The above studies have made good attempts in radar-assisted beam management for mmWave/THz mobile  communications, proving that the sensing information can guide the beam prediction well and reduce the beam training overhead significantly.

It is worth mentioning that conventional approaches in ISAC context mainly focus on single-beam systems, which limits the sensing direction to the same as that of the communication. Directed against this problem, several recent work has raised the idea of conducting \emph{multi-beam design} \cite{DRL-Beam-Management-ISAC-HSR-2022,Multibeam-Joint-Communication-Radar-2019,Multibeam-Design-ISAC-2020}, where the system integrates separate simultaneous beams for C\&S functionalities by considering antenna arrays with multiple elements. L. Yan \emph{et al.} \cite{DRL-Beam-Management-ISAC-HSR-2022} designed a joint C\&S wireless network architecture for high-speed railways, where the remote radio units operating at mmWave bands point two beams at trains for communications and at the target areas for sensing, respectively. In the design, the sensing and communication beams are mutually orthogonal in the spatial domain, but share the same antenna array. To mitigate inter-beam interference, the authors proposed a beam management scheme based on DRL, in which the beamwidth and inter-beam spacings are adaptively adjusted according to the dynamic wireless environment. In \cite{Multibeam-Joint-Communication-Radar-2019}, a multi-beam framework with two analog antenna arrays is presented by J. A. Zhang \emph{et al.} to simultaneously support C\&S. Then a beamforming design approach considering different requirements is proposed, in which stable and high-gain beams for communication and direction-varying beams for environment sensing. Similarly, the multi-beam technique is applied in \cite{Multibeam-Design-ISAC-2020}. The proposed beamforming approach jointly optimizes the beamforming weights of the transmitter and receiver to maximize the sensing performance and mitigate the possible interference from the communication beam, while also ensuring the target beamforming gain for the communication link. It is demonstrated that using the multi-beam technology can significantly reduce leakage and clutter signals. Nevertheless, multi-beam techniques are investigated for full-duplex ISAC systems that are very challenging to implement, particularly for MIMO systems. The main reason is that in a MIMO system, a large number of leakage signals between the transmitter and receiver antennas need to be dealt with simultaneously.

Although the systems of radar sensing and communications may be co-located or even physically integrated, they transmit two different signals that may overlap in the time domain and/or frequency domain. So they need to operate cooperatively to minimize interference with each other.

{\bf{Communication Signal Sensing.}} Conventionally, radio signals are widely used for data transmissions in a wireless communication network. Recently, applying the communication signals to sense the environment in which they propagate has become an interesting topic. The IEEE and 3GPP have put substantial effort into the development of ISAC-related specifications.

$\bullet$ \emph{Wi-Fi Sensing.} Today, Wi-Fi is ubiquitous and widely used in almost all public and private spaces to provide plug-and-play Internet connection. Wi-Fi devices, such as smartphones, tablets, personal computers, televisions, sensors for smart homes, placed in extremely dense and heterogeneous concentration create an excellent opportunity to continuously ``draw'' the surrounding environment using Wi-Fi signals as sensing waveforms. For example, S. Ji \emph{et al.} \cite{SiFall} implemented an RF sensing-based online fall detection system, in which Wi-Fi CSI samples are dynamically processed, segmented, and discriminated to detect ongoing ``falls''.

The IEEE 802.11 working group has formed a new Task Group (TGbf), namely \emph{802.11bf}, to develop a new amendment that defines modifications to the IEEE 802.11 MAC and to the DMG and EDMG PHYs to support \emph{WLAN/Wi-Fi sensing}\footnote{Because of its simplicity, reliability and flexibility, Wi-Fi has gradually become a synonym for WLAN. In the 802.11 standard group, ``Wi-Fi sensing'' can be regarded as the equivalent term of ``WLAN sensing''. The two can be used interchangeably.} in all spectrum bands, including license-exempt frequency bands (sub-7 GHz) as well as its mmWave counterpart (above 45 GHz). In January 2023, TGbf completed a major milestone with the release of IEEE 802.11bf draft D1.0 \cite{IEEE-Std-802.11bf}, which specifies the necessary protocols to enable Wi-Fi sensing. IEEE 802.11bf is the world's first international standard for ISAC, and its standardization is still ongoing at the time of writing this survey. In \cite{An-Overview-80211bf-sensing-2022}, R. Du \emph{et al.} provided a comprehensive overview on the efforts of TGbf through July 2022, including the new use cases, WLAN sensing procedure, candidate technical features, and evaluation methodology.

\begin{figure}[t]
	\centering
	\includegraphics[width=8.7cm]{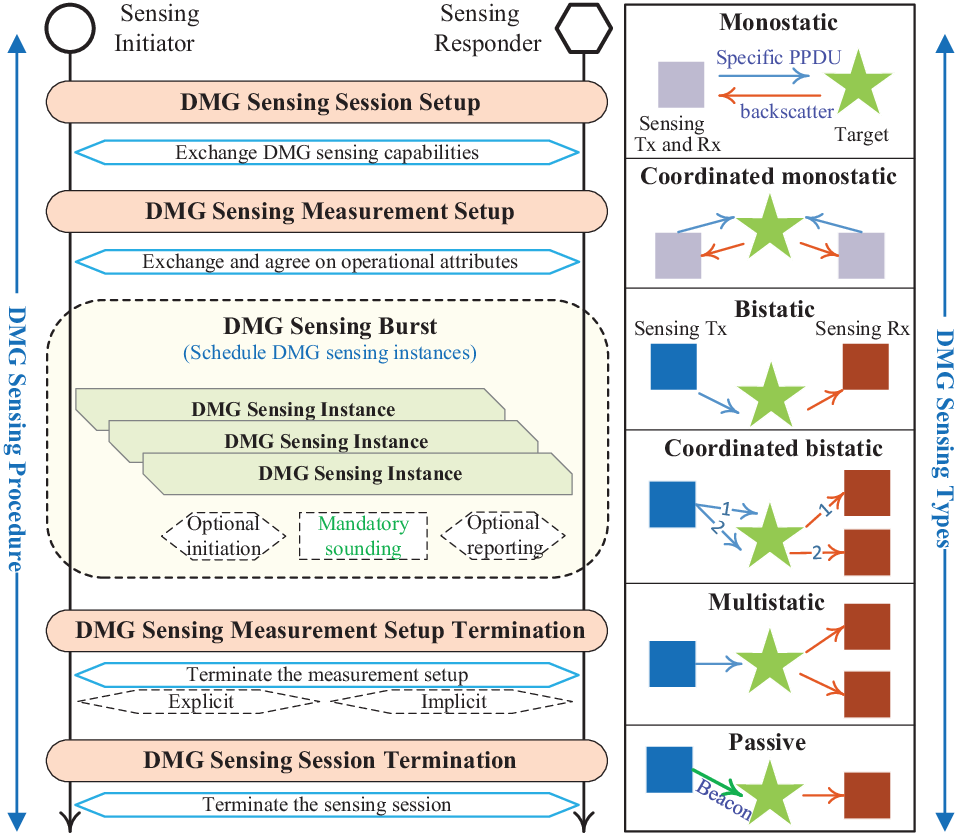}
	\caption{Overview of the DMG sensing procedure and DMG sensing types.}
	\label{fig:DMG-sensing}
\end{figure}

IEEE 802.11bf reuses existing Wi-Fi waveforms and channels defined by earlier IEEE 802.11 standards to support Wi-Fi sensing. It allows recognition of large motions (e.g., whole body motion) using the sub-7 GHz band, and subtle motions (e.g., finger movements or head swing) using the mmWave band to provide higher resolution and improved recognition accuracy. In particular, sensing measurements in the mmWave band, denoted as \emph{DMG sensing procedure}, are built on the standards of IEEE 802.11ad and 802.11ay, which define DMG and EDMG Wi-Fi communications in the 60 GHz band. Compared with sub-7 GHz sensing, DMG sensing is the focus of this paper. As shown in Fig.~\ref{fig:DMG-sensing}, the general DMG sensing procedure entails one or more of the following.

\hangindent 2.0em
\checkmark \emph{DMG sensing session setup.} This phase establishes a sensing session, where sensing-capable STAs can discover each other and exchange the DMG sensing-related capabilities, including the supported sensing types (details are given later) and the roles the STA may assume for each of the supported DMG sensing types.

\hangindent 2.0em
\checkmark \emph{DMG sensing measurement setup.} This process allows a sensing initiator and a sensing responder to exchange and agree on the operational attributes associated with DMG sensing bursts and DMG sensing instances.

\hangindent 2.0em
\checkmark \emph{DMG sensing burst.} A DMG sensing burst is made of a set of scheduled DMG sensing instances, which is the effective time period for measurements.

\hangindent 2.0em
\checkmark \emph{DMG sensing instance.} This is a process of actually performing sensing measurements, which includes an optional initiation phase, a mandatory sounding phase, and an optional reporting phase. In 802.11 context, PHY protocol data units (PPDUs) are used for sensing measurements during the sounding phase, and which PPDU to use depends on the DMG sensing type.

\hangindent 2.0em
\checkmark \emph{DMG sensing measurement setup termination.} This explicitly or implicitly terminates an established DMG sensing measurement setup and releases the allocated resources.

\hangindent 2.0em
\checkmark \emph{DMG sensing session termination.} This terminates the DMG sensing session established between STAs.

Depending on the number of STAs involved in a sensing procedure as well as the roles of each STA, DMG sensing types include monostatic, bistatic, multistatic, monostatic sensing with coordination, bistatic sensing with coordination, and passive sensing. To be specific, \emph{monostatic sensing} involves a single STA, i.e., the sensing transmitter and receiver are the same STA. When the sensing transmitter and receiver are two distinct STAs, the sensing type is said to be \emph{bistatic}. Further, when the sensing transmitter and more than one sensing receivers are distinct STAs, it is \emph{multistatic sensing}. \emph{Coordinated monostatic/bistatic sensing} extends the monostatic/bistatic sensing type. In coordinated monostatic sensing, the transmissions by one or more devices that perform monostatic sensing are coordinated by a sensing initiator (e.g., AP). Similarly, in coordinated bistatic sensing, multiple sensing responders are also coordinated by one sensing initiator. Lastly, if a STA conducts DMG sensing using the received PPDUs (e.g., DMG beacon frames, sector sweep frames) from other STAs that are not necessarily intended for sensing, it is called \emph{passive sensing}.

Prior to the DMG sensing procedure, the beamforming training between the sensing initiator and responder(s) is assumed to be completed in advance. IEEE 802.11bf attempts to reuse the existing (E)DMG protocols as much as possible, making only necessary changes to accommodate sensing. WLAN/Wi-Fi sensing is receiving increasing attention due to the great potential opportunities it creates in applications such as healthcare, home security, enterprise and building automation/management. Up to now, outstanding achievements have been made in using Wi-Fi signals to enhance sensing, but the sensing results have not been fed back to improve communication performance.

$\bullet$ \emph{5G NR Sensing.} In 3GPP NR standards, various communication signals provided specifically for channel estimation can be used as candidate sensing signals \cite{Integrated-Sensing-Communications-2022,Perceptive-Mobile-Networks-2021,5G-PRS-Based-Sensing-2022}, such as demodulation reference signal (DMRS) for both uplink and downlink, sounding reference signal (SRS) for uplink, synchronization signal block (SSB), CSI-reference signal (CSI-RS) and positioning reference signal (PRS) for downlink. To date, although many work states the feasibility of using the standard signals defined in 5G NR for sensing and then for beam prediction and tracking, only a few have achieved practical results. Typically, W. Chen \emph{et al.} proposed two beam alignment approaches exploiting the SSBs and RSs \cite{Enhancing-THz-mmWave-Beam-Alignment-2022,An-ISAC-based-Beam-Alignment-2022}, where the ISAC THz/mmWave network is modeled based on the stochastic geometry. The first \cite{Enhancing-THz-mmWave-Beam-Alignment-2022} is to use SSB for sensing to perceive the blockage and track users, providing timely assistance for beam/cell switches. The SSB sequences covering all directions are periodically transmitted with time multiplexing, which is called an SSB burst. Meanwhile, the authors designed an SSB time-frequency pattern to minimize beam misalignment under given limited resource, which is independent of node density and mobility. On this basis, a joint SSB and RS based sensing scheme is put forward in \cite{An-ISAC-based-Beam-Alignment-2022}, in which the SSB is used for blockage detection and the RS is used for user tracking. Among various standard control signals, the advantages of using SSB and RS to assist beam alignment are as follows. (1) Designed for beam synchronization, the periodical SSBs can cover all directions, thus realizing the omni-directional blockage detection. Meanwhile, SSBs only occupy a small amount of resources in time and frequency domain and are less frequent, which reduces detection overhead and computational complexity. (2) The RSs, e.g., DMRSs, are user-specific and are always transmitted with data payload, which makes them able to provide real-time positioning information and suitable for user tracking. (3) As pilot signals, SSBs are orthogonal over smaller spatial layers, and RSs are typically orthogonal over time and have frequency and spatial domains for different users, which lead to improved sensing performance \cite{Perceptive-Mobile-Networks-2021}. It is still an open topic to explore the design of other specific communication signal-based sensing aided beam management schemes. The above work is a good example for this.

{\bf{C\&S Hybrid Signal Sensing.}} In order to achieve ultra-high throughput requirements, it is essential to make future wireless communication networks coexist harmoniously with radar systems. One of the approaches is that radar sensing is relatively independent of communication in the system, and to design effective interference management algorithms to maintain both C\&S functions by suppressing mutual interference. Another way is to transmit a well-designed signal waveform to achieve the purpose of simultaneous radar sensing and communications. In this survey, signals transmitted in such waveform are referred to as C\&S hybrid signals\footnote{In some papers, the radar-communication dual-functional signal is called ISAC signal. Unlike them, the concept of ISAC in this survey covers a wide range. In addition to the sensing based on dual-functional signals, the communication systems integrating radar sensing, sensing based on communication signals, and sensing based on dedicated sensors are also called ISAC systems.}. The ISAC systems applying C\&S hybrid signals have received a lot of attention recently, mainly because the channel characteristics of radar sensing and mmWave communications are naturally similar. In practice, a large share of mmWave spectrum resources has been preliminarily allocated to radar systems. For example, automotive radar operates in the 76-81 GHz frequency band. It is advisable to use the channel information obtained from radar sensing to promote effective communication design. In \cite{ISAC-signal-Survey-2023}, Z. Wei \emph{et al.} systematically reviewed the literature on C\&S hybrid signals from the perspective of 5G-Advanced and 6G mobile communication systems, including C\&S hybrid signal design, signal processing and signal optimization.

By exploiting the radar functionality of ISAC system with C\&S hybrid signals, the communication beam tracking overheads can be drastically reduced \cite{Radar-Assisted-Predictive-Beamforming-2020,Bayesian-Predictive-Beamforming-2021,Predictive-Beam-Tracking-Cooperative-Sensing-2021,A-Novel-ISAC-2022,ISAC-V2I-beamforming-2022,Vehicular-Connectivity-Complex-Trajectories-2022}.  Most work investigates radar-assisted predictive beam tracking for V2I communications by realizing C\&S  functionalities at the RSU. In \cite{Radar-Assisted-Predictive-Beamforming-2020} and \cite{Bayesian-Predictive-Beamforming-2021}, F. Liu \emph{et al.} employed C\&S hybrid signals in the V2I downlink transmissions, where the echo signals reflected by the vehicles are exploited for tracking and localization. The difference between the two proposed schemes is that the former uses an extended Kalman
filtering method for tracking and predicting the motion parameters of vehicles, while the latter uses a message passing algorithm based on factor graph. Based on these results, the authors further improved their proposed schemes in \cite{ISAC-V2I-beamforming-2022} and \cite{Vehicular-Connectivity-Complex-Trajectories-2022}. As in \cite{Radar-Assisted-Predictive-Beamforming-2020}, Y. Xu \emph{et al.} \cite{Predictive-Beam-Tracking-Cooperative-Sensing-2021} also employed the extended Kalman filter to track the vehicle mobility and predict beam directions, but the cooperative sensing between the vehicle and the RSU is considered to enhance the beam alignment performance. In contrast to most of the existing work that always regards the vehicles as passive targets that could only reflect signals transmitted by the RSU, the scheme presented in \cite{Predictive-Beam-Tracking-Cooperative-Sensing-2021} investigates an intelligent vehicle which can actively cooperate with the RSU by sharing its sensing results. Considering the fact that the paths for radar sensing and communication are potentially mismatched, i.e., the path with the strongest echo power for radar sensing may not be the best path for communication, an ISAC transmission framework based on spatially-spread orthogonal time frequency space modulation is proposed in \cite{A-Novel-ISAC-2022}. By utilizing the previous AoA estimates and radar reflection coefficients, the authors designed beam tracking and AoA estimation algorithms for radar sensing. The aforementioned results have confirmed that V2I systems can avoid frequent feedback between the RSU and vehicles with the aid of the radar functionality built in the RSU, as the uplink feedback from the vehicles to the RSU is replaced by the echo signal. In this sense, the channel/beam information can be extracted from echo signals, and the released uplink resources for feedback can be used to transmit useful data. As a consequence, the predictive beam tracking based on radar functionality has lower overhead and lower latency than the conventional feedback-based approaches.

In ISAC-enabled V2X networks, the transmit beams of RSU and the receive beams of vehicles will be aligned with each other if the estimation and prediction are sufficiently accurate. However, it has to be mentioned that the predictive beam tracking schemes proposed in the literature would face quite a lot of critical challenges that hinder practical implementation. One of the important issues is the strong assumption adopted. For example, in many existing V2X system models, the vehicle is assumed to move along a straight road, which is parallel to the antenna array. This is obviously an ideal scenario. Future beam prediction needs to be based on a more realistic driving behavior model.

{\bf{Hiring Dedicated Sensors.}} The mmWave/THz communication systems can leverage the side information obtained from dedicated sensors, such as camera, LiDAR, and inertial sensors, to reduce the overhead associated with link configuration. For instance, in \cite{Motion-Sensor-Beam-Tracking-2018}, J. Bao \emph{et al.} argued that information from motion sensors (e.g., gyroscope) in smart devices helps mitigate beam misalignment. In \cite{Environment-Aware-Beam-Alignment-2021}, D. Wu \emph{et al.} proposed a training-free beam alignment technique by utilizing the channel knowledge map (CKM), which is a geolocation-based database, together with the user location information. The CKM will be updated when dedicated environment-sensing nodes monitor a significant change in the environment. With the pixel characteristic-based features of satellite images, Y. Yao \emph{et al.} \cite{Joint-Sensing-Communications-DRL-2022} improved user localization accuracy, and then proposed a beam management approach for the cluster groups formed by the clustering algorithm based on UK-medoids.

Among various sensor-aided approaches, the research interests in vision/camera-aided beam management are soaring in recent two years, e.g., \cite{3D-Scene-Beam-Selection-2020,Millimeter-Wave-Cameras-2020,Vision-Aided-Beam-Tracking-DL-2021,Image-Index-2021,Vision-Position-Multi-Modal-Beam-Prediction-2022,Computer-Vision-Beam-Tracking-Real-World-2022,Towards-Real-World-6G-Drone-Communication-2022,Vision-Aided-6G-Blockage-2021,Camera-Sensing-Beam-Tracking-2022,Testbed-Performance-Evaluation-2022,Machine-Learning-Based-Vision-Aided-Beam-Selection-2022}. Encouraged by the developed DL on image processing, vision information extracted from the red-green-blue (RGB) images taken by the cameras at BS and/or mobile station (MS) can be utilized to solve fast beam tracking. A panoramic point cloud from images taken within the cellular coverage area is built in \cite{3D-Scene-Beam-Selection-2020}. This point cloud gives a view of the 3D scattering environment, which is then input to a DNN to predict the optimal beams. In \cite{Millimeter-Wave-Cameras-2020}, BSs equipped with cameras were proposed to employ computer vision and DL techniques to predict mmWave blockage and beam strength. The authors in \cite{Vision-Aided-Beam-Tracking-DL-2021} and \cite{Image-Index-2021} investigated leveraging camera images for better beam prediction, in which the simulation and evaluation are conducted on a synthetic dataset, named vision-wireless (ViWi) dataset\footnote{https://www.viwi-dataset.net/}. In contrast, A. Alkhateeb \emph{et al.} evaluated their vision-aided beam prediction approaches proposed in \cite{Vision-Position-Multi-Modal-Beam-Prediction-2022,Computer-Vision-Beam-Tracking-Real-World-2022,Towards-Real-World-6G-Drone-Communication-2022} on a real-world dataset, namely DeepSense 6G\footnote{https://deepsense6g.net/}. They also developed a vision-aided dynamic blockage prediction solution in \cite{Vision-Aided-6G-Blockage-2021}, and tested the solution on the ViWi framework. A camera-sensing-assisted joint offline and online beam tracking algorithm for connected and automated vehicles is proposed by Q. Zhang \emph{et al.} \cite{Camera-Sensing-Beam-Tracking-2022}, and the hardware testbed is developed in \cite{Testbed-Performance-Evaluation-2022}. Overall, visual data as side information for beam management is driven by two key factors, (1) images are filled with information about the environment they depict, and (2) computer vision has made significant advances in image understanding with the help of DL.

To ensure a full view of the target, cameras in the network must face the direction of the target from a specific angle. Additionally, under harsh lighting conditions, the sensing performance based on vision (camera) will be seriously degraded. This motivates us to investigate other alternatives. Nowadays, high-dimensional sensor information is available on the vehicle side, thanks to the recent surge in autonomous driving technology. LiDAR, for example, is commonly used by autonomous vehicles for high resolution mapping, obstacle detection and road environment recognition. Prior work has shown that LiDAR sensory data can be exploited to reduce beam-selection overhead in V2I networks at no additional cost, e.g.,  \cite{Federated-mmWave-Beam-Selection-LIDAR-2021,Backdoor-Federated-Learning-2022,A-Privacy-Preserved-Split-Learning-2022,LIDAR-Data-DL-2019,LIDAR-Position-Aided-mmWave-Beam-Selection-2022,A-Personalized-Solution-2023,LiDAR-Future-Beam-Prediction-2023}. Moreover, it is widely recognized that data-driven methods can effectively analyze the LiDAR depth map related to mmWave beam quality and process the LiDAR signals as side information for beam searching. In \cite{LIDAR-Data-DL-2019}, the connected vehicle in a V2I scenario leverages its LiDAR data to predict a set of candidate beams conditioned on LoS and NLoS state estimates via a DNN. In \cite{LIDAR-Position-Aided-mmWave-Beam-Selection-2022}, a non-local CNN based beam selection scheme is proposed, where connected vehicles utilize LiDAR measurements along with their location data to reduce the beam search overhead required to establish a reliable communication link with a nearby BS. A RNN model that leverages LiDAR sensory data for beam prediction and tracking is developed in \cite{LiDAR-Future-Beam-Prediction-2023} and evaluated on the DeepSense 6G dataset. In contrast, \cite{LIDAR-Data-DL-2019} and \cite{LIDAR-Position-Aided-mmWave-Beam-Selection-2022} use LiDAR mounted on the vehicles to predict current beams, while \cite{LiDAR-Future-Beam-Prediction-2023} is dedicated to predicting current and future beams using the LiDAR data on the BS side.

In recent years, in addition to the time, space and frequency domains, the semantic domain has been explored to further improve communication efficiency \cite{Deep-Learning-Semantic-Communication,Generalized-Semantic-Communication}. The mmWave channels usually have a sparse structure, and the parameters associated with the structure, such as AOA, AOD, loss and delay, can be regarded as parameter semantics, which can be obtained by radar, GPS, WiFi and other sensing sensors. Meanwhile, key scatters (e.g., the layout, shape, and category) in the propagation environment can be regarded as environment semantics that can be captured through RGB cameras and semantic segmentation techniques. It might be possible to predict beams directly from the channel semantics at both the parameter and environment levels.

{\bf{Summary.}} A critical insight into mmWave/THz beam management is the profound influence of the transmitter and receiver locations, along with the geometry and characteristics of the surrounding environment, on beam selection processes. This implies that possessing knowledge about the environmental context and the precise positioning of the transceivers can substantially enhance beam management, specifically in the facets of beam training, tracking, and prediction. Leveraging the sensing capabilities inherent in ISAC systems presents an effective approach to acquiring this situational awareness. In the context of high-mobility systems such as V2X communication, where the network topology and environmental conditions are in constant flux, radar-based and vision-assisted methodologies emerge as leading contenders for the real-time detection and tracking of cars, pedestrians, road lanes, and barriers. The data procured from radar and vision sensors is subsequently utilized to foresee channel conditions and beam trajectories as users are in motion. Current research provides compelling evidence that the integration of sensing not only elevates the communication rate but also enhances the precision of beam tracking, all while reducing the necessity for extensive training overhead. Nonetheless, previous research employing radar for location tracking, cameras for vision-based observations, and LiDAR technology has predominantly focused on LoS links, with limited capability in addressing NLoS targets. Moreover, there is a pressing need to adapt these proposed methodologies to multi-user environments. For instance, the precision of radar-based positioning can deteriorate in the presence of multiple targets. Furthermore, in ISAC systems that utilize communication signals for sensing purposes, judicious resource distribution across the frequency, time, and spatial domains is critical to striking an optimal balance between sensing accuracy and communication efficacy.

\begin{figure*}[t]
	\centering
	\includegraphics[width=17cm]{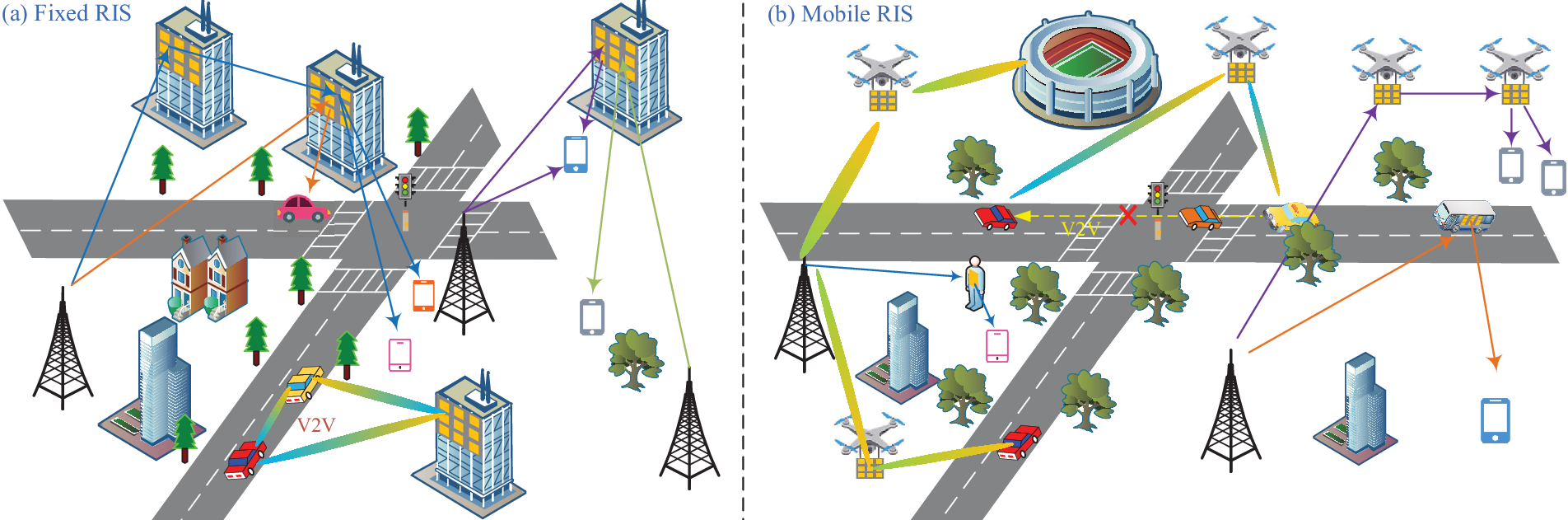}
	\caption{Typical application scenarios of RIS, where (a) fixed RIS deployment and (b) mobile RIS deployment.}
	\label{fig:Use-Case-RIS}
\end{figure*}

\begin{table*}[t]
	\centering
	\caption{Summary of the literature on beam management mechanisms for RIS-assisted systems.}\label{RIS-BM}
	\scalebox{0.8}{
		\begin{tabular}{|c|c|c|c|c|c|c|c|l|}
			\Xhline{0.5pt}
			\Xhline{0.5pt}
			\multicolumn{2}{|c|}{\textbf{Techniques}} &\textbf{Ref.} &\textbf{Year} &\textbf{Frequency} & \textbf{Scenarios} &{\makecell[c]{\textbf{RIS}\\\textbf{Configuration}}} &\textbf{Focus} &\textbf{Performance Metrics}\\
			\Xhline{0.5pt}
			\multicolumn{1}{|c|}{\multirow{23}{*}{\textbf{{\makecell[c]{Beam\\Sweeping}}}}} &{\makecell[l]{$\circ$ Exhaustive\\ $\bullet$ Multi-beam}} &\cite{Fast-Beam-Training-IRS-2020}  &2020 &mmWave &{\makecell[lp{4.2cm}]{An RIS is deployed between a multi-antenna AP and a group of single-antenna users, where the direct AP-user links are blocked.}} &Passive &{\makecell[c]{RIS horizontal\\ beam training}} &{\makecell[l]{$\bullet$ Beam training overhead\\$\bullet$ Beam identification accuracy\\ $\bullet$ Average achievable rate}}\\
			\cline{2-9}
			\multicolumn{1}{|c|}{} &{\makecell[l]{$\circ$ Exhaustive\\ $\bullet$ Partial\\ $\bullet$ Hierarchical}} &\cite{Terahertz-MIMO-RIS-2021}  &2021 &THz &{\makecell[lp{4.2cm}]{A narrowband downlink multi-user massive MIMO system, where each user is served by multiple RISs.}} &Passive &{\makecell[c]{Beam training for\\ both the BS-user link\\ and the reflecting link}} &{\makecell[l]{$\bullet$ Correct detection rate}}\\
			\cline{2-9}
			\multicolumn{1}{|c|}{} &Random &\cite{Joint-Beam-Training-Positioning-RIS-2021}  &2021 &mmWave  &{\makecell[lp{4.2cm}]{Multi-RIS-assisted MIMO system}} &Passive &{\makecell[c]{Joint Beam Training\\ and Positioning}} &{\makecell[l]{$\bullet$ Beam misalignment rate}}\\
			\cline{2-9}
			\multicolumn{1}{|c|}{} &{\makecell[l]{$\circ$ Exhaustive\\ $\bullet$ Distributed}} &\cite{Distributed-Beam-Training-2021}  &2021 &\emph{Null} &{\makecell[lp{4.2cm}]{A multi-antenna BS serves a single-antenna user via a multi-hop LoS link formed by distributed RISs.}} &Passive  &{\makecell[c]{Beam training\\ at the BS and RISs}} &{\makecell[l]{$\bullet$ Effective channel gain}}\\
			\cline{2-9}
			\multicolumn{1}{|c|}{} &Multi-beam &\cite{Multi-Path-Beam-Routing-RIS-2022}  &2022 &\emph{Null} &{\makecell[lp{4.2cm}]{A multi-antenna BS communicates with a single-antenna user via one or more RIS reflection paths.}} &Passive &{\makecell[c]{Active and passive\\ beam selection}} &{\makecell[l]{$\bullet$ Optimized reflection paths \\$\bullet$ Received signal power}}\\
			\cline{2-9}
			\multicolumn{1}{|c|}{} &{\makecell[l]{$\circ$ Exhaustive\\$\circ$ Hierarchical\\ $\bullet$ Multi-beam}} &\cite{Beam-Training-Alignment-RIS-2022}  &2022 &mmWave &{\makecell[lp{4.2cm}]{MIMO system aided by a RIS, where the direct BS-user link is blocked.}} &Passive &{\makecell[c]{BS-RIS-user \\ beam training}} &{\makecell[l]{$\bullet$ Average achievable rate\\$\bullet$ Beam training overhead}}\\
			\cline{2-9}
			\multicolumn{1}{|c|}{} &Hierarchical &\cite{Codebook-design-beam-training-RIS-2022}  &2022 &\emph{Null} &{\makecell[lp{4.2cm}]{A large-scale RIS is deployed between a multi-antenna BS and a single-antenna user, where the direct BS-user link is blocked.}} &Passive &{\makecell[c]{Near-field\\ RIS beam training}} &{\makecell[l]{$\bullet$ Achievable rate\\$\bullet$ Beam training overhead}}\\
			\cline{2-9}
			\multicolumn{1}{|c|}{} &{\makecell[l]{$\circ$ Exhaustive\\$\bullet$ Multi-beam}} &\cite{Fast-Beam-Training-Alignment-IRS-2022}  &2022 &{\makecell[c]{mmWave/\\THz}} &{\makecell[lp{4.2cm}]{An RIS is deployed between a multi-antenna BS and a number of single-antenna users, where all BS-user links are blocked.}} &Passive &{\makecell[c]{Active and passive\\ beam training}} &{\makecell[l]{$\bullet$ Success rate\\$\bullet$ Beamforming gain ratio}}\\
			\Xhline{0.5pt}
			\multicolumn{1}{|c|}{\multirow{23}{*}{\textbf{AI}}} &{\makecell[c]{Conventional ML\\(Genetic algorithm)}} &\cite{Real-Time-Beam-steering}  &2021 &mmWave &{\makecell[lp{4.2cm}]{A reconfigurable meta-surface with beam-steering capabilities has been adopted to track the position of a mobile receiver.}} &Passive &Beam steering &{\makecell[l]{$\bullet$ Outage probability\\ $\bullet$ Channel capacity}}\\
			\cline{2-9}
			\multicolumn{1}{|c|}{} &FCNN &\cite{RIS-Aided-Indoor-Beam-Alignment}  &2021 &5.8 GHz &{\makecell[lp{4.2cm}]{A prototype of wireless communication using the RIS consisting of 256 meta-material antenna elements.}} &Passive &Beam alignment &{\makecell[l]{$\bullet$ CDF of positioning error\\ $\bullet$ CDF of beam angle error}}\\
			\cline{2-9}
			\multicolumn{1}{|c|}{} &RNN &\cite{Deep-Learning-THz-Drones}  &2021 &THz &{\makecell[lp{4.2cm}]{A THz drone network in which a mobile drone is served by a BS and a flying RIS.}} &Passive & Beam prediction &{\makecell[l]{$\bullet$ Prediction accuracy}}\\
			\cline{2-9}
			\multicolumn{1}{|c|}{} &LSTM &\cite{Active-Sensing-learning-2022}  &2022 &mmWave &{\makecell[lp{4.2cm}]{An RIS-assisted system in which a multi-antenna BS serves a single-antenna user.}} &Passive &Beam alignment &{\makecell[l]{$\bullet$ Mean square estimation\\ $\bullet$ Average beamforming gain}}\\
			\cline{2-9}
			\multicolumn{1}{|c|}{} &DDQN &\cite{Enabling-Efficient-Blockage-Aware-Handover}  &2022 &mmWave &{\makecell[lp{4.2cm}]{An RIS-assisted mmWave cellular network with a set of BSs and users, where each UE is equipped with a single omni-directional antenna and BS utilizes a phased array.}} &Passive &{\makecell[c]{Blockage prediction\\(Beam tracking)}} &{\makecell[l]{$\bullet$ Prediction accuracy\\ $\bullet$ Average spectrum efficiency\\ $\bullet$ Handover number}}\\
			\cline{2-9}
			\multicolumn{1}{|c|}{} &{\makecell[c]{Deep\\Residual Network}} &\cite{Low-overhead-Beam-Training-Large-Scale-RIS-2022}  &2022 &mmWave &{\makecell[lp{4.2cm}]{An large-scale RIS is employed between a multi-antenna BS and multiple single-antenna users to assist communications, the direct links are blocked.}} &Passive &Beam training &{\makecell[l]{$\bullet$ Normalizd RIS beam gain \\ $\bullet$ Effective achievable rate}}\\
			\cline{2-9}
			\multicolumn{1}{|c|}{} &FL &\cite{Stratified-Federated-Learning-Beamforming-Design-2023}  &2023 &\emph{Null} &{\makecell[lp{4.2cm}]{A multi-RIS-aided multi-user system with one multi-antenna BS, and each user is equipped with single antenna.}} &Passive &Beamforming &{\makecell[l]{$\bullet$ Spectrum efficiency \\ $\bullet$ Achievable rate }}\\
			\Xhline{0.5pt}
			\multicolumn{1}{|c|}{\multirow{10}{*}{\textbf{Sensing}}} &Dedicated Sensors &\cite{Joint-Hybrid-3D-Beamforming-2022}  &2022 &THz &{\makecell[lp{4.2cm}]{An RIS-assisted THz multi-user
					massive MIMO system, where the BS-user link is blocked.}} &{\makecell[c]{Passive and \\integrated with\\wideband sensors}} &{\makecell[c]{RIS-user beam training}} &{\makecell[l]{$\bullet$ Spectral efficiency\\$\bullet$ Achievable rate}}\\
			\cline{2-9}
			\multicolumn{1}{|c|}{} &Dedicated Sensors &\cite{Computer-Vision-Aided-RIS-Beam-Tracking}  &2022 &5.4 GHz &{\makecell[lp{4.2cm}]{A vision-aided RIS prototype system, in which the BS and user are equipped with single antenna and the LoS path is blocked.}} &{\makecell[c]{Passive and \\attached with a\\ binocular camera}} &{\makecell[c]{RIS beam tracking}} &{\makecell[l]{$\bullet$ SNR variation}}\\
			\cline{2-9}
			\multicolumn{1}{|c|}{} &Dedicated Sensors &\cite{Sensing-Aided-RIS-3GPP-5G}  &2022 &\emph{Null} &{\makecell[lp{4.2cm}]{A standalone RIS is placed to aid the communication between a multi-antenna BS and a single-antenna user, and a blockage exists between the BS and the user.}} &{\makecell[c]{Semi-passive\\and attached with\\cameras}} &RIS beam selection &{\makecell[l]{$\bullet$ Accuracy/recall performance\\ $\bullet$ Achievable rate}}\\
			\Xhline{0.5pt}
			\Xhline{0.5pt}
			\multicolumn{8}{l}{*Note that the symbol ``$\bullet$'' in the ``Techniques'' column indicates the core scheme proposed in the paper.}\\
		\end{tabular}
	}
\end{table*}

\subsection{Beam Management for RIS-Enhanced Systems}
Signal/channel blockage is considered to be a major issue hindering the practical application of mmWave/THz communications. To alleviate it and thus enhance signal coverage, an attractive paradigm is to deploy RISs properly in the system to provide effective reflection paths. Considering different RIS deployment modes, the typical application scenarios are shown in Fig.~\ref{fig:Use-Case-RIS}. Due to the limited scattering of mmWave/THz channels, it is necessary to conduct beam management to fully reap the passive beamforming gain of RIS. Specifically, for the idle users, initial beam training/alignment at the AP/BS as well as at the RIS is required to establish a high-quality link before data transmission. Then fast beam tracking is performed to maintain the links for the connected users. As beam management for RIS-assisted mmWave/THz systems requires jointly manage the beams of the two-hop transmission, AP/BS-RIS and RIS-user, it is much more challenging than that for conventional systems with one-hop transmission topology. This subsection discusses and summarizes the major approaches proposed to address the beam management issue in RIS-assisted mmWave/THz communication systems. We classify the existing schemes into three categories, including methods based on beam sweeping, AI-driven methods, and sensing-aided methods, which are revealed in Table~\ref{RIS-BM}. Detailed descriptions are as follows.

\begin{figure}[t]
	\centering
	\includegraphics[width=8cm]{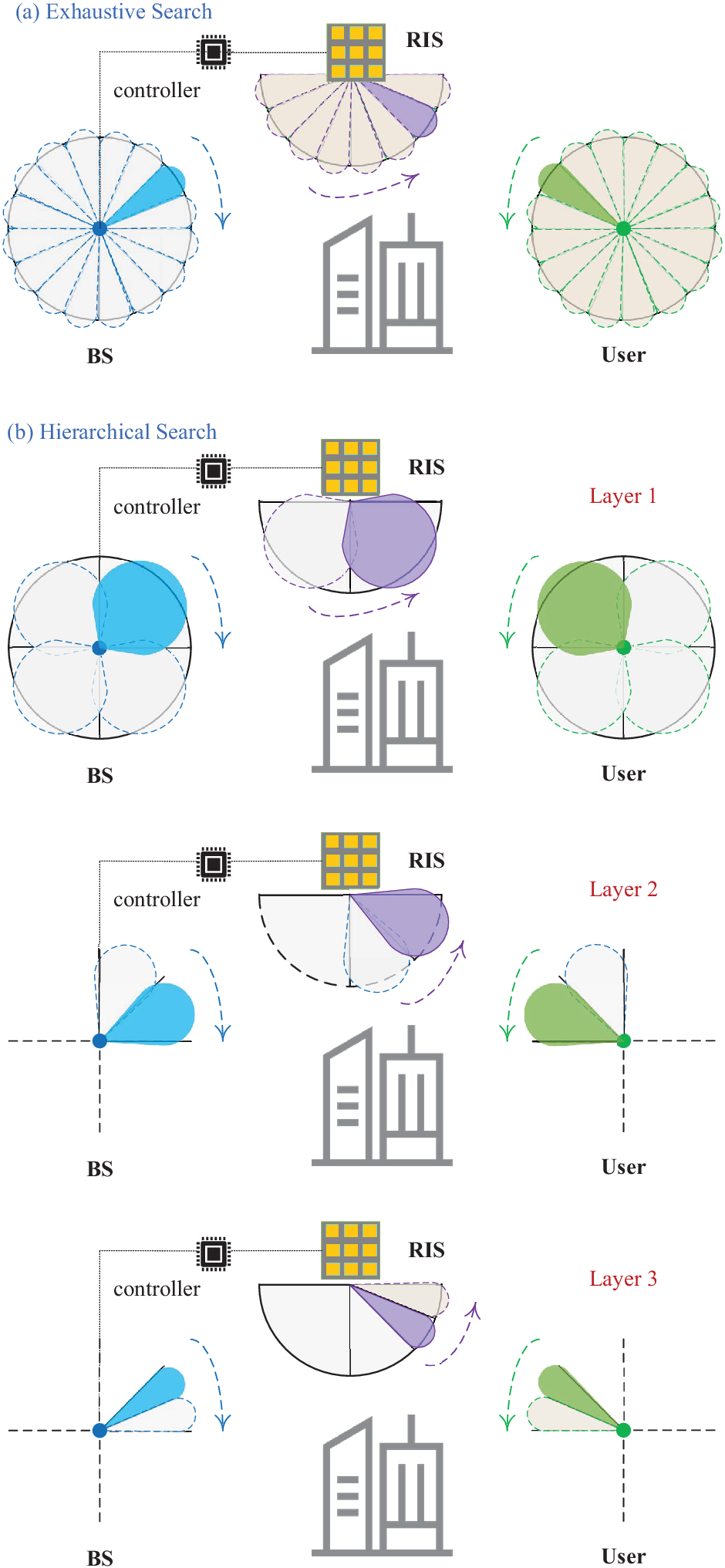}
	\caption{Typical beam search approaches for RIS-assisted systems, where (a) exhaustive search and (b) a 3-layer hierarchical search.}
	\label{fig:Use-Case-RIS-beam-sweeping}
\end{figure}

{\bf{Beam Sweeping-based Methods.}} A general beam training and tracking technique for RIS-assisted mmWave/THz communications is to perform beam sweeping based on pre-designed active/passive beamforming codebooks \cite{Fast-Beam-Training-IRS-2020,Terahertz-MIMO-RIS-2021,Joint-Beam-Training-Positioning-RIS-2021,Distributed-Beam-Training-2021,Multi-Path-Beam-Routing-RIS-2022,Beam-Training-Alignment-RIS-2022,Codebook-design-beam-training-RIS-2022,Fast-Beam-Training-Alignment-IRS-2022}. Each codebook consists of only a finite number of beam directions or patterns. The beam-sweeping-based schemes do not require complex channel estimation, so they are very beneficial for large RIS. Note that the method for the AP/BS-user link cannot be directly applied to the cascaded AP/BS-RIS-user link since RIS itself cannot generate or receive beams like AP/BS or user. In the following, we discuss the extension of conventional beam sweeping techniques in RIS-assisted systems. Two typical approaches are shown in Fig.~\ref{fig:Use-Case-RIS-beam-sweeping}.

$\bullet$ \emph{Exhaustive Search:} As is widely used in WLAN/WPAN, exhaustive beam search \cite{Fast-Beam-Training-IRS-2020,Terahertz-MIMO-RIS-2021,Distributed-Beam-Training-2021,Beam-Training-Alignment-RIS-2022,Fast-Beam-Training-Alignment-IRS-2022} is the most precise and straightforward approach for performing beam training, in which no prior information is known. For RIS-assisted mmWave/THz communications, the optimal beam pair of the direct AP/BS-user link can be estimated by steering beams as previously mentioned. To acquire the best beam combinations for the cascaded link, a natural approach is to exhaustively search all possible $\left| {{{\cal C}_B}} \right|\cdot {\left| {{{\cal C}_R}} \right|}^2 \cdot \left| {{{\cal C}_U}} \right|$ beam tuples/triplets, where ${\cal C}_B$, ${\cal C}_R$ and ${\cal C}_U$ are the predefined training codebooks of AP/BS, RIS and user, respectively. For example, the codebook corresponding to a set of narrow beams can be designed by uniformly quantizing the whole space of interest (e.g., the AoD for AP/BS, the AoA for user, and the relative reflection angle for RIS) \cite{Beam-Training-Alignment-RIS-2022}. With the codebooks, an exhaustive-search-based RIS beam training is as follows: The transmitter (e.g., AP/BS) consecutively sends multiple training symbols, while the RIS sequentially scans its entire reflecting directions over different training symbols, and the receiver (e.g., user) needs to scan the entire AoA space for every combination of transmitting and reflecting beam patterns, so that the optimal beam that achieves the maximum received signal power/SNR will be found. Although straightforward, such method is practically prohibitive for large RIS with massive number of reflecting elements, which incurs excessively high training complexity and time overhead for establishing high quality links. Therefore, the exhaustive-search-based RIS beam training methods described in the existing literature are mainly used to compare their proposed schemes.

$\bullet$ \emph{Hierarchical Search:} As the exhaustive beam search is time consuming, hierarchical multi-resolution beam search \cite{Terahertz-MIMO-RIS-2021,Beam-Training-Alignment-RIS-2022,Codebook-design-beam-training-RIS-2022} has been widely accepted to reduce the complexity. A hierarchical-search-based approach generally realizes beam training based on multi-layer codebooks where a lower-layer codebook consists of wider beams in comparison with higher-layer codebooks, rather than exhaustively searching over all narrow-beam combinations. The beam sweeping procedure for each layer is similar to that of the exhaustive search, except that here we only need to search the range that is identified in the earlier stage. That is, the search range of the latter layer sub-codebook is determined by the optimal codeword searched by the previous layer codebook. Therefore, the performance of hierarchical search is highly dependent on training codebook design and the spatial resolution increases with the number of layers. B. Ning \emph{et al.} \cite{Terahertz-MIMO-RIS-2021} presented a cooperative beam training scheme that combines \emph{partial search} at RIS and ternary-tree hierarchical search at BS and users. In particular, only a subset of candidate RIS phase shift solutions needs to be searched, which helps to reduce the complexity. To realize the ternary-tree search, two codebooks are designed, namely tree dictionary codebook and phase shift deactivation codebook. Thanks to low cost and low power consumption, RIS is more likely to develop into super-scale RIS for future 6G communications to efficiently boost the system capacity. X. Wei \emph{et al.} \cite{Codebook-design-beam-training-RIS-2022} point out that the scatters are more likely to be in the near-field region of the super-scale RIS and the far-field codebook \cite{Fast-Beam-Training-IRS-2020,Terahertz-MIMO-RIS-2021} mismatches the near-field channel model. The existing far-field beam training scheme will cause severe performance loss in the super-scale RIS assisted near-field communications. To solve this problem, they designed a hierarchical near-field codebook which consists of several different levels of sub-codebooks determined by different sampling ranges and sampling steps. Based on the codebook, the corresponding beam training scheme is further proposed.

$\bullet$ \emph{Multi-beam Search:} Conventional beam training usually searches over possible beam directions by a single beam. The single-beam search is practically challenging for systems assisted by RIS with large reflecting elements that can generate pencil-like beams, because it requires a large number of beam directions in the training codebook to cover the space of interest. To reduce the single-beam training time, multi-beam training has attracted the attention of some researchers \cite{Fast-Beam-Training-IRS-2020,Multi-Path-Beam-Routing-RIS-2022,Beam-Training-Alignment-RIS-2022,Fast-Beam-Training-Alignment-IRS-2022}. For instance, a multi-beam sweeping method based on grouping-and-extracting was proposed by C. You \emph{et al.} \cite{Fast-Beam-Training-IRS-2020}. They divided the RIS reflecting elements into multiple sub-arrays and designed their multi-beam codebook to steer different beam directions simultaneously over time. Then, the user can detect its optimal RIS beam direction via simple comparison of the received signal power/SNR. For simplicity, this work assumes that the AP-RIS link and the RIS vertical beamforming has aligned, and then only focuses on the horizontal beam training between the RIS and the user. Nevertheless, in practice, the information of RIS location may not be available to the BS, in which case one needs to perform a joint BS-RIS-user beam training. W. Mei \emph{et al.} \cite{Multi-Path-Beam-Routing-RIS-2022} proposed a more general multi-path beam routing scheme by exploiting  active beam splitting and passive beam combining techniques. Specifically, the BS sends the user's information signal via multiple orthogonal active beams pointing towards different RISs, and then these beamed signals are subsequently reflected by selected RISs via their cooperative passive beamforming in different paths, and finally coherently combined at the user. In addition to the active multi-beam, the passive multi-beam has also been considered by P. Wang \emph{et al.} \cite{Beam-Training-Alignment-RIS-2022,Fast-Beam-Training-Alignment-IRS-2022}. That is, they proposed to let both the BS and the RIS form multiple pencil beams simultaneously and steer them towards different directions. It is worth mentioning that whether passive multi-beam training is effective in practice needs further verification, as the inter-beam interference, the achievable passive beamforming gain, and the codebook size at RIS may increase the overall training complexity.

$\bullet$ \emph{Other Advanced Methods:} Instead of exhaustively or sequentially searching over the combinations of active and passive beam patterns, several other beam training methods have been tried to reduce the complexity. For instance, W. Wang \emph{et al.} \cite{Joint-Beam-Training-Positioning-RIS-2021} proposed to break down beam training of multi-RIS assisted mmWave MIMO into two mathematically equivalent sub-problems. They further perform random beamforming and maximum likelihood estimation to jointly estimate AoA and AoD of the dominant path in each sub-problem. A beam training method with combined offline and online distributed beam training is proposed in \cite{Distributed-Beam-Training-2021}, by exploiting the (nearly) time-invariant BS-RIS and inter-RIS channels and the cooperative training among the BS and RISs' controllers. In order to provide a zero overhead training procedure for RIS-assisted single-input multiple-output (SIMO) orthogonal frequency-division multiplexing (OFDM) system, K. Chen-Hu \emph{et al.} \cite{Differential-Data-Aided-Beam-Training} investigated a differential-data-aided beam training combined with a codebook, relying on data transmission and reception based on non-coherent demodulation.

{\bf{AI-driven Methods.}} The AI technique is a powerful technology that has gained significant interest in wireless communications due to its learning ability and huge search space. AI may constitute efficient approaches for leveraging the potential benefits of RIS-empowered smart radio environment. For RIS, especially for large RIS with massive number of reflecting elements, channel estimation or reflection beam training will face the following challenges. (1) If all RIS elements are passive, prohibitive training overhead will be incurred. (2) In the case of fully-digital or analog-digital hybrid RIS architectures, hardware complexity and power consumption is not feasible in practice \cite{Enabling-Large-RIS-DL-2021}. In this context, RIS combined with AI has been proposed as a potential solution to improve the coverage and spectrum efficiency for 6G wireless communications.

In recent years, RIS-assisted communication has adapted different AI techniques, especially ML, for performance enhancement. For instance, the joint design of transmit beamforming at BS and phase shifting at reflecting RIS using DRL to maximize the sum rate of a multi-user downlink MISO system was studied by C. Huang \emph{et al.} \cite{RIS-Multiuser-MISO-DRL-2020}. Specifically, an algorithm based on DDPG neural network is developed to solve the optimization problem, in which the joint design is obtained through trial-and-error interactions with the environment by observing predefined rewards, in the context of continuous state and action. In \cite{Millimeter-Wave-Communications-IR-DRL-2022}, Q. Zhang \emph{et al.} proposed a DRL approach to learn the optimal RIS reflection and maximize the rate expectation of the RIS-aided downlink MISO transmission with imperfect CSI, where an iterative learning algorithm based on quantile regression is developed to model the probability distribution of rate. Compared to the conventional or traditional model-based approaches, AI-based technology can help better extract the inherent relationship between input and output signals and enable more reliable channel estimation, phase shift configuration and beamforming. M. A. S. Sejan \emph{et al.} \cite{ML-RIS-2022} provided a comprehensive overview of the state-of-the-art on ML/DL-based RIS-enhanced wireless communications, and classified ML applied in RIS for phase shift configuration and beamforming into DL, RL, SL, unsupervised learning, and FL. They concluded that the AI/ML-based approaches have a comparable performance to conventional methods while reducing computational complexity.

\begin{figure}[t]
	\centering
	\includegraphics[width=8.5cm]{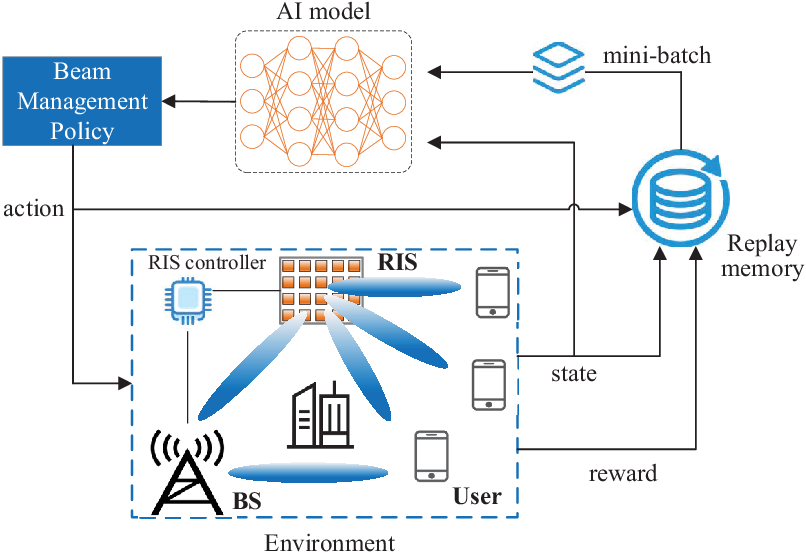}
	\caption{DRL-based beam management in RIS-aided communication system.}
	\label{fig:RIS-DRL}
\end{figure}

Back to beam management, in order to alleviate the heavy overhead of the passive beam alignment in RIS-assisted communications, different AI techniques are recently introduced for beam tracking under various wireless communication environments \cite{Real-Time-Beam-steering,RIS-Aided-Indoor-Beam-Alignment,Deep-Learning-THz-Drones,Active-Sensing-learning-2022,Enabling-Efficient-Blockage-Aware-Handover,Low-overhead-Beam-Training-Large-Scale-RIS-2022}. A typical DRL approach is shown in Fig.~\ref{fig:RIS-DRL}. For instance, C. Rizza \emph{et al.} \cite{Real-Time-Beam-steering} proposed a Genetic algorithm to integrate a reconfigurable meta-surface as computing unit and perform real-time beam steering functions in a mobile scenario. C. Xiao \emph{et al.} \cite{RIS-Aided-Indoor-Beam-Alignment} constructed an indoor wireless communication prototype for verifying the passive beamforming using RIS and proposed an architecture of adaptive beam alignment and indoor positioning \cite{RF-VLCP} by using DNN. The beam alignment is realized by using the relationship between an energy array image and the beam direction of the RIS. To integrate a flying RIS into THz drone communications, N. Abuzainab  \emph{et al.} \cite{Deep-Learning-THz-Drones} considered a RNN solution for beam prediction and proactive hand-off based on the prior observations of drone location/beam trajectories. F. Sohrabi \emph{et al.} \cite{Active-Sensing-learning-2022} developed an adaptive active and reflection beam alignment strategy to the high-dimensional analog channel sensing problems by using a LSTM framework. L. Jiao \emph{et al.} \cite{Enabling-Efficient-Blockage-Aware-Handover} proposed an RIS-assisted handover scheme by leveraging DDQN to deal with the frequent mmWave channel blockages, where the DRL agent manages to reduce the cumulative handover overhead by jointly adjusting beamformers and RIS phase shifts. W. Liu \emph{et al.} \cite{Low-overhead-Beam-Training-Large-Scale-RIS-2022} proposed two DL-based near-field beam training schemes for large-scale RIS-assisted communication systems, where deep residual networks are employed to determine the optimal near-field RIS codeword.

While AI can improve communication performance for RIS-assisted systems in different ways, the privacy of user data has often been ignored in previous work. Against this background, a privacy-preserving paradigm combining FL with RIS in the mmWave/THz communication system needs to be designed. L. Li \emph{et al.} \cite{Enhanced-RIS-FL-2020} proposed two FL-based mmWave wireless communication scenarios. One is RIS-assisted outdoor mmWave communication, where the user participating in communication under the FL framework is regarded as a client/agent that trains a local model, and the controller of RIS is regarded as the central server for local model aggregation. The second scenario is multi-RIS-assisted IoT communication, where an AP connected to multiple RISs is regarded as the central server while each RIS as client. The former applies FL to train the optimal DNN model of the mapping between the user channels and the optimal RIS configuration matrix through distributed learning, so as to effectively protect the privacy of users while maximizing the user achievable rate. In the IoT communication, FL is used to optimize multiple RISs in parallel under the protection of private CSI, so as to achieve the optimal sum rate of combined signals, namely the superposition of all RISs' signals. In \cite{Stratified-Federated-Learning-Beamforming-Design-2023}, H. Min \emph{et al.} proposed an unsupervised stratified FL-based beamforming optimization approach for heterogeneous RIS-aided multi-user communication systems, of which the aim was to maximize the achievable sum rate by jointly optimizing the active beamforming at BS and the passive phase shifts at RISs. It allows users to train NNs collaboratively in an unsupervised manner by exploiting internal correlations within the observed channel data while avoiding the effects of heterogeneity of distributed datasets.

Although AI-based beam management for mmWave communications has been studied extensively, there are limited solutions for RIS-assisted mmWave and THz communication scenarios. Since RIS beam alignment for mmWave/THz communications usually requires high computational complexity, the scheme of combining RIS-assisted system with AI is worth further exploration.

{\bf{Sensing-aided Methods.}} In recent years, we have witnessed a great deal of effort to design solutions that integrate sensing capabilities into communication systems, i.e., ISAC. More recently, RIS-assisted ISAC has emerged and is attracting increasing attention in both academia and industries. A. M. Elbir \emph{et al.} \cite{The-Rise-IRS-ISAC-2022} investigated the role of RIS in the future ISAC paradigms. It presents various RIS-assisted radar and ISAC models that leverage RIS to provide flexibility for dynamic and accurate beamforming, resulting in advantages such as improved energy-efficiency, spectral efficiency, signal coverage, parameter estimation, and interference suppression.

While there is a rich literature on RIS-assisted wireless communications, the RIS-assisted ISAC remains relatively unexamined. By considering a basic MIMO ISAC system model, S. P. Chepuri \emph{et al.} \cite{ISAC-RIS-2022} showcased that joint sensing and communications designs are mostly beneficial when the respective channels are coupled, and that RISs can facilitate the efficient control of this beneficial coupling. L. Wang \emph{et al.} \cite{JCAS-RIS-mmWave-2022} studied the potential of RISs in the OFDM ISAC system at mmWave bands, where the hybrid beamforming and RIS phase shifts are jointly designed to guarantee the functionalities of both communication for the mobile users and sensing of the targets. J. Zuo \emph{et al.} \cite{NOMA-RIS-ISAC-2022} proposed a RIS-NOMA-ISAC system, in which the RIS beampattern gain for the radar target is adopted as a sensing metric. The objective is to maximize the minimum beampattern gain by jointly optimizing active beamforming, power allocation coefficients and passive beamforming. H. Luo \emph{et al.} \cite{Joint-Beamforming-Design-RIS-ISAC} investigated the joint active and passive beamforming design for an RIS-assisted ISAC system, with the aim of maximizing the achievable sum-rate of the communication users, while satisfying the constraints of beampattern similarity for radar sensing, the RIS restriction and the transmission power budget. Z. Zhu \emph{et al.} \cite{Resource-Allocation-RIS} focused on an RIS-aided ISAC system for the multi-user downlink scenario in the mmWave band, and formulated a sum-rate maximization problem by jointly optimizing the radar signal covariance matrix, the communication beamforming vector, and the RIS phase shift. X. Song \emph{et al.} \cite{Joint-Transmit-Reflective-Beamforming-RIS-ISAC} studied the joint transmit and reflective beamforming design in an RIS-assisted ISAC system with one communication user and multiple potential sensing targets at the NLoS areas of the BS. The objective is to maximize the RIS minimum beampattern gain towards the desired sensing angles by jointly optimizing the transmit information and sensing beamforming at the BS and the reflective beamforming at the RIS. To minimize the total transmit power at the BS in an RIS-aided radar-communication system, M. Hua \emph{et al.} \cite{Joint-Active-Passive-Beamforming-Design-HuaMeng} jointly optimized the active beamformers at the BS, i.e., communication beamformers and radar beamformers, and the phase shifts at the RIS, subject to the minimum SINR required by communication users, the minimum SINR required by the radar, and the cross-correlation pattern design.

The aforementioned studies indicate that for RIS-assisted ISAC systems under various settings, both sensing and communications can experience improved performance when the RIS is properly deployed and designed. In particular, the beamforming design in RIS-assisted ISAC scenarios needs to simultaneously consider the mobile users for communications, the targets for sensing, and the phase configuration of RIS. It is noteworthy that, to date, the research on beam management for RIS-assisted ISAC at mmWave and THz bands is still a gap to be filled, to the best of our knowledge.

So far, there are only a few attempts made on beam management based on sensing for RIS-assisted systems. In \cite{Joint-Hybrid-3D-Beamforming-2022}, the ultra-wideband sensors are integrated into the RIS, and the user location information obtained by the sensors is exploited to expedite the channel estimation and beamforming processes for RIS-assisted THz multi-user massvie MIMO systems. In fact, among various sensing means, computer vision may be a suitable option for the design of RIS reflection coefficient in RIS-assisted mmWave/THz communication systems \cite{Computer-Vision-Aided-RIS-Beam-Tracking}, because (1) complex feedback links can be replaced by cheap cameras while keeping RIS low cost, (2) camera can effectively and quickly locate users within the LoS range, and (3) the acquisition of visual information does not occupy the communication frequency bands, thus saving spectrum resources. For instance, M. Ouyang \emph{et al.} \cite{Computer-Vision-Aided-RIS-Beam-Tracking} proposed a computer vision-based approach to aid RIS for dynamic beam tracking, where a binocular camera is used to obtain visual information, and the object detection and stereo vision algorithms are used to calculate the user coordinates with respect to the RIS. With the user coordinates, RIS quickly identifies the desired reflected beam direction and then adjusts the reflection coefficients according to the pre-designed codebook. Unlike most prior work using fully passive RIS, S. Jiang \emph{et al.} \cite{Sensing-Aided-RIS-3GPP-5G} investigated the feasibility of enabling 3GPP 5G transparent/standalone RIS operation, where the RIS is equipped with sparse active antenna elements what is known as the semi-passive RIS architecture. To guide the RIS beam selection in the proposed standalone RIS operation, the authors developed a ML framework and a neural network architecture that leverage the visual data captured by cameras installed at the RIS to predict the user-sider candidate beam set.

{\bf{Summary.}} To maximize the benefits of passive beamforming in RIS-enhanced mmWave/THz communication systems, sophisticated beam management for the dual-hop transmission architecture is essential yet poses significant challenges, as the RIS lacks the capability to autonomously generate and decode beams. Traditional exhaustive search based beam management methods, while comprehensive, lead to exorbitant training overheads. Hierarchical search strategies can mitigate this overhead effectively, yet they suffer from a key limitation: employing broader beams in the initial stages of the search can substantially diminish beamforming gains. Additionally, employing a multi-beam search approach in a practical setting may confront issues such as escalated inter-beam interference, diminished passive beamforming effectiveness, and an expansion in the required codebook size. Owing to their rapid response capabilities, AI-based algorithms are highly suited for mobile communications, such as V2X, where high mobility poses significant challenges necessitating real-time beam steering. As RIS-assisted ISAC gains prominence, the design of beamforming has garnered considerable attention. However, the research on beam management, particularly for ISAC systems operating in the mmWave and THz bands, remains an area in urgent need of further development. Recent innovations include the integration of computer vision (cameras) into RIS-assisted communication systems, aiming to reduce beam training overhead and eliminate the need for additional feedback links. The synergy of RIS-assisted systems with AI and/or ISAC presents a promising avenue for the advancement of future beam management, warranting extensive exploration. Additionally, while the standalone (semi-passive) RIS architecture aids in channel estimation, it does come with the trade-off of increased costs.

\begin{table*}[t]
	\centering
	\caption{Comparison of the general overhead of the surveyed beam management techniques.}\label{BM-overhead}
	\scalebox{0.9}{
		\begin{tabular}{|l|l|l|l|}
			\Xhline{0.5pt}
			\Xhline{0.5pt}
			\textbf{Technologies} & \textbf{Overhead} & \textbf{Pros \& Cons} &\textbf{Parameters}\\
			\Xhline{0.5pt}
			Exhaustive Search&${N_T} \cdot {N_R}$ &Simple and effective, but time-consuming. &{\multirow{11}{*}{\makecell[l]{$N_T$: Number of transmit beams;\\$N_R$: Number of receive beams;\\ $M$: Number of search layers;\\$N_{T,m}$: Number of transmit beams at layer $m$; \\$N_{R,m}$: Number of receive beams at layer $m$; \\ $n_T$: Number of simultaneous beams;\\ $N_{T,c}$: Number of promising candidate beams. }}} \\
			\cline{1-3}
			Hierarchical Search&$\sum\nolimits_{1 \le m \le M} {{N_{T,m}}} {N_{R,m}}$ &{\makecell[lp{7cm}]{Reduces the number of beams that need to be measured, but may experience significant misalignment error propagation.}} & \\
			\cline{1-3}
			Multi-beam Search&$\left\lceil {{{{N_T}} \mathord{\left/
						{\vphantom {{{N_T}} {{n_T}}}} \right.
						\kern-\nulldelimiterspace} {{n_T}}}} \right\rceil  \cdot {N_R}$ &{\makecell[lp{7cm}]{Efficiency can be improved, but the beam alignment accuracy may be reduced due to inter-beam interference.}}& \\
			\cline{1-3}
            AI-based Method &${N_{T,c}} \cdot {N_R}$ &{\makecell[lp{7cm}]{Enhances alignment accuracy and reduces optimization time, but heavily dependent on the quality and quantity of data, with generalization issues in complex environments.}}& \\	
            \cline{1-3}
            Sensing-based Method &${N_{T,c}} \cdot {N_R}$ &{\makecell[lp{7cm}]{Enhances resource utilization efficiency and communication quality, but may increase system complexity and heavily rely on the accuracy of sensing information.}}& \\ 		
			\Xhline{0.5pt}
			\Xhline{0.5pt}
		\end{tabular}
	}
\end{table*}

\subsection{Lessons Learned: Summary and Insights}
Table~\ref{BM-overhead} provides a summary of the key characteristics of existing mmWave and THz beam management techniques discussed in Section III-B, C and D. Then, some lessons are shared, as follows:

{\bf{Lesson 1.}} Lessons learned for beam management based on AI in the context of wireless communication systems include:  \emph{1) Adaptability and Flexibility:} AI-based beam management systems must be adaptable to changing environmental conditions and flexible enough to adjust beamforming parameters dynamically. This adaptability allows the system to optimize performance in real time, considering factors such as mobility, interference, and multipath propagation. \emph{2) Data-Driven Optimization:} Leveraging large-scale data analysis and machine learning techniques, AI can optimize beam management based on historical data and real-time feedback. The system should be designed to continuously learn from its operational environment to improve decision-making and adapt to new scenarios. \emph{3) Complexity Management:} AI-based beam management introduces algorithmic complexity. It's important to carefully manage the computational overhead associated with AI algorithms, ensuring that the system remains efficient and responsive, especially in resource-constrained environments such as mobile devices or IoT devices. \emph{4) Robustness and Reliability:} AI-based beam management systems need to be robust against uncertainties in the environment, such as dynamic channel conditions, varying user distributions, and unexpected interference. Robust algorithms and models are essential to ensure reliable performance under diverse operating conditions. \emph{5) Interoperability and Standards:} When deploying AI-based beam management solutions, interoperability and standardization play crucial roles. Lessons learned highlight the importance of developing systems that adhere to industry standards, enabling seamless integration with existing network infrastructure and equipment from different vendors. \emph{6) Ethical Considerations:} As AI becomes increasingly integrated into critical systems, including beam management, it's important to address ethical considerations such as privacy, fairness, and transparency. Ensuring that AI algorithms operate ethically and transparently is an important aspect of system design and deployment. \emph{7) Human-in-the-Loop Integration:} Human expertise remains crucial in overseeing and interpreting the decisions made by AI-based beam management systems. Integrating human expertise into the loop can help validate AI-generated insights, interpret complex situations, and handle edge cases that AI may struggle to address. \emph{8) Security and Resilience:} AI-based beam management systems should be designed with security and resilience in mind. Lessons learned emphasize the need for robust cybersecurity measures to protect AI models and the communication infrastructure against potential threats and attacks. Overall, the lessons learned emphasize the need for a holistic approach to integrating AI into beam management, considering technical, operational, ethical, and regulatory aspects to ensure successful deployment and operation in real-world environments. Ongoing research and practical experience will continue to shape best practices in this rapidly evolving field.

{\bf{Lesson 2.}} Lessons learned from sensing-assisted beam management, which integrates various types of sensor data to optimize wireless communication systems, include the following: \emph{1) Enhanced Environmental Awareness:} Sensing-assisted beam management provides a comprehensive view of the wireless communication environment by leveraging data from sensors such as radar, LiDAR, or environmental monitors. This heightened environmental awareness enables more informed beamforming decisions, taking into account factors like user locations, interference sources, and dynamic obstacles. \emph{2) Dynamic Adaptation to Environmental Changes:} Lessons learned emphasize the importance of real-time adaptation to changes in the environment. Sensing-assisted systems should be able to dynamically adjust beamforming parameters based on sensor feedback, enabling efficient response to variations in user distribution, mobility patterns, and channel conditions. \emph{3) Optimization through Multi-Modal Sensor Fusion:} Integrating data from diverse sensors allows for multi-modal sensor fusion, which can provide a more complete understanding of the communication environment. Lessons learned highlight the benefits of combining data from different sensors to create a holistic view that improves the accuracy and reliability of beam management decisions. \emph{4) Robustness in Challenging Conditions:} Sensing-assisted beam management systems need to exhibit robust performance in challenging conditions, such as high interference, multipath propagation, or rapidly changing environments. Lessons learned stress the importance of designing algorithms and decision-making processes that maintain performance under adverse circumstances. \emph{5) Resource-Efficient Operation:} Efficient use of sensor resources is crucial for practical implementation. Lessons learned underscore the need to balance the collection and processing of sensor data with the associated energy and computational costs, ensuring that sensing-assisted beam management remains viable for resource-constrained devices and networks. \emph{6) Integration of ML and Signal Processing:} Sensing-assisted beam management often involves the integration of ML and signal processing techniques to interpret sensor data and make informed decisions. Lessons learned highlight the value of combining these disciplines to extract actionable insights from sensor information. \emph{7) Security and Privacy Considerations:} Lessons learned emphasize the importance of addressing security and privacy concerns \cite{DCN-WANG2020281} associated with sensor data. Sensing-assisted systems must incorporate measures to protect sensitive information and ensure that sensor data is utilized in a secure and privacy-preserving manner. In summary, lessons learned from sensing-assisted beam management highlight the potential for leveraging sensor data to enhance environmental awareness, enable dynamic adaptation, and improve the robustness of beamforming strategies in wireless communication systems. By integrating sensor information effectively, sensing-assisted beam management can contribute to optimized performance and improved user experience in diverse operational scenarios.

{\bf{Lesson 3.}} Lessons learned about beam management in RIS-assisted systems include: \emph{1) Adaptive Beam Steering:} RIS-assisted systems can dynamically steer transmitted or reflected beams to adapt to changing channel conditions and user requirements. Lessons learned highlight the need for adaptive beam steering algorithms that leverage real-time feedback to optimize signal paths and coverage areas. \emph{2) Interference Mitigation:} Beam management in RIS-assisted systems aims to mitigate interference and enhance signal quality by intelligently manipulating signal reflections. Lessons learned underscore the potential for RIS-based interference mitigation techniques to improve spectral efficiency and overall network performance. \emph{3) CSI Utilization:} Effective beam management in RIS-assisted systems relies on accurate CSI to optimize beamforming and phase control. Lessons learned emphasize the value of robust CSI acquisition and utilization algorithms for informed beam management decisions. \emph{4) Energy-Efficient Beamforming:} Lessons learned stress the importance of developing energy-efficient beam management strategies in RIS-assisted systems to minimize power consumption while maintaining high-quality communication links. Optimizing beamforming for energy efficiency is a key consideration in RIS-enabled networks. \emph{5) ML and Optimization:} Leveraging ML and optimization algorithms plays a crucial role in intelligent beam management within RIS-assisted communication systems. Lessons learned underscore the potential for ML techniques to enhance beamforming performance through data-driven decision-making. \emph{6) Spectrum Efficiency and Capacity Optimization:} Beam management in RIS-assisted systems aims to improve spectrum efficiency and overall network capacity through intelligent signal reflection and steering. Lessons learned highlight the potential for RIS-enabled beam management to unlock additional capacity and enhance spectral utilization. \emph{7) Integration with Network Control:} Effective beam management in RIS-assisted systems requires seamless integration with network-level control and management functions. Lessons learned stress the importance of coordinating beam management decisions with broader network optimization objectives to achieve holistic performance improvements. \emph{8) Real-World Deployment Challenges:} Practical deployment of RIS-assisted systems necessitates addressing real-world challenges related to hardware complexity, calibration, and scalability. Lessons learned underscore the need to develop practical solutions that overcome deployment obstacles and ensure the viability of RIS-enabled beam management in diverse deployment scenarios. In summary, lessons learned about beam management in RIS-assisted communication systems highlight the potential for intelligent signal reflection, adaptive beam steering, interference mitigation, and energy-efficient optimization. By leveraging advanced beam management techniques within RIS-enabled networks, it becomes possible to achieve significant performance enhancements and address key communication challenges.

{\bf{Lesson 4.}} By effectively integrating these technologies, it is possible to create a holistic beam management system that leverages AI's decision-making capabilities, ISAC/sensing systems' environmental awareness, and RIS's reconfigurability to optimize beamforming and communication performance in dynamic and complex wireless environments. This integration can lead to enhanced spectral efficiency, improved coverage, and better quality of service. Here are some strategies for their integration: \emph{1) Joint Decision Making:} Establish a framework where AI algorithms, sensing systems, and RIS controllers work collaboratively to make joint decisions on beamforming and resource allocation. AI can analyze historical data and real-time feedback to optimize beamforming parameters, while sensing systems provide environmental information for adaptive beam steering, and RIS controllers adjust the reflective surfaces to optimize the propagation environment. \emph{2) Dynamic Adaptation:} Develop a system that allows for dynamic adaptation based on the changing communication environment. AI algorithms can continuously learn and adapt to the environment, sensing systems can provide real-time feedback on channel conditions, and RIS can dynamically adjust its configuration to optimize signal propagation. This dynamic adaptation ensures that the beam management system remains effective in different scenarios. \emph{3) Resource Allocation Optimization:} Utilize AI's optimization capabilities to efficiently allocate resources such as beam resource, spectrum resource, and power levels. Sensing-assisted information can provide valuable input for AI algorithms to make informed decisions, and RIS controllers can dynamically adjust the reflective properties to optimize resource allocation. \emph{4) Learning from Each Other:} Establish a learning framework where each component learns from the others. AI algorithms can learn from the feedback provided by sensing systems and RIS controllers, gaining insights into the impact of the environment on beamforming. Similarly, RIS controllers can adjust their configurations based on the optimized beamforming decisions made by AI algorithms.

\section{Challenges and Open Issues}
In order to efficiently implement 6G mmWave and THz communications with the importance of using AI, sensing (ISAC) and RIS, several potential technical concerns of beam management are discussed below.

\subsection{AI-Empowered 6G Framework}
While AI techniques in wireless networks are already being discussed in 5G, we anticipate that 6G deployments will be much denser in terms of the number of small BSs/APs and users, and more heterogeneous in terms of integration of different technologies and application characteristics, with stricter performance requirements compared to 5G. This is calling for deeper research on how AI can enhance 6G. Let's next identify some common and specific characteristics that are being considered.

{\bf{Collaborative Edge AI.}} Given the requirements of emerging 6G, collaborative edge AI is expected to be an indispensable component in 6G. By executing AI models directly at network edges, edge training and inference can provide AI services with low latency and high reliability by reducing computing, communication, storage, and engineering resources. The environment state in most real-world mobility applications, like traffic scenarios, is partially observable. Collaborative AI provides an opportunity to combine partial observations across multiple network edges or tasks to gather more information. The paradigms of collaborative AI, like communication or collaboration, can explicitly or implicitly utilize this opportunity to improve performance. To tame privacy leakages and adversarial attacks, various collaborative edge AI models and architectures have been proposed, such as FL (i.e., a server-client network architecture with data partitioning among edge devices) and split learning (i.e., an architecture with model parameters partitioned among edge servers and edge devices). By leveraging and learning the joint beam dynamics of multiple users, collaborative edge AI can help design flexible and low-latency beam management strategies to adapt to dynamic environmental fluctuations. This topic is still being explored. The optimal tradeoff between model efficiency and communication overhead is one of the challenges in implementing collaborative edge AI-based beam management that must be further studied.

{\bf{Sensing AI.}} One function of AI in wireless networks is to use sensors to sense and collect data from the physical environments, followed by smart data analytic using AI techniques. This function can be termed as sensing AI that well supports sensing tasks over wireless networks such as people and object detection, localization, and motion and activity recognition. Among them, localization aimed at determining the geographic coordinates of network nodes and objects is important for efficient beam management. Some important works are highlighted with the exploitation of sensing AI for localization tasks. Typically, a continuous wireless localization process can be formulated as an MDP, which can then be derived by applying DRL. By estimating the location of target users in the complex dynamic networks using sensing AI, fast beam tracking can be implemented.

{\bf{Model Generalization.}} Generalization is a term used to describe a model's ability to react to previously unseen, new data. The optimal beam angle for mmWave and THz communications highly depends on the propagation environment. Naturally, it is best for each node (i.e., BS or UE) to collect its own data for model training, which comprises the specific environmental features for optimizing the local model. Nevertheless, collecting huge amounts of training data is too costly or simply not possible. Fortunately, many of these environmental factors share similar properties (e.g., for the LOS path), while others such as scatterers vary with the environment. In these contexts, TL can be adopted for applying the knowledge extracted from one environment to another, so that exciting performance can be achieved even with only a limited training dataset. Up to now, there are only a few exploratory achievements in introducing TL into beam management process. As a further emerging technology, meta-learning \cite{Meta-Learning-2020}, or \emph{learning to learn}, can be adopted for further enhancing the model's generalization capability. Its potential in beam management has not yet been explored. With the ability to generalize, it would likely be very useful for reducing the burden of beam management in complex environments.

{\bf{Life Cycle Management.}} 3GPP TSG RAN is studying on AI/ML for NR air interface. One aspect worth exploring in the AI/ML framework is life cycle management (LCM) \cite{R1-2301586}. The necessity and general components in AI/ML LCM were discussed in 3GPP TSG RAN WG1 (RAN1) meetings. In RAN1\#110bis-e meeting, RAN1 made much progress on the different components of LCM, especially for model selection, activation, deactivation, switching, fallback, and model monitoring. In RAN1\#111 meeting, RAN1 decided to study both functionality-based and model-ID-based LCM. An AI/ML functionality refers to beam management, CSI compression, positioning, etc. In RAN1\#112 meeting, RAN1 discussed common and use case specific aspects in LCM, including data collection, model control and model monitoring, model ID, model identification, and model transfer/delivery. Typically, RAN1 also agreed to study model monitoring at least for the purposes of model control and model update and elaborated on how model monitoring interacts with model control for the use case of CSI compression and beam management. The AI/ML deployment and LCM process is an exciting journey filled with many subtleties and complexities. Despite the preliminary victories in standardization, LCM in AI/ML-empowered beam management has yet to be addressed in concrete scenarios, which is a topic worthy of attention in the future.

\subsection{ISAC-Enabled 6G Framework}
ISAC technology can realize blockage detection and user positioning with the help of sensing, and thus provides instructions for fast beam alignment and tracking. Several key challenges are as follows.

{\bf{Radar-Type ISAC Implementation.}} The ISAC systems integrating radar-type sensing may be divided into the following two categories. (1) Loose integration. C\&S functions are physically integrated in one system, where two sets of dedicated hardware components and/or two different signal waveforms overlapped or separated in time, frequency or spatial domains are used by them. (2) Tight integration. By sharing the majority of hardware, the C\&S functions are more firmly integrated and delivered by a common signal waveform. The first ISAC system can be implemented with simple modifications to the existing system infrastructure. The key to enabling the smooth operation of individually deployed C\&S functions is to develop effective interference management techniques. The second ISAC system reduces the equipment size, hardware cost and power consumption while improving the spectral efficiency. In such a system, flexible and reconfigurable C\&S hybrid signals are crucial to improve the efficiency of both communications and radar sensing. To this end, it is necessary to carry out ISAC signal design (e.g., design of waveform, frame structure and transmission mode), ISAC signal processing (single-node sensing and multi-node collaborative sensing), and ISAC signal optimization in the space-time-frequency domains.

{\bf{DMG Sensing.}} For the DMG sensing procedure defined by IEEE 802.11bf, which is still an ongoing project at the time of writing, there are various research directions that can be identified for further study, such as sensing security and privacy, multi-band cooperative sensing, sensing in spectrum sharing environments, integrating sensing and data transmissions. It appears to have no effect on beam training/tracking in the protocol specification, because they are relatively independent in the time domain. But in practice, from a beam management standpoint, the sensing measurement result (e.g., an R-D-A map), which provides an ``image'' of the surrounding environment, can help reduce the overhead associated with beam scanning and alignment. Work on this exciting topic remains blank. An example of future beam management aided by Wi-Fi sensing is to use the high-resolution sensing to solve the problem of misalignment of extremely narrow beams on the smartphone side due to tiny movements of the hand. The Wi-Fi sensing report for small motion could well guide beam selection in the high directional link, with the possibility of significantly improving robustness.

{\bf{Collaborative Sensing.}} Single-node sensing has a low-cost and simple structure, but the range of C\&S is limited, which makes it difficult to sense the NLoS targets. To make up for this deficiency, collaborative sensing \cite{Collaborative-Sensing-2022} has drawn increasing attention, which involves and collaborates with a multitude of sensors spatially scattered in a geographical area. It can be used for sensing the environment with obstacles or 3D positioning imaging, and then for real-time beam alignment and tracking. Obviously, leveraging multiple sensors to cover the same target point enhances the sensing robustness, but it has some problems such as high cost, complex structure and difficult data synchronization. Additionally, most researchers focus on homogenous and static sensors installed in specific locations. It is still an open issue how to collaborate multiple sensors, especially mobile and/or heterogeneous sensors, to perform sensing tasks with low time complexity. Heterogeneous sensors with distinct types, such as cameras and radar, for collaborative sensing, also can be regarded as \emph{multi-modal sensing}. Data fusion for original  multimodal information with dimension mismatch is critical for sensing data processing and interpretation. AI is the preferred technology for processing multimodal data to achieve target recognition and decision-making. Moreover, to meet the key requirements of personalization, data privacy and communication efficiency for intelligent sensing applications, \emph{federated sensing} \cite{Federated-Sensing-2021} based on FL comes into public view. In this context, more devoted efforts need to be made in collaborative sensing-aided beam management to obtain training-free beam prediction.

{\bf{Supported by RIS.}} RIS has shown great potential in improving beamforming gain and reducing interference in wireless communications. It is also expected to benefit the ISAC by providing better sensing coverage, accuracy and resolution. Although limited, the research on RIS-assisted ISAC is on the rise. RIS is not only conducive to improving the accuracy and resolution of sensing and positioning by providing an additional path to observe the target with LoS path from a different angle, but also contributes a feasible new solution to the challenging problem of target sensing without LoS connections. To fully reap these benefits, a key task will be to properly deploy RISs according to the beamforming requirements for C\&S, and optimize RIS phase shift matrix in real time by considering the sensing performance metrics and co-channel interference of C\&S. In return, an alternative approach can be provided for the thorny beam management problem posed by the passive nature of RIS by employing the sensing function to measure the link parameters related to communications. Research on this promising subject is sorely lacking.

\subsection{RIS-Enhanced 6G Framework}
The beam management for traditional one-hop transmission system is already a time-consuming and challenging task, and it will be even worse under the two-hop paradigm, which is unacceptable for 6G mmWave/THz communications, especially for mobile scenarios. Despite the initial attempts at two-hop beam management design, the cost is still overwhelming, which calls for designing more efficient mechanisms. Meanwhile, the investigation on the more challenging scenarios of RIS-assisted mmWave/THz systems is still limited.

{\bf{Support for Mobility.}} For RIS-assisted systems, the vast majority of current research is aimed at designing RIS passive beamforming to increase the achievable rate, while the research on beam management issues that greatly affects the performance of mmWave/THz communications is far from enough. The existing beam management mechanisms for conventional one-hop communication topology are difficult to directly extend to the new two-hop topology due to the passive of RIS. The lack of RF chains results in the inability of RIS to sense signal, which further complicates beam training for the paths assisted by RIS. Besides, to adapt to time-varying channels due to user mobility, RIS passive beam training should be jointly coordinated with the active beam training on the AP/BS and/or users. Although some beam management solutions can be used to support certain RIS beamforming performance for RIS-assisted mmWave/THz communications, they remain large channel estimation complexity and training overhead/delay that scale with the number of RIS reflecting elements and the BS/user antennas, making it hard for these systems to support mobile scenarios, especially high-mobility scenarios. As discussed before, having sensing capability at RIS may be a good option for beam tracking. With the sensing capability, RIS can obtain rich information about the position and mobility pattern of user as well as the environment layout, from which RIS can infer its future user-side beam from the current/previous beam sequence.

{\bf{Multi-Cell Multi-RIS.}} The mmWave/THz communication network featured by dense or even ultra-dense small cells is promoted as one of the technical trends to meet the high capacity requirements of future mobile networks. Besides, to fully tap the potential of RIS, a number of RISs can be deployed in the cell to provide more path diversity to bypass the dense and scattered obstacles in a complex environment. In this context, for beam management procedure, the user needs to jointly determine and coordinate the connected RIS(s) and AP(s)/BS(s), which is quite challenging. The limited beam management schemes for RIS-assisted systems usually focus on single cell scenarios, especially those with only one RIS, but rarely consider multi-cell scenarios with dense RISs. There is a hard way to go from single-cell single-RIS to future multi-cell multi-RIS beam management. Meanwhile, most of the existing beam management mechanisms focus on beam training and alignment when users move in a limited area, while few discuss \emph{beam switching} (or seamless \emph{handover}) in high mobility applications, where link blockage may occur frequently when user moves between different cells. When it comes to the RIS-assisted network, more frequent handovers between RISs in the same cell or between different cells are required to ensure the link quality of users in real time. The conventional handover mechanisms may not be suitable for the new network paradigm, and more flexible and effective mechanisms are needed to deal with multi-point beam handover.

{\bf{Multi-User and MIMO.}} In general, an AP/BS can serve multiple users simultaneously by allocating different time, frequency, and beam resources to different users. Most existing work on beam management for RIS-assisted systems focuses on the single-user scenario. It is important to extend that to general multi-user scenarios. Furthermore, future users in 6G mmWave/THz communication systems are most likely to be equipped with multiple antennas, but currently the majority of studies are on single-antenna users. Efficient beam management against unfavorable propagation characteristics for RIS-assisted multi-user multi-antenna scenario is very challenging, which is still an open problem that requires future research consideration. To serve multiple users simultaneously, one RIS reflecting beam with this ability may be designed, or the RIS can be divided into multiple sub-arrays to serve different users with different reflecting beams. In addition, despite the aforementioned new challenges of integrating multiple RISs to mmWave/THz communications, a notable advantage is that the path parameter estimation can be cross verified. For example, three accurate estimates of AoA/AoD, associated with other essential information, e.g., direction of arrays, can yield the location of user, and the location information will in turn reproduce the path parameters \cite{Joint-Beam-Training-Positioning-RIS-2021}. In this way, the path parameters of multi-RIS-assisted mmWave/THz MIMO can be enhanced according to their geometric relationship.

{\bf{Bring in Active Ability.}} The passivity of fully-passive RIS makes beam management process difficult to handle. Happily, to overcome the fundamental limitations posed by RIS passivity, different kinds of RISs are currently under research and design for wireless communication systems, such as semi-passive RIS composed of passive reflectors and active sensors \cite{Enabling-Large-RIS-DL-2021,Sensing-Aided-RIS-3GPP-5G,Tensor-Algebraic-Channel-Estimation}, relay-type RIS in which a few elements serve as active relays \cite{Hybrid-Relay-Reflecting-2022}, active RIS that actively reflect signals with amplification \cite{Active-RIS-vs-Passive-2023}, and reconfigurable distributed antennas and reflecting surfaces (RDARS) motivated by distributed antenna system \cite{RDARS-2023}. Unlike conventional fully-passive RISs, one common design philosophy for these RIS variants is to introduce some active features while maintaining low-cost and low-energy-consumption benefits. As such, a remarkable active gain to RIS-assisted systems can be provided. The active ability may also enable efficient beam management for the cascaded channel. For instance, with the proposed RDARS architecture, the passivity is less restrictive to beam management than with a fully-passive RIS, because beam training with RIS elements programmed in connected (i.e., active) mode can be readily used for that in reflection mode. In this regard, the potential of using the active ability of the newly designed RISs to deal with its passive beam management is worth exploring.

\subsection{THz Beam Management Towards 6G}
It must be pointed out that most existing beam management schemes are aimed at mmWave communication scenarios. Although THz beam management is similar to that of mmWave in terms of features, basics, techniques, and challenges, more in-depth and targeted research is necessary. It is because that, to compensate the higher propagation loss than mmWave, much narrower beams are required for THz communications to obtain more benefits from the beamforming gain, which further complicates the beam management process. The research on this topic is still in its infancy.

In addition to the open issues listed in subsections \emph{A-C}, THz beam management in 6G wireless networks also faces some unique issues. One of the reasons is that, compared with mmWaves, THz communication systems are more likely to use extremely large-scale massive MIMO and ultra-large-scale RIS, which incurs the {\bf{near-field effect}}, making beam management more challenging. For example, for the most commonly used codebook-based beam training in the near-field, both the direction and distance from the transmitter to the receiver need to be considered in the codebook design, which is different from that in the far-field, and the training overhead is extremely high. In practice, the near-field effect has a significant impact on the performance of beam management. Unfortunately, this important new problem has not been well studied in the literature. Most of the current contributions are based on the far-field, yet the far-field codebooks cannot match near-field channels. That is to say, the far-field beam training will cause serious performance loss of near-field THz communications in extremely large-scale massive MIMO or ultra-large-scale RIS-assisted systems. In this context, it is crucial to deeply consider near-field effects in THz beam management.

\section{Conclusions}
Key challenges for the feasibility of mmWave and THz communication systems are the rapid channel fluctuation and frequent deafness (beam misalignment) between the communication endpoints. In this regard, it is important to properly design efficient initial access and tracking strategies to periodically identify the optimal beam pair with which a transmitter (e.g., BS) and a receiver (e.g., mobile terminal) can communicate. The recent developments and future challenges in beam management algorithms have been investigated in this paper. Specifically, various state-of-the-art beam management technologies for AI-empowered, ISAC-enabled, and RIS-enhanced 6G mmWave and THz communication networks have been comprehensively surveyed.

We presented a classification of papers on AI-based beam management according to their training modes (independent or collaborative), AI paradigms (SL or RL), and AI models, and summarized the key contributions of these solutions. We broadly divided the existing beam management approaches in ISAC systems into four major categories based on the sensing technologies employed, namely, radar sensing, communication signal sensing, C\&S hybrid signal sensing, and hiring dedicated sensors. In highly mobile systems such as V2X, where the network topology and surroundings are time-varying, sensing-aided techniques are envisioned as most promising candidates to detect and track cars in real-time, so as to facilitate beam management process. In addition, we carefully discussed the extensibility and limitations of the existing beam management solutions of conventional mmWave and THz systems toward the RIS-assisted new paradigm, and concluded that the AI-driven and sensing-aided beam management frameworks will play an essential role for the implementation of RIS-enhanced networks. Finally, several potential technical concerns of future beam management were discussed. This survey may serve as an enlightening guideline for the research work of mmWave and THz communications towards 6G.

\section*{Common Abbreviation List}

\begin{description}
	\item[AoA] \qquad\, Angle of Arrival
	\item[AoD] \qquad\, Angle of Departure
	\item[AC] \qquad\, Actor-Critic
	\item[AI] \qquad\, Artificial Intelligence
	\item[AP] \qquad\, Access Point
	\item[BFT] \qquad\, Beamforming Training
	\item[BS] \qquad\, Base Stations
	\item[CNN] \qquad\, Convolutional Neural Network
	\item[CSI] \qquad\, Channel State Information
	\item[DDPG]  \qquad\, Deep Deterministic Policy Gradient
	\item[DL] \qquad\, Deep Learning
	\item[DMG] \qquad\, Directional Multi-Gigabit
	\item[DQN] \qquad\, Deep Q-Network
	\item[DRL] \qquad\, Deep Reinforcement Learning
	\item[FL] \qquad\, Federated Learning
	\item[IoT] \qquad\, Internet of Things
	\item[ISAC] \qquad\, Integrated Sensing and Communication
	\item[LoS] \qquad\, Line-of-Sight
	\item[LSTM] \qquad\, Long Short-Term Memory
	\item[mmWave] \qquad\, Millimeter wave
	\item[MAB] \qquad\, Multi-Armed Bandit
	\item[MAC] \qquad\, Medium Access Control
	\item[MDP] \qquad\, Markov Decision Process
	\item[MIMO] \qquad\, Multiple-Input Multiple-Output
	\item[MISO] \qquad\, Multiple-Input Single-Output
	\item[MU-MIMO] \qquad\, Multi-user MIMO
	\item[ML] \qquad\, Machine Learning
	\item[NLoS] \qquad\, Non-Line-of-Sight
	\item[NR] \qquad\, New Radio
	\item[OFDM] \qquad\, Orthogonal Frequency-Division Multiplexing
	\item[POMDP] \qquad\, Partially Observable MDP
	\item[RAN] \qquad\, Radio Access Network
	\item[RF] \qquad\, Radio Frequency
	\item[RIS] \qquad\, Reconfigurable Intelligent Surface
	\item[RL] \qquad\, Reinforcement Learning
	\item[RSU] \qquad\, Roadside Unit
	\item[SL] \qquad\, Supervised Learning
	\item[STA] \qquad\, Station
	\item[SU-MIMO] \qquad\, Single-user MIMO
	\item[SVM] \qquad\, Support Vector Machine
	\item[THz] \qquad\, Terahertz
	\item[TL] \qquad\, Transfer Learning
	\item[TRP] \qquad\, Transmission-Reception Point
	\item[UAV] \qquad\, Unmanned Aerial Vehicle
	\item[UE] \qquad\, User Equipment
	\item[URLLC] \qquad\, Ultra-Reliable Low-Latency Communication
	\item[V2I] \qquad\, Vehicle-to-infrastructure
	\item[V2X] \qquad\, Vehicle-to-everything
	\item[WLAN] \qquad\, Wireless Local Area Network
	\item[WPAN] \qquad\, Wireless Personal Area Network
\end{description}

\ifCLASSOPTIONcaptionsoff
  \newpage
\fi

\bibliographystyle{IEEEtran}
\bibliography{reference}

\begin{IEEEbiography}[{\includegraphics[width=1in,height=1.25in,clip,keepaspectratio]{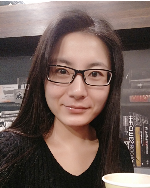}}]{Qing Xue} received the B.E. degree in communication engineering from University of Jinan in 2011 and the Ph.D. degree in information and communication engineering from Southwest Jiaotong University in 2018. She joined the School of Communications and Information Engineering, Chongqing University of Posts and Telecommunications, in 2018 as a Lecturer. From Dec. 2019 to Jan. 2024, she was a post-doctoral fellow with the National Key Laboratory of Wireless Communications, University of Electronic Science and Technology of China. From Dec. 2021 to Nov. 2023, she was also a post-doctoral fellow with the State Key Laboratory of Internet of Things for Smart City, University of Macau, under the Macao Young Scholars Program. Her research interests include millimeter wave communications, intelligent wireless networking, and resource management in mobile networks.
\end{IEEEbiography}
\begin{IEEEbiography}[{\includegraphics[width=1in,height=1.25in,clip,keepaspectratio]{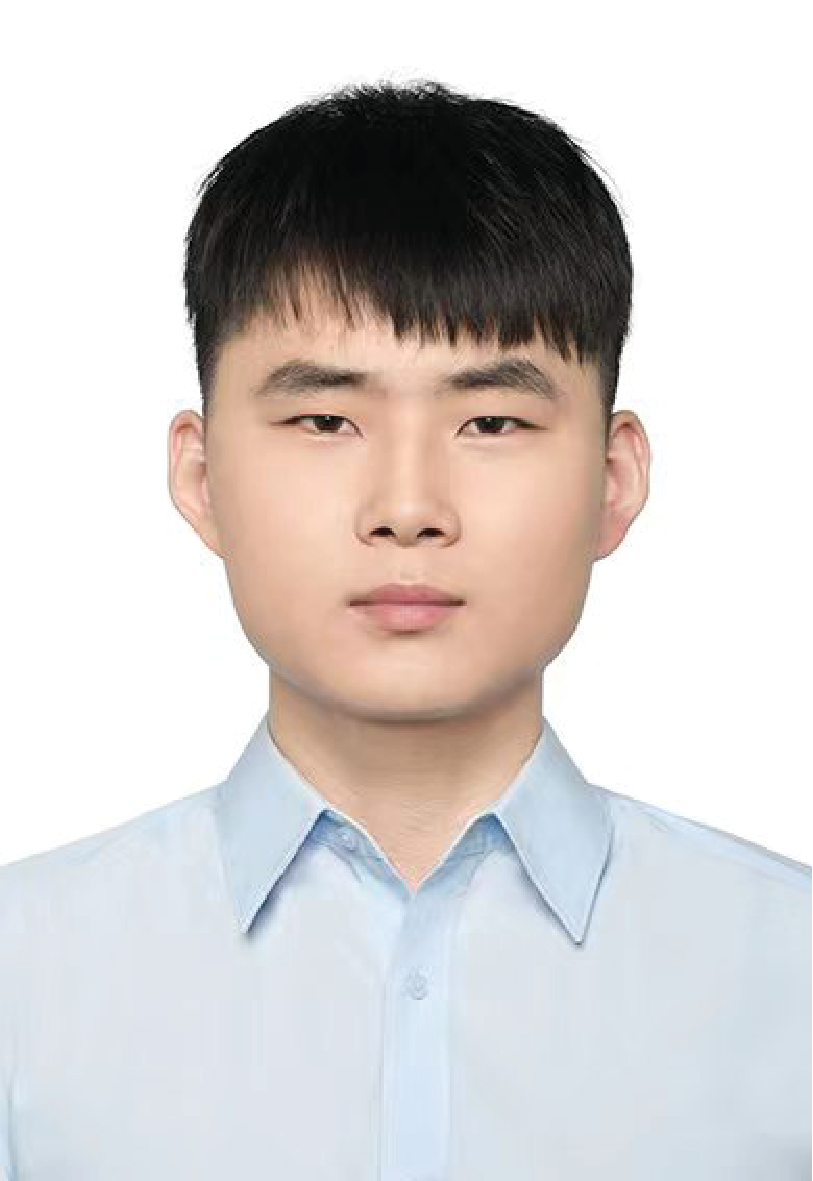}}]{Chengwang Ji} received the B.S. degree in communication engineering from Jilin University, Changchun, China, in 2022. He is currently pursuing the Ph.D. degree with the State Key Laboratory of Internet of Things for Smart City, Department of Electrical and Computer Engineering, University of Macau, Macau, China. His main research interests include RIS, mmWave communication, beam management, federated learning.
\end{IEEEbiography}
\begin{IEEEbiography}[{\includegraphics[width=1in,height=1.25in,clip,keepaspectratio]{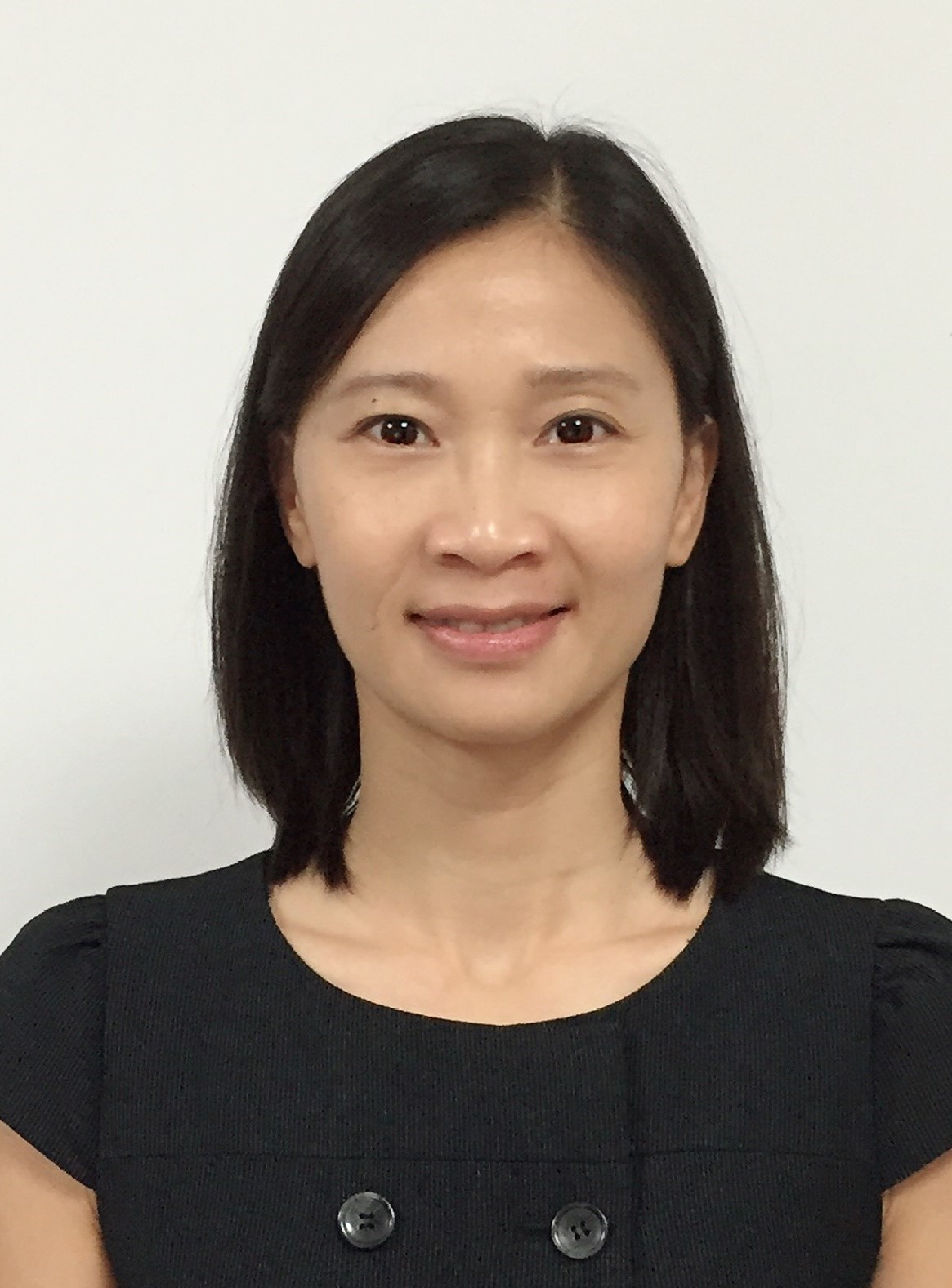}}]{Shaodan Ma} received the double Bachelor's degrees in science and economics and the M.Eng. degree in electronic engineering from Nankai University, Tianjin, China, in 1999 and 2002, respectively, and the Ph.D. degree in electrical and electronic engineering from The University of Hong Kong, Hong Kong, in 2006. From 2006 to 2011, she was a post-doctoral fellow at The University of Hong Kong. Since August 2011, she has been with the University of Macau, where she is currently a Professor. Her research interests include array signal processing, transceiver design, localization, integrated sensing and communication, mmwave communications, massive MIMO, and machine learning for communications. She was a symposium co-chair for various conferences including IEEE ICC 2021, 2019 \& 2016, IEEE GLOBECOM 2016, IEEE/CIC ICCC 2019, etc. She has served as an Editor for IEEE Transactions on Wireless Communications (2018-2023), IEEE Transactions on Communications (2018-2023), IEEE Wireless Communications Letters (2017-2022), IEEE Communications Letters (2023), and Journal of Communications and Information Networks (2021-present).
\end{IEEEbiography}
\begin{IEEEbiography}[{\includegraphics[width=1in,height=1.25in,clip,keepaspectratio]{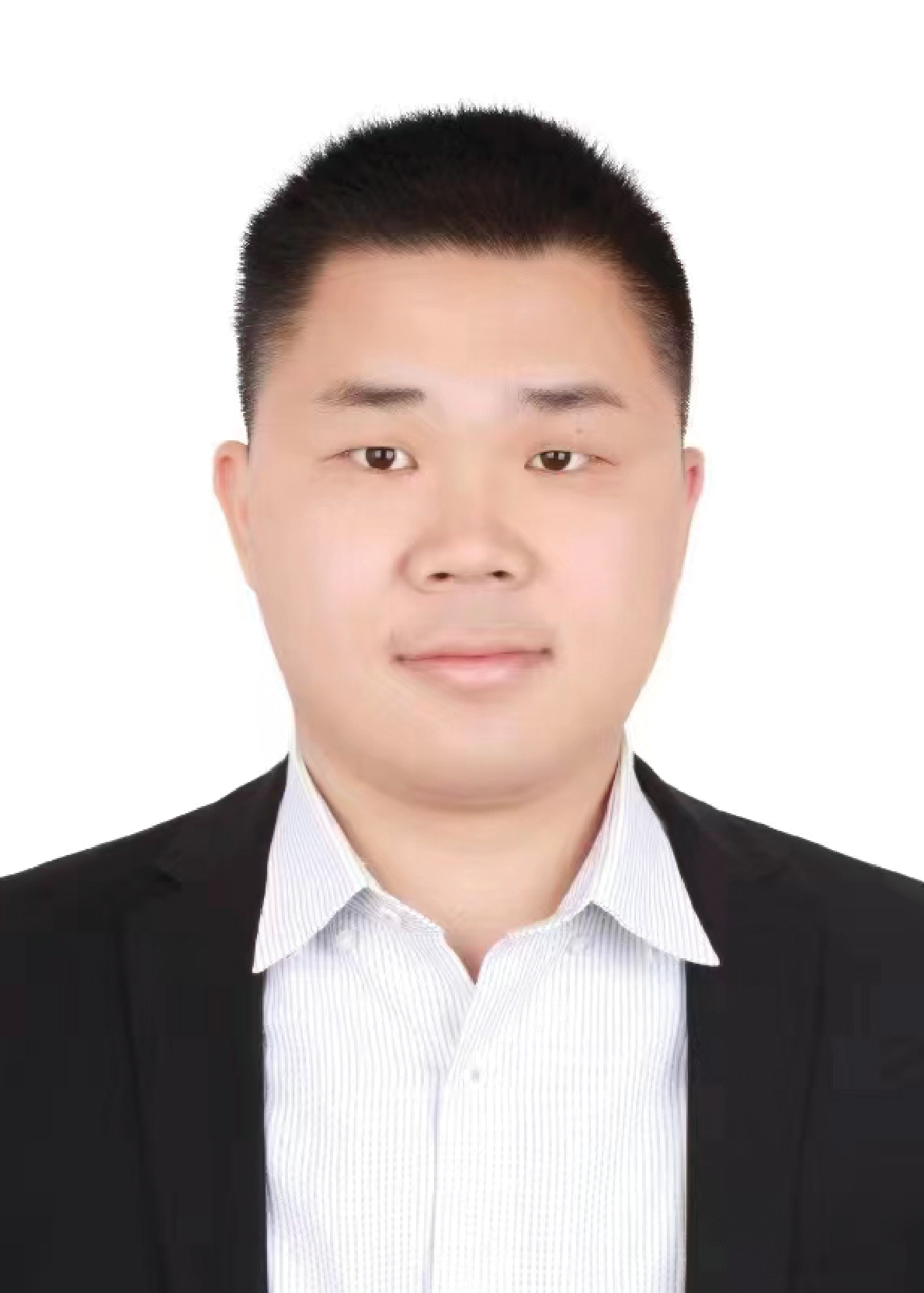}}]{Jiajia Guo (S'21-M'23)} received the B.S. degree from the Nanjing University of Science and Technology, Nanjing, China, in 2016, the M.S. degree from the University of Science and Technology of China, Hefei, China, in 2019, and the Ph.D. degree in information and communications engineering from Southeast University, Nanjing, China, in 2023. His current research interests include AI-native air interface, reconfigurable intelligent surfaces, and massive MIMO.	
\end{IEEEbiography}
\begin{IEEEbiography}[{\includegraphics[width=1in,height=1.25in,clip,keepaspectratio]{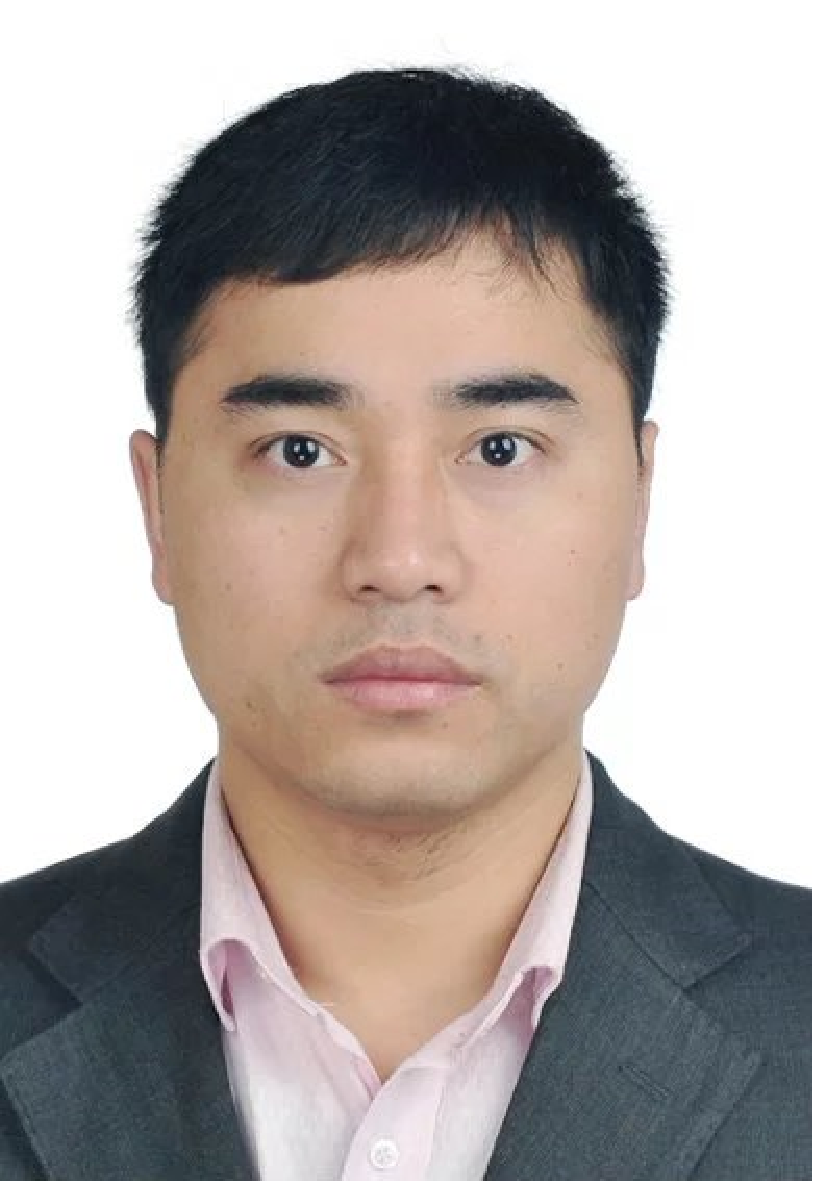}}]{Yongjun Xu} received the Ph.D. degree (Hons.) in communication and information system from Jilin University, Changchun, China, in 2015. He is an Associate Professor in Chongqing University of Posts and Telecommunications, Chongqing, China. He is also a Bayu Youth Scholar of Chongqing. From December 2018 to December 2019, He was a Visiting Scholar in the Utah State University, Logan, UT, USA. His recent interests include C-V2X, heterogeneous networks, resource allocation, intelligent reflecting surface, energy harvesting, backscatter communications, etc. He has published more than 100 IEEE Journal/Conference papers and received the Outstanding Doctoral Thesis of Jilin Province in 2016. He serves as an editor for Physical Communication, EURASIP Journal on Wireless Communications and Networking, Digital Communications and Networks, Chongqing YouDian Daxue XueBao, and also a reviewer for IEEE Transactions on Wireless Communications, IEEE Transactions on Industrial Electronics, IEEE Transactions on Communications, IEEE Transactions on Vehicular Technology, IEEE Communications Letters, etc. He serves on the Chair of ICCT 2023, ICCC 2020, WCSP 2021, and also a TPC member of many IEEE international conferences, such as Globecom, ICC, ICCT, WCNC, ICCC, and CITS.
\end{IEEEbiography}
\begin{IEEEbiography}[{\includegraphics[width=1in,height=1.25in,clip,keepaspectratio]{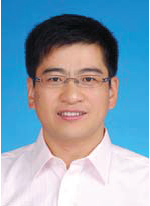}}]{Qianbin Chen} received the B.S. degree from Sichuan University, Chengdu, China, in 1988, and the Ph.D. degree in electrical engineering from the University of Electronic Science and Technology of China, Chengdu, in 2006. He joined the School of Communications and Information Engineering, Chongqing University of Posts and Technology, Chongqing, China, where he is currently a Professor. He has been working in the areas of wireless and mobile networking for more than 20 years. He has authored more than 120 international journals and conference articles. His research interests include wireless communication, network theory, and multi-media technology.	
\end{IEEEbiography}
\begin{IEEEbiography}[{\includegraphics[width=1in,height=1.25in,clip,keepaspectratio]{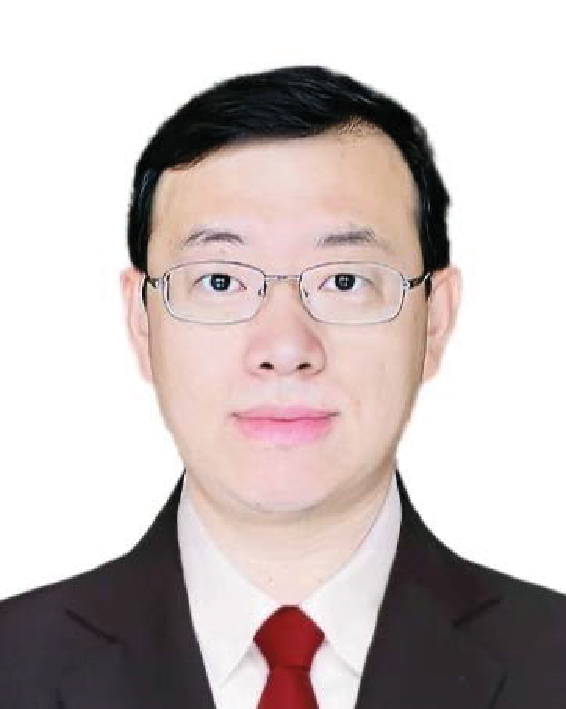}}]{Wei Zhang (S'01-M'06-SM'11-F'15)} received the Ph.D. degree from The Chinese University of Hong Kong in 2005. Currently, he is a Professor at the School of Electrical Engineering and Telecommunications, the University of New South Wales, Sydney, Australia. His current research interests include 6G communications and networks. He was elevated to Fellow of the IEEE in 2015 and was an IEEE ComSoc Distinguished Lecturer in 2016-2017. He is Vice President of IEEE Communications Society.

Within the IEEE ComSoc, he has taken many leadership positions including Member-at-Large on the Board of Governors (2018-2020), Chair of Wireless Communications Technical Committee (2019-2020), Vice Director of Asia Pacific Board (2016-2021), Editor-in-Chief of IEEE Wireless Communications Letters (2016-2019), Technical Program Committee Chair of APCC 2017, ICCC 2019, ICCC 2024 and WCNC 2025, Award Committee Chair of Asia Pacific Board and Award Committee Chair of Technical Committee on Cognitive Networks. Currently, he serves as an Area Editor of the IEEE Transactions on Wireless Communications and the Editor-in-Chief of Journal of Communications and Information Networks. Previously, he served as Editor of IEEE Transactions on Communications, IEEE Transactions on Wireless Communications, IEEE Transactions on Cognitive Communications and Networking, and IEEE Journal on Selected Areas in Communications – Cognitive Radio Series.
\end{IEEEbiography}
\end{document}